\documentclass[seceqn]{elsart}
\journal{Nuclear Physics A}

\usepackage{graphicx}
\usepackage{amsmath}
\usepackage{amssymb}
\usepackage{mathrsfs}
\usepackage{closedcases}
\usepackage{citesort}

\renewcommand{\arraystretch}{1}
\newenvironment{tablenotes}{\par\footnotesize}{}

\newcommand{\<}[1]{\hspace{-0.11111em}#1\hspace{-0.11111em}}
\newcommand{\rtrim}[1]{#1\hspace{-0.11111em}}

\DeclareRobustCommand{\grp}[1]{\mathrm{#1}}

\DeclareRobustCommand{\symbolopensquare}{\resizebox{0.7em}{!}{$\square$}}

\DeclareRobustCommand{\symbolopencircle}{\raisebox{-0.3ex}{\resizebox{0.9em}{!}{$\circ$}}}

\DeclareRobustCommand{\Lhat}{\hat{L}}
\DeclareRobustCommand{\Lambdahat}{\hat{\Lambda}}
\DeclareRobustCommand{\Lambdahatbf}{{\hat{\Lambda}_{BF}}}
\DeclareRobustCommand{\jhat}{\hat{j}}
\DeclareRobustCommand{\Sigmahat}{\hat{\Sigma}}
\DeclareRobustCommand{\Jhat}{\hat{J}}
\DeclareRobustCommand{\Qhat}{\hat{Q}}
\DeclareRobustCommand{\Galphaalpha}[1]{(\alpha\times\alpha)^{(#1)}}
\DeclareRobustCommand{\Galphapi}[1]{(\alpha\times\tilde\pi)^{(#1)}}
\DeclareRobustCommand{\Gaa}[1]{(a^\dagger\times\tilde{a})^{(#1)}}
\DeclareRobustCommand{\Gdd}[1]{(d^\dagger\times\tilde{d})^{(#1)}}
\DeclareRobustCommand{\Gsdds}[1]{(s^\dagger\times\tilde{d}+d^\dagger\times\tilde{s})^{(#1)}}

\DeclareRobustCommand{\grpu}[1]{\grp{U}(#1)}
\DeclareRobustCommand{\grpsu}[1]{\grp{SU}(#1)}
\DeclareRobustCommand{\grpsp}[1]{\grp{Sp}(#1)}

\DeclareRobustCommand{\grpso}[1]{\grp{SO}(#1)}
\DeclareRobustCommand{\grposp}[1]{\grp{OSp}(#1)}
\DeclareRobustCommand{\grpspin}[1]{\grp{Spin}(#1)}
\DeclareRobustCommand{\grpe}[1]{\grp{E}(#1)}

\DeclareRobustCommand{\grpspinb}[1]{\grp{Spin}_{B}(#1)}
\DeclareRobustCommand{\grpspinf}[1]{\grp{Spin}_{F}(#1)}
\DeclareRobustCommand{\grpspinbf}[1]{\grp{Spin}_{BF}(#1)}

\DeclareRobustCommand{\grpub}[1]{\grp{U}_{B}(#1)}

\DeclareRobustCommand{\grpufbrack}[1]{\grp{U}_{F}[#1]}
\DeclareRobustCommand{\grpubrack}[1]{\grp{U}[#1]}

\DeclareRobustCommand{\nut}{\nu_\triangle}
\DeclareRobustCommand{\nutt}{\tilde\nu_\triangle}
\DeclareRobustCommand{\nuttp}{\tilde\nu_\triangle'}

\DeclareRobustCommand{\sixj}[6]{
\left\lbrace\begin{array}{ccc}#1&#2&#3\\#4&#5&#6\end{array}\right\rbrace
}
\DeclareRobustCommand{\ISF}[6]{
\left(\begin{array}{cc|c}#1&#2&#3\\#4&#5&#6\end{array}\right)
}
\DeclareRobustCommand{\ISFBF}[4]{\ISF{(#1,0)}{(\frac12,\frac12)}{(#3,\frac12)}{#2}{\frac32}{#4}}

\begin{document}
%bibliographystyle{apsrevm}

\begin{frontmatter}

\title{Analytic descriptions for transitional nuclei near the critical point}

\author{M. A. Caprio\hspace{-2pt}\corauthref{cor}}\hspace{-1.0ex}  
% spacing compensation is for undesired spacing added before \corauthref
% and by intervening commands before \and
\corauth[cor]{Corresponding author.}
\ead{mark.caprio@yale.edu}
\and
\author{F. Iachello}
\address{Center for Theoretical Physics, Sloane Physics Laboratory, 
Yale University, New Haven, Connecticut 06520-8120, USA}

\begin{abstract}
Exact solutions of the Bohr Hamiltonian with a five-dimensional square
well potential, in isolation or coupled to a fermion by the
five-dimensional spin-orbit interaction, are considered as examples of
a new class of dynamical symmetry or Bose-Fermi dynamical symmetry.
The solutions provide baselines for experimental studies of even-even
[$\grpe{5}$] and odd-mass [$\grpe{5|4}$] nuclei near the critical
point of the spherical to deformed $\gamma$-unstable phase transition.
\end{abstract}

% suppress "Key words:" heading
\makeatletter
\def\@keywordheading{}
\makeatother  


\end{frontmatter}

\section{Introduction}
\label{secintro}

Dynamical symmetries have played an important role in the spectroscopy
of nuclei, in particular in the description of their collective
properties.  A dynamical
symmetry~\cite{iachello1979:dynsymm,iachello2006:liealg} occurs when
the Hamiltonian is constructed from the Casimir operators of the Lie
algebras in a subalgebra chain ($G\supset G' \supset
G''\supset\cdots$).  Dynamical symmetries yield good quantum numbers,
labeling the irreducible representations of $G\supset G' \supset
G''\supset\cdots$, and make possible analytic solutions for the
eigenvalues, eigenstates, and related observables.  They thereby allow
simple and straightforward comparison with experiment.  Most
applications of dynamical symmetries have been in the context of
algebraic models, such as the interacting boson model (IBM) for
nuclei~\cite{iachello1987:ibm}.  The algebraic structure of this model
is based upon $\grpu{6}$, and three dynamical symmetries arise,
involving the subalgebras $\grpu{5}$, $\grpso{6}$, and $\grpsu{3}$.
These symmetries can be related to the geometric description of
nuclei~\cite{bohr1952:vibcoupling,bohr1998:v2,eisenberg1987:v1},
through the geometry $\grpu{6}/[\grpu{5}\<\otimes\grpu{1}]$ associated
with
$\grpu{6}$~\cite{ginocchio1980:ibm-classical,dieperink1980:ibm-classical,bohr1980:ibm-coherent}.
They are seen to describe \textit{limiting} cases of geometric
structure involving the quadrupole degree of freedom: spherical
oscillator [$\grpu{5}$], $\gamma$-soft rotor [$\grpso{6}$], and
axially symmetric rotor [$\grpsu{3}$] structure.  However, the
\textit{intermediate} situations between the structural limits are
often of greatest interest, both for applications to actual
transitional
nuclei~\cite{scholten1978:ibm-u5-su3,casten1998:152sm-beta} and in the
study of phase transitions between the structural
limits~\cite{feng1981:ibm-phase}.  For these situations, results can
usually only be obtained by numerical diagonalization of the
Hamiltonian.

Recently, it has been suggested~\cite{iachello2000:e5} that there is a
class of dynamical symmetry which can be useful in the analysis of
experimental data near the critical point of a phase transition.
These symmetries cannot easily be formulated in algebraic models but
arise naturally in models based upon the consideration of differential
equations, such as the geometric description of nuclei with the Bohr
Hamiltonian~\cite{bohr1952:vibcoupling,bohr1998:v2,eisenberg1987:v1}.
These symmetries are defined as dynamical symmetries in the
traditional sense~\cite{iachello1979:dynsymm,iachello2006:liealg}, in
that the Hamiltonian is written in terms of the Casimir operators of a
chain of Lie algebras $G\supset G' \supset G''\supset\cdots$, but the
operators are now differential operators defined on a coordinate space
restricted to a \textit{bounded} domain.

A simple application of this approach is to the spectroscopy of nuclei
near the critical point of the second order phase transition between
the $\grpu{5}$ and $\grpso{6}$ structural limits.  The resulting
description, denoted $\grpe{5}$, is simply the solution of the Bohr
Hamiltonian for a five-dimensional square well
potential~\cite{wilets1956:oscillations}, which is an idealized form
of the potential near the critical point of the phase
transition~\cite{iachello2000:e5}.  The first part of this article
(after a brief summary of some prerequisite results for the
five-dimensional square well potential) describes the algebraic
properties of the $\grpe{5}$ dynamical symmetry and extends the
calculations for electromagnetic matrix elements beyond the basic
results reported in Ref.~\cite{iachello2000:e5}. The calculations
include electromagnetic operators involving higher order terms and
make use of a recent construction of the angular wave functions
(hyperspherical harmonics) in terms of orthogonalized products of
generating
functions~\cite{rowe2004:tractable-collective,rowe2004:spherical-harmonics}.

While dynamical symmetries in their simplest form involve purely
bosonic or purely fermionic realizations of Lie algebras, dynamical
symmetries can also be formulated for mixed bosonic and fermionic
systems.  Bose-Fermi dynamical
symmetries~\cite{iachello1980:supersymm} have mostly been applied
within the framework of the interacting boson fermion model (IBFM) for
nuclei~\cite{iachello1991:ibfm}. For a nucleus consisting of a core
coupled to a fermion distributed among orbitals with angular momenta
$j$, the algebraic structure is based on the Lie algebra
$\grpub{6}\<\otimes\grpufbrack{\sum_j(2j+1)}$, which can then be
embedded in the superalgebra
$\grpubrack{6|\sum_j(2j+1)}$~\cite{balantekin1981:ibfm-u64-os-pt}.
The dynamical Bose-Fermi symmetries of the IBFM have been very useful
in the description of odd-mass nuclei, but the symmetries have only
been applicable to nuclei at structural limits.  For odd-mass nuclei,
even more than for even-even nuclei, intermediate situations between
the structural limits are of greatest interest, both for application
to specific nuclei and for the study of the influence of coupled
fermions on nuclear phase transitions.

In the second part of this article, the dynamical symmetry approach to
describing structure near the critical point of a phase transition is
extended to the odd-mass Bose-Fermi system.  The simplest case of such
a Bose-Fermi dynamical symmetry arises for a particle with $j\<=3/2$
coupled to the quadrupole degrees of freedom by a five dimensional
spin-orbit interaction, leading to an $\grpe{5}\otimes\grpu{4}$
algebraic structure.  Here we discuss the solution of the Bohr
Hamiltonian with a square well potential in the presence of such a
coupling.  The calculation of electromagnetic observables is
considered in detail.  Preliminary results were reported in
Ref.~\cite{iachello2005:geomsuper}, where the notation $\grpe{5|4}$
was adopted for the model.  A byproduct of the calculation is a
systematic method for construction of the necessary
$\grpspin{5}\supset\grpspin{3}$ isoscalar factors, relevant not only
to the situation discussed here but to all problems involving
$\grpspin{5}$ coupling schemes.

The $\grpe{5}$ model has provided the basis for both experimental
investigations and further theoretical
developments~\cite{casten2000:134ba-e5,arias2001:134ba-e5,frank2001:104ru-e5,zhang2001:gcm-random,zamfir2002:102pd-beta,zhang2002:108pd-e5,arias2003:quartic-e5,bonatsos2004:e5-monomial,marginean2006:58cr-transfer-e5,fossion2006:hfb-pt-e5-x5}.
Preliminary experimental and theoretical studies involving $\grpe{5|4}$ have
also been carried
out~\cite{alonso2005:ibfm-gsoft-phase,fetea2006:135ba-beta-e54}.  This
article is meant primarily to provide a baseline for further study,
both experimental and theoretical, of nuclei near the critical point
of the spherical to deformed $\gamma$-unstable phase transition.

\section{Transitional symmetry for the $\gamma$-independent potential}
\label{sece5}

\subsection{Hamiltonian and solution}
\label{sece5soln}

We begin by considering an analytic solution
describing the structure of even-even nuclei near the critical point
of the transition between spherical and deformed $\gamma$-soft
structure.  The resulting $\grpe{5}$ model provides a benchmark for
the basic features of nuclear structure soft to both $\gamma$ and
$\beta$ deformation.  It was presented in part in
Ref.~\cite{iachello2000:e5}.  We consider it here in further detail
both since it provides a basis for comparison with experimental data
(\textit{e.g.},
Refs.~\cite{casten2000:134ba-e5,arias2001:134ba-e5,frank2001:104ru-e5,zamfir2002:102pd-beta,zhang2002:108pd-e5})
and since it provides the foundation for the description of
transitional odd-mass nuclei considered in Sec.~\ref{sece54}.

For this analysis, we consider the Bohr
Hamiltonian~\cite{bohr1952:vibcoupling,bohr1998:v2,eisenberg1987:v1},
\begin{equation}
\label{eqnHppv}
H = \frac{\hbar^2}{2B} \tilde\pi \cdot \tilde\pi + V(\alpha).
\end{equation}
Here $\alpha_\mu$ ($\mu\<=0,\pm1,\pm2$) are the quadrupole
deformation coordinates, $\pi_\mu$ are the conjugate momenta,
and
standard spherical tensor notation~\cite{deshalit1963:shell} is used, with $\tilde{T}^{(\lambda)}_\mu\<\equiv(-)^{\lambda-\mu}T^{(\lambda)}_{-\mu}$.
The quadrupole coordinates may be written in terms of the intrinsic
deformation variables $\beta$ and $\gamma$ and the Euler angles
$(\vartheta_1,\vartheta_2,\vartheta_3)$ as
\begin{equation}
\label{eqnalpha}
\alpha_\mu =
\beta\bigl[\mathscr{D}^{(2)}_{\mu0}(\vartheta)\cos\gamma
+\frac{1}{\sqrt{2}}[\mathscr{D}^{(2)}_{\mu2}(\vartheta)+\mathscr{D}^{(2)}_{\mu-2}(\vartheta)]\sin\gamma\bigr]
.
\end{equation}
The Hamiltonian, reexpressed in terms of these variables, is
\begin{multline}
\label{eqnH}
H=- \frac{\hbar^2}{2B} \Biggl[
\frac{1}{\beta^4}
\frac{\partial}{\partial \beta}
\beta^4  \frac{\partial}{\partial \beta}
+
\frac{1}{\beta^2}
\left(
\frac{1}{\sin 3\gamma} 
\frac{\partial}{\partial \gamma} \sin 3\gamma \frac{\partial}{\partial \gamma}
 - \frac{1}{4}
\sum_\kappa \frac{\hat{L}_\kappa^{\prime2}}{\sin^2(\gamma -
\frac{2}{3} \pi \kappa
)}
\right)
\Biggr]
\\
+V(\beta,\gamma),
\end{multline}
where the $\hat{L}_\kappa^\prime$ are the intrinsic frame angular
momentum operators.  We consider here specifically the case in which
the potential is $\gamma$-independent, \textit{i.e.},
$V(\beta,\gamma)\<=V(\beta)$.  A $\gamma$-independent potential
suffices to describe spherical oscillator structure, deformed
$\gamma$-soft structure~\cite{wilets1956:oscillations}, and the
transition between them.  This range of structural possibilities
corresponds to the $\grpu{5}$--$\grpso{6}$ transition in the
interacting boson model (IBM)~\cite{arima1976:ibm-u5}.

The Hamiltonian~(\ref{eqnH}) with $\gamma$-independent potential is
invariant under rotations in the five-dimensional space of the
coordinates $\alpha_\mu$.  The resulting $\grpso{5}$ symmetry yields the
five-dimensional analogue of the central force problem. A
separation of ``radial'' ($\beta$) and ``angular'' ($\gamma$ and
$\vartheta$) variables can be carried out in the standard
way~\cite{wilets1956:oscillations,rakavy1957:gsoft}.  The
eigenfunctions are of the form
\begin{equation}
\Phi(\beta,\gamma,\vartheta)=f(\beta)\Psi(\gamma,\vartheta).
\end{equation}
The angular and radial factors satisfy the equations
\begin{gather}
\label{eqnangular}
\Biggl[
-\frac{1}{\sin 3\gamma} 
\frac{\partial}{\partial \gamma} \sin 3\gamma \frac{\partial}{\partial \gamma}
+ \frac{1}{4}
\sum_\kappa \frac{\hat{L}_\kappa^{\prime2}}{\sin^2(\gamma -
\frac{2}{3} \pi \kappa
)}
\Biggr]
\Psi(\gamma,\vartheta)
=
\Lambda \Psi(\gamma,\vartheta)\\
\intertext{and}
\label{eqnradial}
\Biggl[
-\frac{1}{\beta^4}
\frac{\partial}{\partial \beta}
\beta^4  \frac{\partial}{\partial \beta}
+
\frac{\Lambda}{\beta^2}
+V(\beta)
\Biggr]
f(\beta)
=\varepsilon f(\beta),
\end{gather}
related by the separation constant $\Lambda$.  [Here, and for the
remainder of this article, we set $\hbar^2/(2B)\<=1$.  This can always
be achieved by transformation to the reduced eigenvalues
$\varepsilon'=2B\varepsilon/\hbar^2$ and reduced potential
$V'(\beta)=2BV(\beta)/\hbar^2$.]

The solutions for the ``angular'' wave functions,
$\Psi_{\tau\nutt{}LM_L}(\gamma,\vartheta)$, are common to all
$\gamma$-soft problems and are well
known~\cite{bes1959:gamma,rowe2004:tractable-collective,rowe2004:spherical-harmonics}.
They can be constructed explicitly as functions of the form
\begin{equation}
\label{eqnPsiDseries}
\Psi_{\tau\nutt{}LM_L}(\gamma,\vartheta)=\sum_{\substack{K=0\\\text{even}}}^L
F_{\tau\nutt{}LK}(\gamma)\phi^L_{M_LK}(\vartheta),
\end{equation}
where
$\phi^L_{MK}(\vartheta)\<\equiv[(2L+1)/(16\pi^2(1+\delta_K))]^{1/2}[\mathscr{D}^{(L)}_{MK}(\vartheta)+(-)^L\mathscr{D}^{(L)}_{M-K}(\vartheta)]$
is a symmetrized, normalized combination of $\mathscr{D}$ functions,
and the $F_{\tau\nutt{}LK}(\gamma)$ are polynomials involving
trigonometric functions of
$\gamma$~\cite{rowe2004:spherical-harmonics}.
Their eigenvalues in~(\ref{eqnangular}) are $\Lambda\<=\tau(\tau+3)$, for
$\tau\<=0,1,\ldots$.  (The quantum numbers are defined more fully below.)

The radial equation (\ref{eqnradial}) posesses analytic solutions (or
dynamical symmetries) only for a limited set of potentials.  It is
equivalent to the usual one-dimensional Schr\"odinger equation with an
effective radial potential term $\rtrim \propto 1/\beta^2$.  This
can easily be seen by 
transformation to the ``auxiliary'' radial wave function
$\varphi(\beta)\<\equiv\beta^{-2}f(\beta)$, which satisfies
\begin{equation}
\label{eqnradialse}
\left[-\frac{\partial^2}{\partial\beta^2}
+\frac{\Lambda+2}{\beta^2}+V(\beta)\right]\varphi(\beta)=\varepsilon\varphi(\beta).
\end{equation}
This equation posesses analytic solutions~\cite{fluegge1971:qm} for
only two classes of potential, the generalized oscillator potential
\begin{equation}
\label{eqndavidson}
V(\beta)=a\beta^2+b\beta^{-2},
\end{equation}
which has been used in a study of the $\grpu{5}$--$\grpso{6}$
transition by Elliott, Evans, and
Park~\cite{elliott1986:gsoft-davidson}, and the generalized Coulomb
potential
\begin{equation}
\label{eqnkratzer}
V(\beta)=a\beta^{-1}+b\beta^{-2}.
\end{equation}

The free potential, $V(\beta)\<=0$, occurs as a limiting case
($a\<=b\<=0$) of either potential~(\ref{eqndavidson})
or~(\ref{eqnkratzer}).  The free potential, though exceedingly simple,
is of special interest for the description of structure near the
critical point between spherical and deformed structure.  A potential
$V(\beta)$ which is flat with respect to $\beta$ allows
the nucleus to assume either a spherical ($\beta\<=0$) or deformed
($\beta\<>0$) shape with minimal energy penalty.  An idealized
approximation near the critical point is thus the five-dimensional infinite square well
potential~\cite{wilets1956:oscillations},
\begin{equation}
\label{eqnsquarewell}
V(\beta)=\begin{cases}0&\beta\leq\beta_W\\\infty&\beta>\beta_W ,\end{cases}
\end{equation}
which is just the free potential on the bounded domain
$\beta\<\leq\beta_W$.  Solution proceeds exactly as for the classic
three-dimensional spherical well problem~\cite{fluegge1971:qm}.  The
radial equation~(\ref{eqnradial}) or~(\ref{eqnradialse}) is, for
$V(\beta)\<=0$, equivalent to the Bessel equation (see
Refs.~\cite{fluegge1971:qm,iachello2000:e5} for details).  It is
satisfied by functions
$f(\beta)\<\propto\beta^{-3/2}J_\nu(\varepsilon^{1/2}\beta)$, with
Bessel function order
\begin{equation}
\label{eqnnu}
\nu=(\Lambda+9/4)^{1/2},
\end{equation}
or, in terms of $\tau$, $\nu(\tau)\<=\tau+3/2$.
The boundary condition arising from the well wall
[$f(\beta_W)\<=0$] restricts the eigenvalue to
\begin{equation}
\label{eqneigen}
\varepsilon_{\xi\tau}=\left(\frac{x_{\nu(\tau),\xi}}{\beta_W}\right)^2,
\end{equation}
where $x_{\nu,\xi}$ is the $\xi$th zero of
$J_\nu$~\cite{abramowitz1965}.  The resulting radial eigenfunctions
are
\begin{equation}
\label{eqnf}
f_{\xi\tau}(\beta)=C_{\xi\tau}\beta^{-3/2}J_{\nu(\tau)}\left(\frac{x_{\nu(\tau),\xi}\beta}{\beta_W}
\right).
\end{equation}
The normalization integral with respect to the metric
$\beta^4d\beta$~\cite{bohr1998:v2} can be evaluated
explicitly~\cite{gradshteyn1994:table}, giving
\begin{equation}
C_{\xi\tau}=\beta_W^{-5/2}[-J_{\nu(\tau)-1}(x_{\nu(\tau),\xi})J_{\nu(\tau)+1}(x_{\nu(\tau),\xi})]^{-1/2}.
\end{equation}

\begin{figure}
\begin{center}
\includegraphics[width=0.9\hsize]{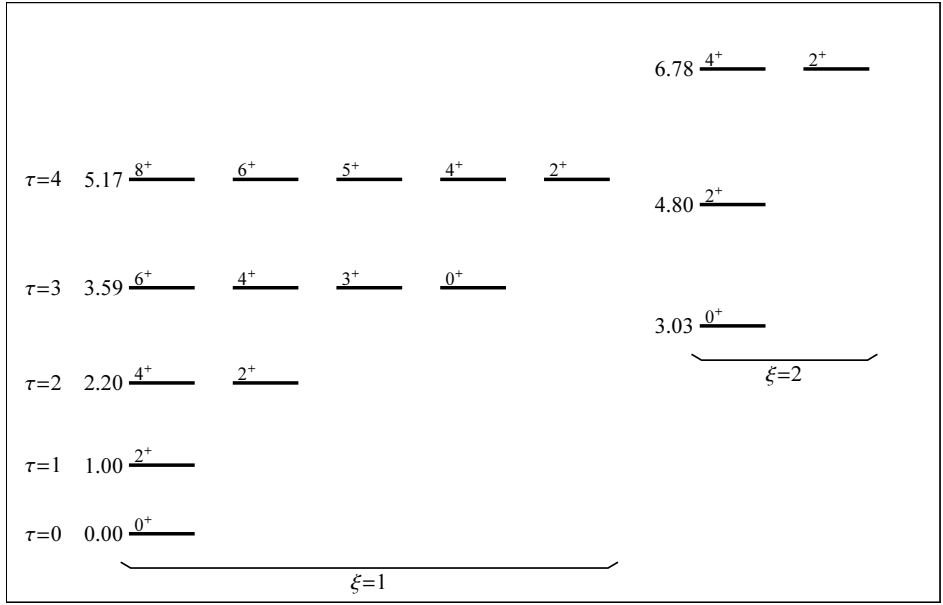}
\end{center}
%%\vspace{-12pt}
\caption
{Level scheme for the $\grpe{5}$ model, showing the excitation energies of the
lowest $\tau$ multiplets of the first two $\xi$ families.  All
energies are normalized to $E(2^+_{1,1})$.
\label{FigCoreLevels}
}
\end{figure}

As usual for
problems with $\grpso{5}$ symmetry, each value of $\tau$ yields a
multiplet of degenerate states of various angular momenta $L$,
according to the $\grpso{5}\supset\grpso{3}$ branching
relations~\cite{rakavy1957:gsoft,kemmer1968:so5-irreps-1,arima1979:ibm-o6,rowe2004:spherical-harmonics}.
The angular momentum contents of the lowest $\grpso{5}$
representations $(\tau,0)$ are summarized for convenience in
Table~\ref{TabSO5Branching}.  For $\tau\<\geq6$, the same angular
momentum can occur more than once within an $\grpso{5}$
representation, and an additional multiplicity index, here chosen as
$\nutt$~\cite{footnote-multiplicity}, is needed to distinguish between
angular wave functions with otherwise identical quantum numbers.
The eigenstates of the $\grpe{5}$
Hamiltonian are fully specified by the angular quantum numbers
($\tau$, $\nutt$, $L$, and $M_L$) together with the radial quantum number $\xi$
($\xi\<=1,2,\ldots$).  The eigenstates may thus be denoted $\lvert
\xi \tau \nutt L M_L \rangle$ or, more concisely,
$L^+_{\xi,\tau}$ as in Ref.~\cite{iachello2000:e5}.  
The levels, quantum numbers, and excitation energies obtained for the
$\grpe{5}$ model are summarized in Fig.~\ref{FigCoreLevels}.  
% ---
% LaTeXTable style Elsart
\begin{table}
\caption{Angular momentum contents of the symmetric representations $(\tau,0)$ of $\grpso{5}$, for $\tau\<\leq6$.}
\label{TabSO5Branching}
\vspace{1ex}
\begin{tabular}{rl}
\hline
$\tau$&$L$\\
\hline
$0$&$0$\\
$1$&$2$\\
$2$&$4~~2$\\
$3$&$6~~4~~3~~0$\\
$4$&$8~~6~~5~~4~~2$\\
$5$&$10~~8~~7~~6~~5~~4~~2$\\
$6$&$12~~10~~9~~8~~7~~6~~6~~4~~3~~0$\\
\hline
\end{tabular}
\end{table}

% ---
% ---
% ---

\subsection{Classification of states}
\label{sece5class}

Exactly solvable problems are usually characterized by the presence of
dynamical symmetries (\textit{e.g.},
Refs.~\cite{iachello1994:dynsymm,iachello2006:liealg}).
Therefore, an interesting question is the extent to which the square
well problem is amenable to group theoretical treatment.

We begin by recognizing that the Hamiltonian~(\ref{eqnHppv}) can be
written, within the well boundary ($\beta\<\leq\beta_W$), as the
five-dimensional free Hamiltonian
\begin{equation}
\label{eqnHfree}
H = \frac{\hbar^2}{2B} \tilde\pi \cdot \tilde\pi.
\end{equation}
The five-dimensional rotation algebra $\grpso{5}$, already discussed,
has as its generators the ``five-dimensional angular momentum
operators'' $\Lambdahat_{ij}$, ten in
number~\cite{chen1989:group,eisenberg1987:v1}.  Here it is most useful
to
reexpress these in Racah (or spherical tensor) form, as
\begin{equation}
\label{eqnLambdahat}
\Lambdahat^{(\lambda)}_\mu \equiv i\sqrt{2}(\alpha\times\tilde\pi)^{(\lambda)}_\mu,
\end{equation}
where $\lambda\<=1$ and $3$.  
The Casimir operator of $\grpso{5}$ is
$C_2[\grpso{5}]\<=2\Lambdahat\circ\Lambdahat\<=
2[\Lambdahat^{(1)}\cdot \Lambdahat^{(1)} +
\Lambdahat^{(3)}\cdot\Lambdahat^{(3)}]$, with eigenvalue
$2\tau(\tau+3)$, where an open dot indicates the $\grpso{5}$ scalar
product.  The ordinary three-dimensional angular momentum operators
$\Lhat_\mu$, which generate $\grpso{3}$, are
\begin{equation}
\label{eqnLhat}
\Lhat_\mu \equiv i\sqrt{10}(\alpha\times\tilde\pi)^{(1)}_\mu=\sqrt{5}\Lambdahat^{(1)}_\mu.
\end{equation}
The
$\grpso{3}$ Casimir operator is simply $C_2[\grpso{3}]\<=
2\Lhat\cdot \Lhat$, with eigenvalue $2L(L+1)$.
The algebras $\grpso{5}$ and $\grpso{3}$ can furthermore be embedded
in the Euclidean algebra $\grpe{5}$.  This is the (non-semisimple)
algebra which generates translations and rotations in five dimensions.
The generators, in spherical tensor form, are the five linear momentum
components $\tilde\pi_\mu$ taken together with the ten angular
momentum components $\Lambdahat^{(\lambda)}_{\mu}$.  The free
Hamiltonian~(\ref{eqnHfree}) is, naturally, invariant under both
translations and rotations in five-dimensional space and is thus
recognized as the Casimir operator
\begin{equation}
C_2[\grpe{5}]=\tilde\pi\cdot\tilde\pi.
\end{equation}

A dynamical symmetry is a situation in which the Hamiltonian is
constructed from the Casimir operators of the algebras in a subalgebra
chain~\cite{iachello1979:dynsymm,iachello1994:dynsymm,iachello2006:liealg}.
The eigenvalues of the Casimir operators provide good quantum numbers
for the eigenstates.  The algebras just described form the subalgebra
chain $\grpe{5}\supset\grpso{5}\supset\grpso{3}\supset\grpso{2}$.  The
\textit{free} Hamiltonian~(\ref{eqnHfree}) describes a dynamical symmetry
situation for this algebra chain, as $H\<=C_2[\grpe{5}]$.  The
\textit{square well} Hamiltonian~(\ref{eqnHppv}) therefore also
describes a dynamical symmetry, \textit{provided} that all operators
are restricted to the bounded domain defined by $\beta\<\leq\beta_W$.
Note that Casimir operators of the subalgebras $\grpso{5}$ and
$\grpso{3}$ can also be included in the Hamiltonian, as discussed in
Sec.~\ref{sece5gen}.
The Bessel functions form bases for the representation
of the Euclidean algebras, as discussed in
Ref.~\cite{miller1968:lie-special-chap3}.  In the present
five-dimensional case, the wave functions
$\Phi(\beta,\gamma,\vartheta)\<\propto\beta^{-3/2}J_{\tau+3/2}(\varepsilon^{1/2}\beta)\Psi_{\tau\nutt{}LM_L}(\gamma,\vartheta)$
are characterized by the $\grpe{5}$ quantum number $\langle
C_2[\grpe{5}]\rangle\<=\varepsilon$.  For the free problem
$\varepsilon$ may vary continuously ($\varepsilon\<\geq0$).  For the square well problem, the
node condition at the boundary instead restricts $\varepsilon$ to
discrete values~(\ref{eqneigen}).
To summarize, the free or square well Hamiltonians are both dynamical
symmetry situations, characterized by the subalgebra
chain
%%\begin{equation}
%%\label{eqnchaine5}
%%\begin{array}{ccccccc}
%%\grpe{5}&\supset&\grpso{5}&\supset&\grpso{3}&\supset&\grpso{2},\\
%%\varepsilon&&(\tau,0)&\nutt&L&&M_L
%%\end{array}
%%\end{equation}
\begin{equation}
\label{eqnchaine5}
\Biggl\lvert
\begin{array}{ccccccc}
\grpe{5}&\supset&\grpso{5}&\supset&\grpso{3}&\supset&\grpso{2}\\
\varepsilon&&(\tau,0)&\nutt&L&&M_L
\end{array}
\Biggr\rangle ,
\end{equation}
and the eigenstates are members of representations with the quantum
numbers indicated.

However, some important practical differences
should be noted between the present $\grpe{5}$ dynamical symmetry and
others encountered in either
geometric~\cite{wybourne1974:groups} or
algebraic~\cite{iachello1987:ibm,iachello2006:liealg} problems.  
Many of the calculational simplicities usually associated
with a dynamical symmetry arise from the presence of a
\textit{spectrum generating algebra}, which provides ladder
operators allowing the various states in the model to be constructed
from each other.
For the generalized oscillator potential~(\ref{eqndavidson}) and
generalized Coulomb potential~(\ref{eqnkratzer}), in any number of
dimensions, the spectrum generating algebras are known.  Namely, in
the present five-dimensional case, the algebras are $\grpsp{10,R}$ or
$\grpu{5,1}$ for the oscillator
potential~\cite[p.~290]{wybourne1974:groups} and $\grpso{6,2}$ for the
Coulomb potential~\cite[p.~184]{iachello2006:liealg}.  For the square well
problem, the boundary condition makes construction of the ladder
operators considerably more involved.  The spectrum generating algebra
for the square well has been explicitly constructed only in one
dimension, where it is $\grpsu{1,1}\<\approx\grpso{2,1}$, in
a quantum number dependent
realization~\cite{kais1986:square-well-algebraic,antoine2001:square-well-poeschl-teller-coherent,lemus2003:square-well-algebraic}.
An explicit construction in more than one dimension remains to be
done.

\subsection{Electromagnetic transition strengths}
\label{sece5trans}

\subsubsection{General properties}

Electromagnetic transition strengths play an essential role in
establishing the extent to which the present description applies to
nuclei near the phase transition.  The various terms contributing to
the electromagnetic transition operators transform as $\grpso{5}$
tensors, and the $\grpe{5}$ eigenstates posess $\grpso{5}$ tensor
character as well.  Consequently, the pattern of transitions between
levels is determined by $\grpso{5}$ selection rules.  Typically,
either the transition between two levels is forbidden for the leading
order term in the transition operator but allowed for the next higher
order term or \textit{vice versa}.  It is thus especially important
in the present problem to consider terms beyond leading order in the
transition operators.

\subsubsection{$E2$ transitions}

Let us first consider electric quadrupole ($E2$) transitions.  The
general $E2$ operator may be expanded in terms of the coordinates
$\alpha$~(\ref{eqnalpha}) as
\begin{equation}
\label{eqncorete2}
T^{(E2)}= t_{21} \alpha+ t_{22} \Galphaalpha{2} +\cdots.
\end{equation}
The operator with first and second order terms is analogous to the
$E2$ operator commonly used in the IBM.  Microscopic derivations
within the context of the IBM~\cite{otsuka1978:ibm2-shell-details}
suggest that the contribution of the second order term is of the same
order of magnitude as that of the first order term, as do
phenomenological studies of actual nuclei~\cite{warner1983:cqf}.
Transition strengths are related to the reduced matrix elements by
\begin{equation}
B(E2;\xi\tau L\rightarrow\xi' \tau' L')=\frac{1}{2L+1} \langle
\xi'\tau'L' \Vert T^{(E2)} \Vert \xi \tau L \rangle ^2, 
\end{equation}
and quadrupole moments by
\begin{equation}
eQ(\xi\tau L)= \sqrt\frac{16\pi}{5} \left[
\frac{L(2L-1)}{(2L+1)(L+1)(2L+3)} \right]^{1/2}
 \langle
\xi \tau L  \Vert T^{(E2)} \Vert \xi \tau L \rangle.
\end{equation}

The explicit expression for the leading order term $\alpha$ in terms
of Bohr variables is given in~(\ref{eqnalpha}), while the second
order term $(\alpha\times\alpha)^{(2)}$ is given by
\begin{equation}
\label{eqnalphaalpha2}
(\alpha\times\alpha)^{(2)}_\mu =
\sqrt\frac{2}{7}\beta^2\bigl[- \cos2\gamma \mathscr{D}^{(2)}_{\mu0}(\vartheta)
+\frac{1}{\sqrt{2}}\sin2\gamma[\mathscr{D}^{(2)}_{\mu2}(\vartheta)+\mathscr{D}^{(2)}_{\mu-2}(\vartheta)]\bigr]
.
\end{equation}
Each of these operators, $\alpha$ or $(\alpha\times\alpha)^{(2)}$, factors
into a radial part ($\beta$ or $\beta^2$) and an angular part
(involving $\gamma$ and $\vartheta$).  Its
reduced matrix element between $\grpe{5}$ eigenstates is thus the product of a radial matrix
element and an angular reduced matrix element.  
The radial matrix element 
\begin{equation}
\label{eqnradialme}
\int_0^{\beta_W} \beta^4 d\beta f_{\xi'\tau'} (\beta) \beta^m
f_{\xi\tau} (\beta)
\end{equation}
is evaluated by numerical integration, using the wave functions from
Eq.~(\ref{eqnf}).  [The radial wave functions
obtained for different values of the well width $\beta_W$ are related by a
dilation in the radial coordinate $\beta$, as seen from~(\ref{eqnf}),
and consequently radial matrix elements scale as
$\beta_W^m$.]   Once the angular wave functions
$\Psi_{\tau\nutt{}LM_L}(\gamma,\vartheta)$ are explicitly constructed
as in Eq.~(\ref{eqnPsiDseries}), evaluation of the angular matrix
elements is straightforward.  For a general spherical tensor operator
$f^{(\lambda)}_\mu(\gamma,\vartheta)\<=\sum_{\substack{\kappa=0\\\text{even}}}^\lambda
f^\lambda_\kappa(\gamma)\phi^{\lambda}_{\mu\kappa}(\vartheta)$,
the
reduced matrix element is~\cite{caprio2005:axialsep}
\begin{multline}
\label{eqngammatensorme}
\langle\Psi_{\tau^\prime \nutt^\prime L^\prime } \Vert f^{(\lambda)}(\gamma,\vartheta) \Vert \Psi_{\tau \nutt L }\rangle
\\=\frac{1}{4\pi}
\left[(2L +1)(2\lambda+1)\right]^{1/2}
\sum_{\substack{K^\prime ,\kappa,K \\\text{even}}}
\left[\frac{1+\delta_{K^\prime }}{(1+\delta_{K
})(1+\delta_{\kappa})}\right]^{1/2}
\\\times
\left[ (L K \lambda\kappa | L^\prime K^\prime ) 
+ \begin{closedcases}
(-)^\lambda(L K \lambda\bar\kappa | L^\prime  K^\prime ) & K \geq\kappa\\
(-)^{L }(L \bar K \lambda\kappa | L^\prime  K^\prime ) & K \leq\kappa
  \end{closedcases}
\right]\\\times
\left[\int \lvert\sin 3 \gamma\rvert d \gamma F_{\tau^\prime \nutt^\prime L^\prime K^\prime }^*(\gamma) f^\lambda_\kappa(\gamma)  F_{\tau \nutt L K }(\gamma) \right]
,
\end{multline}
where we follow Racah's normalization convention~\cite{edmonds1960:am}
for the Wigner-Eckart theorem.  For the transition operators
considered here, the integrals only involve polynomials in
trigonometric functions and may therefore be evaluated exactly.  The
resulting angular matrix elements are tabulated in
Appendix~\ref{appcorerme}.  Evaluation of the angular matrix elements
can also be carried out using algebraic
methods~\cite{arima1979:ibm-o6}.
% ---
\begin{figure}
\begin{center}
\includegraphics[width=0.9\hsize]{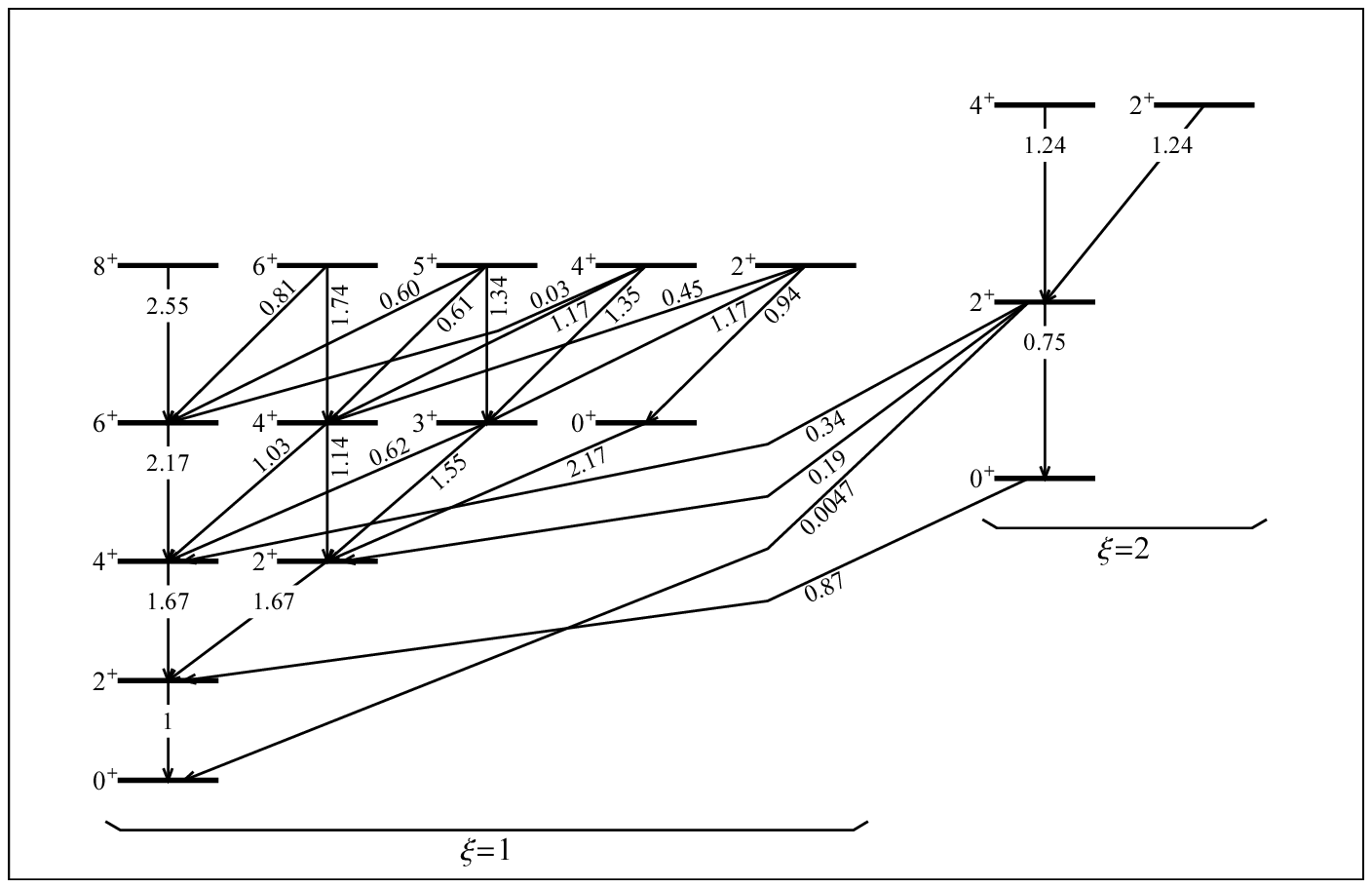}
\end{center}
%%\vspace{-12pt}
\caption
{$B(E2)$ strengths induced by the operator $t_{21} \alpha$, for the
$\grpe{5}$ model.  Transitions obey the selection rule $\Delta
\tau\<=1$.  All transition strengths are normalized relative to
$B(E2;2^+_1\rightarrow0^+_1)$.  For absolute values, 
strengths should be multiplied by $(0.07453) t_{21}^2 \beta_W^2$.
\label{FigCoreAlpha}
}
\end{figure}
% ---
% ---
\begin{figure}
\begin{center}
\includegraphics[width=0.9\hsize]{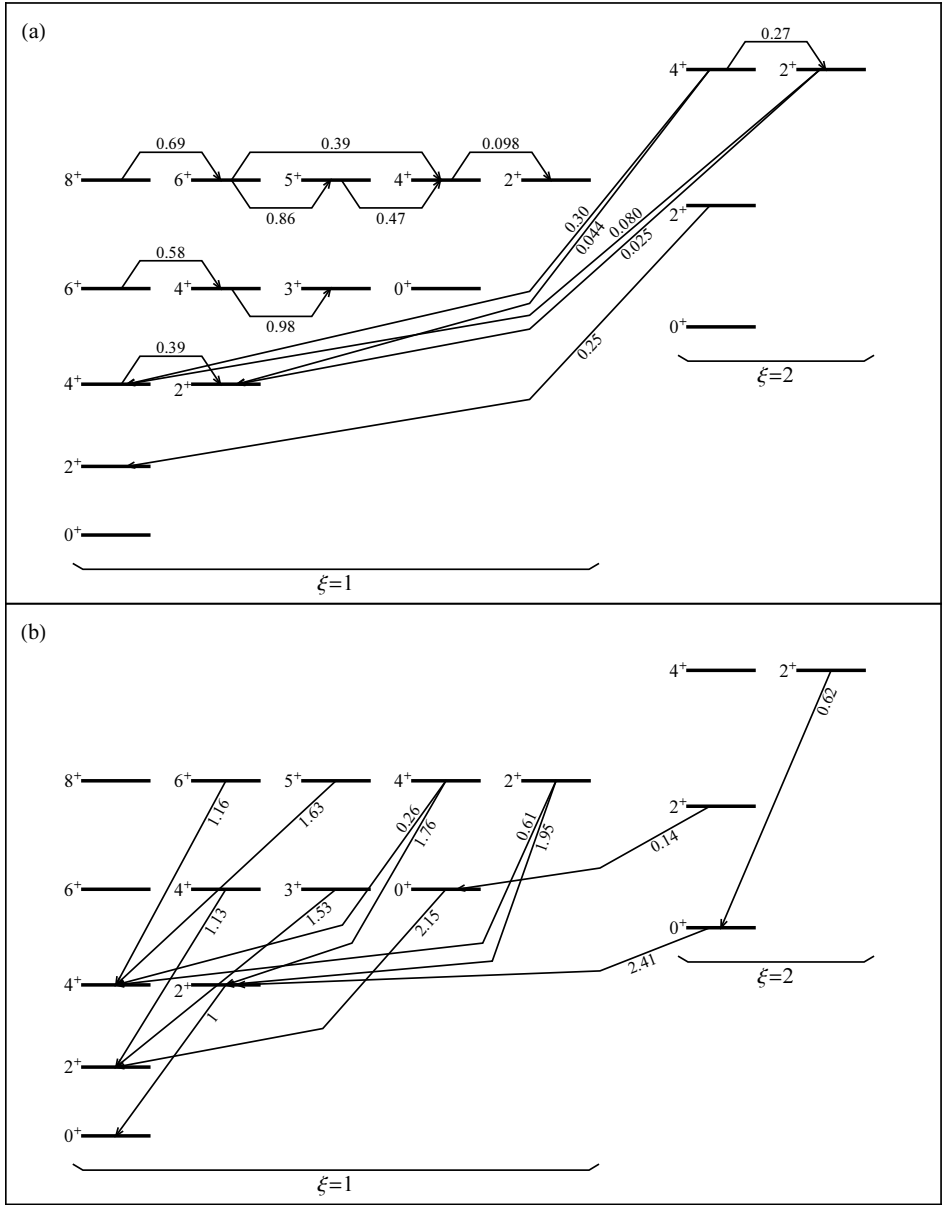}
\end{center}
%%\vspace{-12pt}
\caption
{$B(E2)$ strengths induced by the operator $t_{22}
(\alpha\times\alpha)^{(2)}$, for the $\grpe{5}$ model: (a)~transitions
with $\Delta \tau\<=0$ and (b)~transitions with $\Delta \tau\<=\pm2$.
All transition strengths are normalized relative to $B(E2;2^+_2\rightarrow0^+_1)$.  For
absolute values,  strengths should be multiplied by $(0.009754)
t_{22}^2 \beta_W^4$.
\label{FigCoreAlphaAlpha2Trans}
}
\end{figure}
% ---
% ---
\begin{figure}
\begin{center}
\includegraphics[width=0.9\hsize]{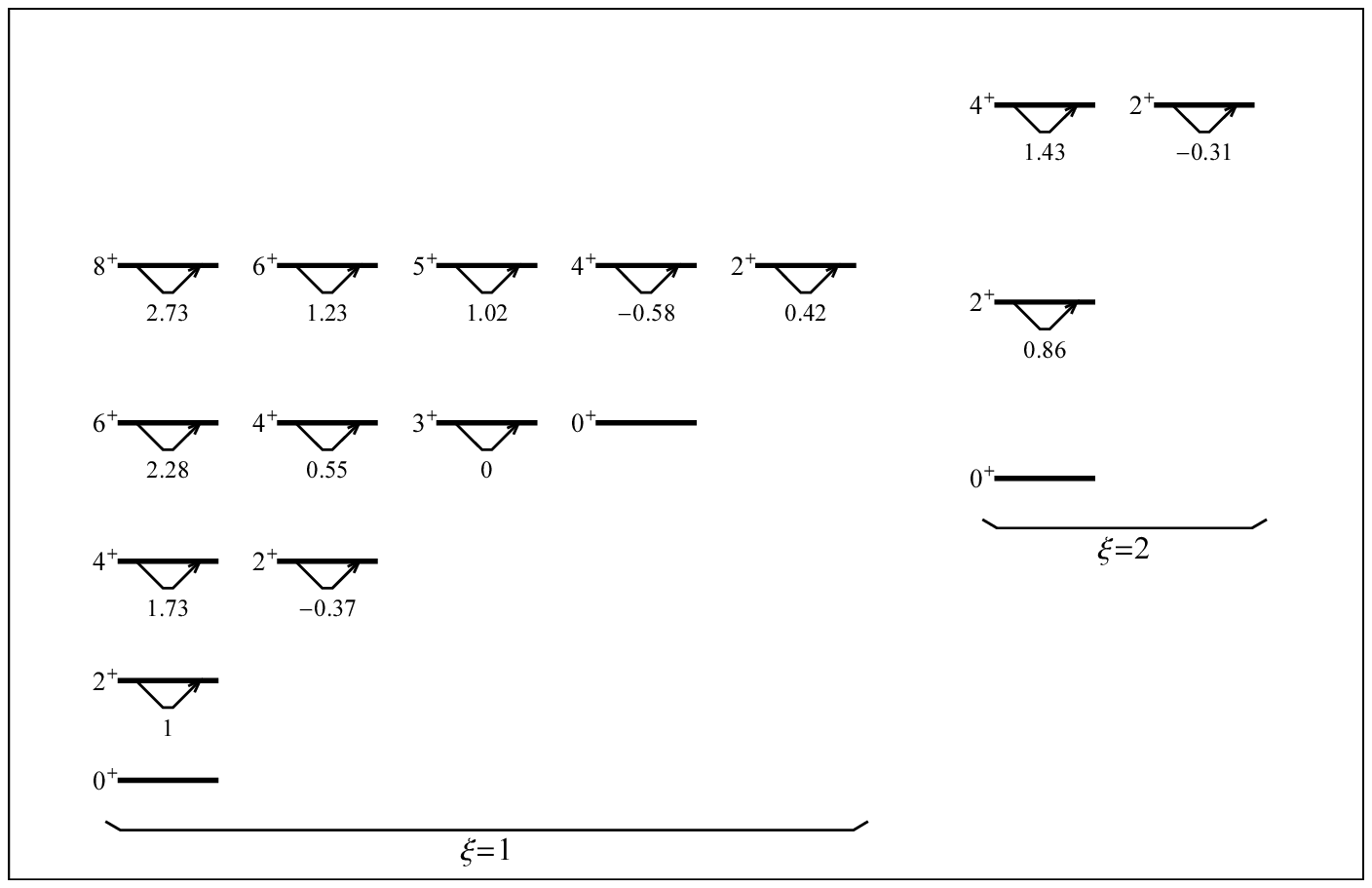}
\end{center}
%%\vspace{-12pt}
\caption
{Quadrupole moments induced by the operator $t_{22}
(\alpha\times\alpha)^{(2)}$, for the $\grpe{5}$ model.  All moments
are normalized relative to $Q(2^+_1)$.  For absolute values, moments should be
multiplied by $(0.2124) t_{22} \beta_W^2$.
\label{FigCoreAlphaAlpha2Moment}
}
\end{figure}
% ---

The operator $\alpha$ is proportional to the angular wave function
$\Psi_{102}(\gamma,\vartheta)$. It thus transforms as a $(1,0)$ tensor
under $\grpso{5}$ and can only connect states related by
$\Delta\tau\<=\pm1$.  The transition strengths induced by $\alpha$
are shown in Fig.~\ref{FigCoreAlpha}.  All quadrupole moments vanish
if only this leading-order contribution to the $E2$ operator is
considered.  

The operator $(\alpha\times\alpha)^{(2)}$ is instead proportional to
the angular wave function $\Psi_{202}(\gamma,\vartheta)$.  It thus
transforms as a $(2,0)$ tensor under $\grpso{5}$ and connects states
related by $\Delta\tau\<=0$ or $\pm2$.  Consequently, it can yield
nonzero quadrupole moments.  The transition strengths induced by
$(\alpha\times\alpha)^{(2)}$ are shown in
Fig.~\ref{FigCoreAlphaAlpha2Trans}, and the quadrupole moments are
shown in Fig.~\ref{FigCoreAlphaAlpha2Moment}.  Note that the second
order $E2$ operator in the $\grpe{5}$ model was also considered in
Ref.~\cite{arias2001:134ba-e5}, but a modified radial dependence
$\rtrim\propto1/(1+\beta^2)$ was
used for the transition operator in that work, so slightly different
results are obtained.

The operators $\alpha$ and $(\alpha\times\alpha)^{(2)}$ contribute to
entirely distinct sets of transitions due to the selection rules in
the $\grpe{5}$ problem.  Consequently, interference between them never
occurs.  The parameter $t_{21}$ determines the overall normalization
of all $\Delta\tau\<=\pm1$ transition strengths.  Predictions for all
strength ratios of $\Delta\tau\<=\pm1$ transitions are parameter
independent within the $\grpe{5}$ model.  Similarly, $t_{22}$
determines the overall normalization of all $\Delta\tau\<=0$ or $\pm2$
transition strengths and the quadrupole moments.  The predictions are
discussed further in Sec.~\ref{sece5expt}.

\subsubsection{$M1$ transitions}

Magnetic dipole transitions arise from differences between the proton
and neutron contributions to collective motion.  They are therefore
most naturally treated in a model which distinguishes proton and
neutron degrees of
freedom~\cite{otsuka1978:ibm2-shell-details,maruhnrezwani1975:gcm-pn}.
However, within the Bohr picture, an effective $M1$ operator may be
expressed in terms of $\alpha$ and $\tilde\pi$
as~\cite{eisenberg1987:v1,footnote-tm1-real}
\begin{equation}
\label{eqncoretm1}
\begin{aligned}
T^{(M1)}&= t_{10} [i\Galphapi{1}] + t_{11}
[\alpha\times i\Galphapi{1}]^{(1)}+\cdots\\
&=t_{10} \tfrac1{\sqrt{10}}  \Lhat + t_{11}\tfrac1{\sqrt{10}}
(\alpha\times\Lhat)^{(1)}+\cdots.
\end{aligned}
\end{equation}
This is closely analogous to the effective $M1$ operator used in the
IBM~\cite{iachello1987:ibm}.  Transition strengths are given by
\begin{equation}
B(M1;\xi\tau L\rightarrow\xi' \tau' L')=\frac{1}{2L+1} \langle
\xi'\tau'L' \Vert T^{(M1)} \Vert \xi \tau L \rangle ^2, 
\end{equation}
and magnetic dipole moments by 
\begin{equation}
\mu(\xi\tau L)= \sqrt\frac{4\pi}{3} \left[ \frac{L}{(2L+1)(L+1)}
\right]^{1/2}
 \langle
\xi\tau L \Vert T^{(M1)} \Vert \xi \tau L \rangle.
\end{equation} 
% ---
\begin{figure}
\begin{center}
\includegraphics[width=0.9\hsize]{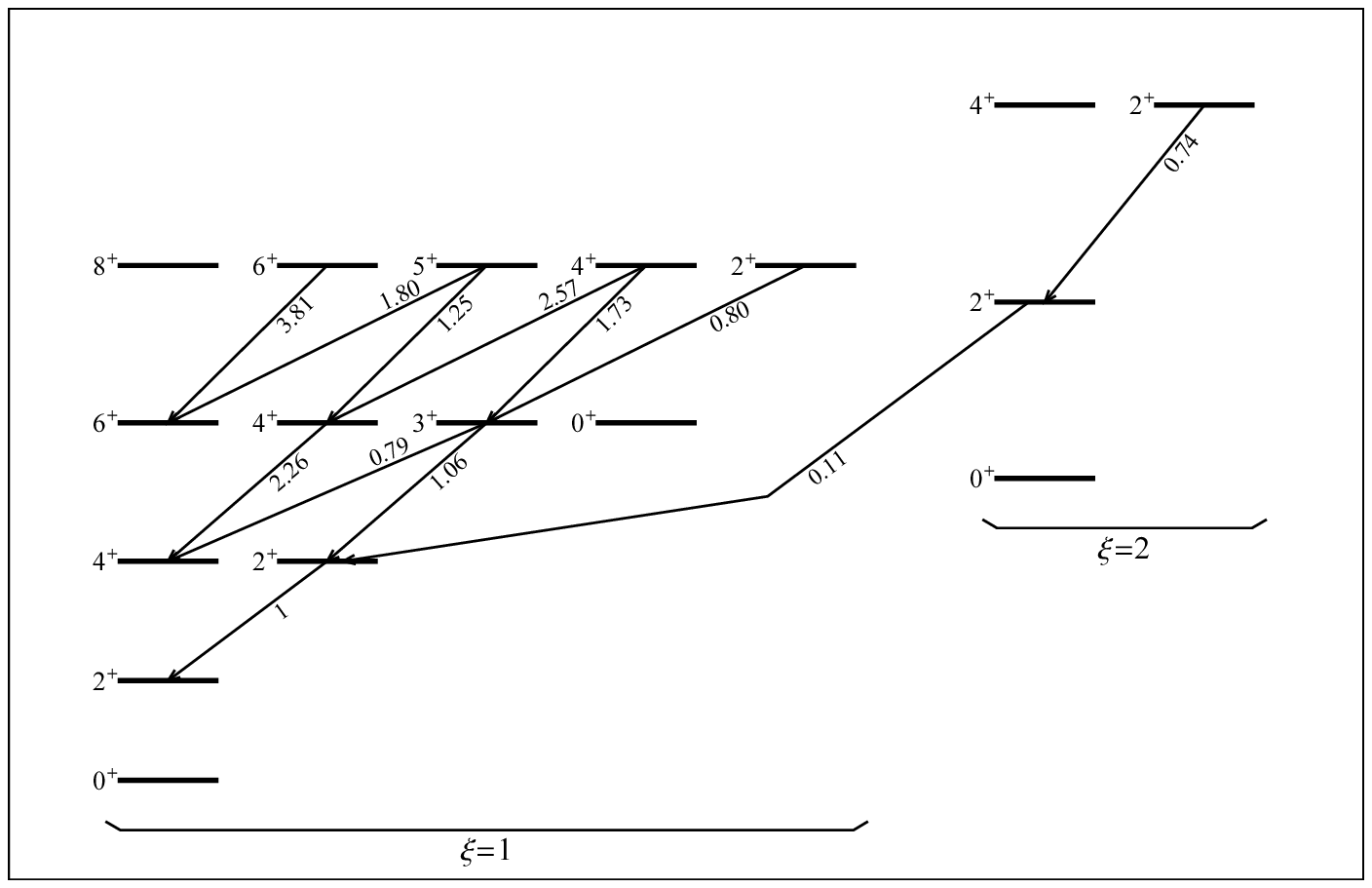}
\end{center}
%%\vspace{-12pt}
\caption
{$B(M1)$ strengths induced by the operator $t_{11}
[\alpha\times(\alpha\times\pi)^{(1)}]^{(1)}$, for the $\grpe{5}$
model.  Transitions obey the selection rule $\Delta
\tau\<=1$.  All transition strengths are normalized relative to
$B(M1;2^+_2\rightarrow2^+_1)$.  For absolute values, 
strengths should be multiplied by $(0.02620) t_{11}^2 \beta_W^2$.
\label{FigCoreAlphaAlphaPi11}
}
\end{figure}
% ---

Since the leading order term in $T^{(M1)}$ is simply proportional to
the collective angular momentum operator $\Lhat$, it is diagonal in
the eigenfunctions and cannot induce transitions.  It yields
magnetic moments which depend only upon $L$,
\begin{equation}
\mu(\xi\tau L)=\sqrt\frac{4\pi}{3}t_{10}L=g_BL,
\end{equation}
where $g_B\<\equiv(4\pi/3)^{1/2}t_{10}$, as usual in the collective picture.  However, the higher order
$M1$ operator $[\alpha\times\Lhat]^{(1)}$ does yield nonvanishing
transitions.  The matrix elements follow from those already calculated above
for $\alpha$ according to the Racah reduction
formula~\cite{edmonds1960:am}, giving
\begin{multline}
\langle \xi'\tau' L' \Vert (\alpha\times\Lhat)^{(1)} \Vert \xi \tau L
\rangle \\
=
-\frac{1}{\sqrt{40}}[(L-L'+2)(-L+L'+2)(L+L'-1)(L+L'+3)]^{1/2}
\\\times
\langle \xi'\tau' L' \Vert \alpha \Vert \xi \tau L
\rangle. 
\end{multline}
The operator follows the same $\Delta\tau\<=\pm1$ selection rule as
$\alpha$ (but with the more restrictive angular momentum selection
rule appropriate to a dipole operator). The resulting transition
strengths are shown in Fig.~\ref{FigCoreAlphaAlphaPi11}.

\subsubsection{$E0$ transitions}

The collective electric monopole operator is of the form
\begin{equation}
\label{eqncorete0}
\begin{aligned}
T^{(E0)}&= t_{00} \Galphaalpha{0} +\cdots\\
&= t_{00} \frac{1}{\sqrt{5}}\beta^2+\cdots.
\end{aligned}
\end{equation}
Transition strengths are usually quoted
as squared $\rho(E0)$ values, defined by~\cite{wood1999:e0}
\begin{equation}
\begin{aligned}
(eR^2) \rho(E0;\xi\tau L\rightarrow\xi' \tau' L)&=\frac{1}{\sqrt{2L+1}} \langle
\xi'\tau'L \Vert T^{(E0)} \Vert \xi \tau L \rangle \\
&= \langle
\xi'\tau'LL \vert T^{(E0)} \vert \xi \tau LL \rangle,
\end{aligned}
\end{equation}
where $R$ is the nuclear radius.  The leading order
operator $\beta^2$ consists only of a radial factor, so evaluation of
the matrix elements involves only the radial
integral~(\ref{eqnradialme}).  Transitions occur only between states
with identical angular quantum numbers.  Selected transition strengths
are listed in Table~\ref{TabCoreMonopole}.
% ---
% LaTeXTable style Elsart
\begin{table}
\caption{Selected $\rho^2(E0)$ strengths for the $\grpe{5}$ model.  All $\rho^2(E0)$ strengths are normalized relative to $\rho(E0;0^+_{\xi=2}\rightarrow0^+_1)$.  For absolute values, strengths should be multiplied by $(0.006343) t_{02}^2 \beta_W^4 / (eR^2)^2$.}
\label{TabCoreMonopole}
\vspace{1ex}
\begin{tabular}{rrrrrrr}
\hline
$\xi$&$\tau$&$J$&$\xi'$&$\tau'$&$J'$&\multicolumn{1}{c}{$\rho^2(E0;J\rightarrow J')$}\\
\hline
$2$&$0$&$0$&$1$&$0$&$0$&1.0000\\
$2$&$1$&$2$&$1$&$1$&$2$&0.9234\\
$2$&$2$&$2$&$1$&$2$&$2$&0.8426\\
$2$&$2$&$4$&$1$&$2$&$4$&0.8426\\
\hline
\end{tabular}
\end{table}

% ---

\subsection{Comparison with experimental data}
\label{sece5expt}

Since $\grpso{5}$ symmetry is present throughout the transition
between spherical and deformed $\gamma$-soft structure, comparison of
the $\grpe{5}$ predictions with experimental data is not simple.  Many
of the gross spectroscopic features, including the level multiplet
structure and electromagnetic branching ratios, follow from the
$\grpso{5}$ symmetry and therefore persist throughout the transition.
In this section, we therefore consider the distinguishing observables
which do vary along the $\grpu{5}$--$\grpso{6}$ transition.  

The most fundamental energy ratios from a spectroscopic viewpoint are
\begin{align}
E(4^+_{1,2})/E(2^+_{1,1})&=E(2^+_{1,2})/E(2^+_{1,1})=2.20
\intertext{and}
\label{eqncoreenergyratiovib}
E(0^+_{2,0})/E(2^+_{1,1})&=3.03.
\end{align}
These place the yrast
$E(4^+)/E(2^+)$ ratio intermediate between the $\grpu{5}$ value of $2$
and the $\grpso{6}$ value of $2.5$.  They also provide an estimate for
the excitation energy of the first radial ($\xi$) excitation.
The $B(E2)$ strength ratios involving these same levels are
\begin{align}
\frac{
B(E2;4^+_{1,2}\rightarrow2^+_{1,1})
}{
B(E2;2^+_{1,1}\rightarrow0^+_{1,0})
}
&=
\frac{
B(E2;2^+_{1,2}\rightarrow2^+_{1,1})
}{
B(E2;2^+_{1,1}\rightarrow0^+_{1,0})
}
=
1.68
\intertext{and}
\frac{
B(E2;0^+_{2,0}\rightarrow2^+_{1,1})
}{
B(E2;2^+_{1,1}\rightarrow0^+_{1,0})
}
&=
0.68.
\end{align}
% ---
\begin{figure}
\begin{center}
\includegraphics[width=0.6\hsize]{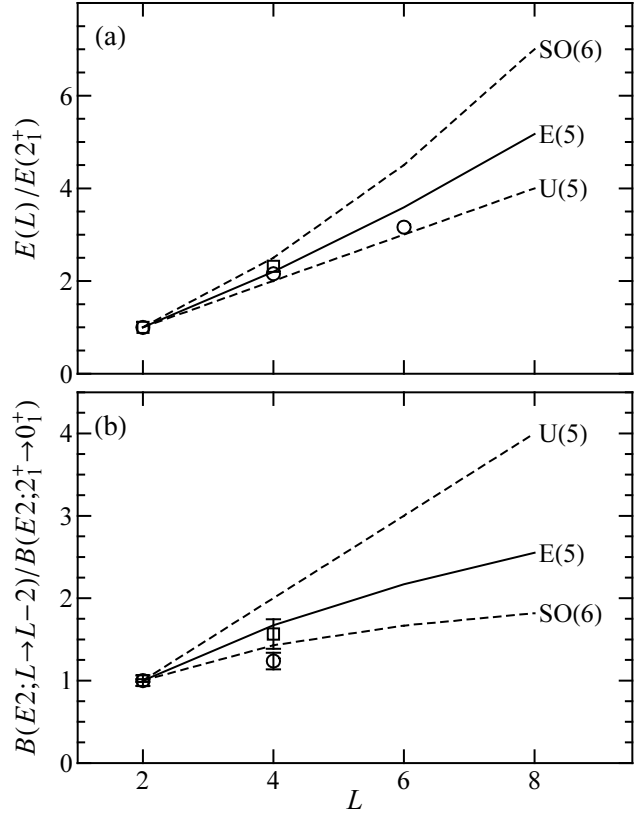}
\end{center}
%%\vspace{-12pt}
\caption
{Yrast (a)~energies and (b)~$B(E2)$ strengths as functions of angular
momentum in the
$\grpe{5}$ model.  The experimental data for
$^{132}\mathrm{Xe}$~(\symbolopencircle) and
$^{134}\mathrm{Ba}$~(\symbolopensquare)~\cite{nds2005:132,nds2004:134}
are indicated for comparison, as are the predictions for the
$\grpu{5}$ and $\grpso{6}$ dynamical symmetries in the
large boson number limit (dashed curves).
Energies are normalized to $E(2^+_1)$ and $B(E2)$ strengths to
$B(E2;2^+_1\rightarrow0^+_1)$.
\label{FigCoreYrast}
}
\end{figure}
% ---

The properties of the yrast states as a function of angular momentum
are also among the most relevant observables.  They vary continuously
along the $\grpu{5}$--$\grpso{6}$ transition, and they are among
the most feasible observables to measure.  The yrast excitation energy
ratios $E(L^+)/E(2^+_{1,1})$ are shown in Fig.~\ref{FigCoreYrast}(a).
The yrast $B(E2)$ ratios
$B[E2;L^+\rightarrow(L-2)^+]/B(E2;2^+_{1,1}\rightarrow0^+_{1,0})$ are
shown in Fig.~\ref{FigCoreYrast}(b).  Casten and
Zamfir~\cite{casten2000:134ba-e5} have suggested that
$^{134}\mathrm{Ba}$ is well described by the $\grpe{5}$ predictions.
The measured yrast observables for $^{134}\mathrm{Ba}$ and $^{132}\mathrm{Xe}$,
as well as the predictions for the $\grpu{5}$ and $\grpso{6}$ limits
of the transition, are shown for comparison in
Fig.~\ref{FigCoreYrast}.

A distinguishing feature of the evolution along the transition from
spherical to deformed structure is the increase in energy of the
radial excitation, specifically the $0^+$ head of the excited family
of levels.  For the $\grpu{5}$ limit, $E(0^+)/E(2^+_1)\<=2$.  The
$\grpe{5}$ prediction is $E(0^+)/E(2^+_1)\<=3.03$, as noted above.
The prediction for true ``rigidly'' $\beta$-deformed, $\gamma$-soft
structure would be $E(0^+)/E(2^+_1)\rightarrow\infty$.  For more
realistic situations, the result is dependent upon the stiffness of
the potential~\cite{wilets1956:oscillations} in the geometric picture
or upon the $\grpso{6}$ Hamiltonian coefficients (and boson number) in
the algebraic picture~\cite{arima1979:ibm-o6}.  There is thus no firm
limiting $\grpso{6}$ value for this ratio from theory alone.  Rather,
the limit must be considered empirically, suggesting
$E(0^+)/E(2^+_1)\<\approx4$.

An important consideration in experimental determination of the
properties of the radial excitation is the need to distinguish between
the $0^+_{2,0}$ state (radial or $\xi$ excitation) and the $0^+_{1,3}$
state (member of the $\tau\<=3$ multiplet of the ground state family)
which may lie nearby in energy~\cite{zamfir2002:102pd-beta}.  In many
of the nuclei of experimental interest, the situation is further
complicated by the presence of intruder
configurations~\cite{zamfir2002:102pd-beta}.

Several nuclei, including various Ru, Pd, and Ba isotopes, have been
considered as candidates for description by the $\grpe{5}$ model.
Detailed comparisons between theory and experiment may be found in
Refs.~\cite{casten2000:134ba-e5,arias2001:134ba-e5,frank2001:104ru-e5,zamfir2002:102pd-beta,zhang2002:108pd-e5}.
It has been suggested that $\grpe{5}$ structure may also be present in
light nuclei~\cite{marginean2006:58cr-transfer-e5}.

\subsection{Generalizations of the $\grpe{5}$ description}
\label{sece5gen}

A simple solution such as that described so far can provide a
qualitative description of structure near the critical point of the
$\grpu{5}$-$\grpso{6}$ transition, but it is naturally limited as a
model for detailed description.  Here we summarize 
generalizations of the $\grpe{5}$ description addressing some of the most
significant physical issues.

\begin{figure}
\begin{center}
\includegraphics[width=0.9\hsize]{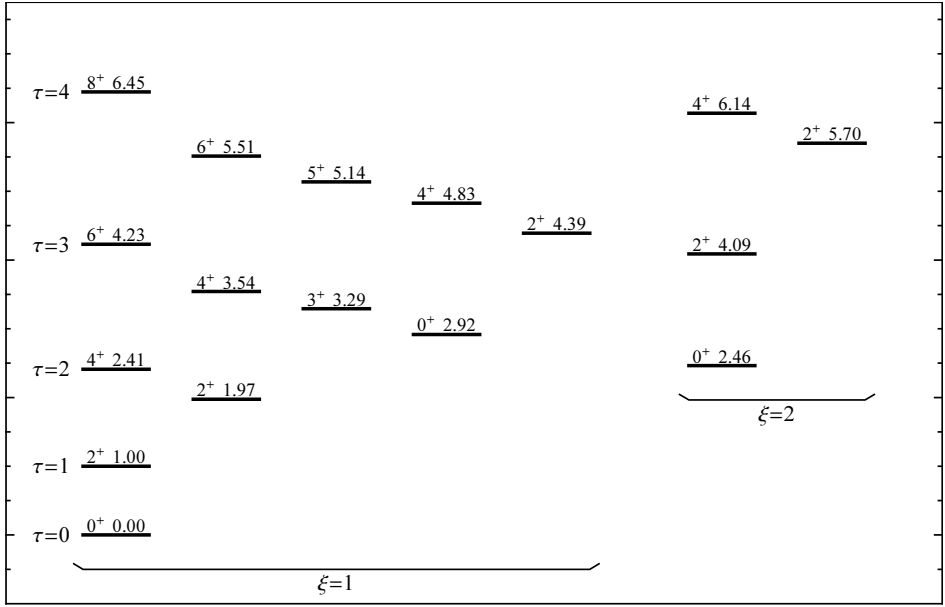}
\end{center}
%%\vspace{-12pt}
\caption
{Level scheme for the for the $\grpe{5}$ model with angular momentum
degeneracy breaking term $k'\Lhat\cdot\Lhat$, for $k'\<=1/2$.  All
energies are normalized to $E(2^+_1)$.
\label{FigCoreLevelsSplit}
}
\end{figure}

\begin{figure}
\begin{center}
\includegraphics[width=0.6\hsize]{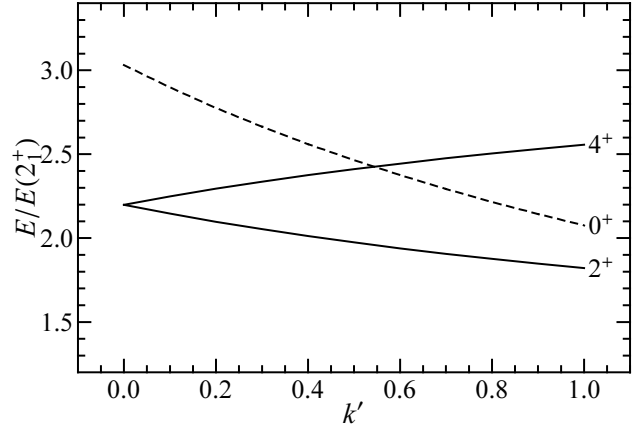}
\end{center}
%%\vspace{-12pt}
\caption
{Evolution of energies as functions of the angular momentum degeneracy
breaking parameter $k'$, for the $\grpe{5}$ model with degeneracy
breaking term $k'\Lhat\cdot\Lhat$.  Energies are shown for the
$\xi\<=1$ levels $4^+_1$ and $2^+_2$ (solid curves) and the $\xi\<=2$
family head $0^+_{\xi=2}$ (dashed curve).  Energies are normalized to
$E(2^+_1)$.
\label{FigCoreEvolnkp}
}
\end{figure}

Empirically, it is found that $\grpso{5}$ level multiplets are often
significantly split in energy, but in a fashion which depends
monotonically upon the angular momentum.  The splitting can be
reproduced trivially by the introduction of a term proportional to
$C_2[\grpso{3}]$ in the Hamiltonian, which leaves the wave functions
unchanged and perturbs the energies as $L(L+1)$.  This is the usual
dynamical symmetry approach, in which the Hamiltonian is simply written
as a sum of Casimir operators of a chain of subalgebras (see
Sec.~\ref{sece5class}), which for chain~(\ref{eqnchaine5}) gives
\begin{equation}
\label{eqnHe5splitcasimir}
H=C_2[\grpe{5}]+ c'' C_2[\grpso{5}] + c' C_2[\grpso{3}],
\end{equation}
or, explicitly in terms of the linear and angular momentum operators,
\begin{equation}
\label{eqnHe5split}
H = \tilde\pi \cdot \tilde\pi +V(\beta)
+ k'' {\Lambdahat\circ\Lambdahat}
+ k' {\Lhat\cdot\Lhat}.
\end{equation}
An example level scheme is shown in Fig.~\ref{FigCoreLevelsSplit}.
The evolution of energies with respect to the splitting parameter
$k'$ is shown in Fig.~\ref{FigCoreEvolnkp}.  Transition strengths are
unaffected by the additional terms.
% ---
% ---
% ---
% ---
% ---
\begin{figure}
\begin{center}
\includegraphics[width=0.6\hsize]{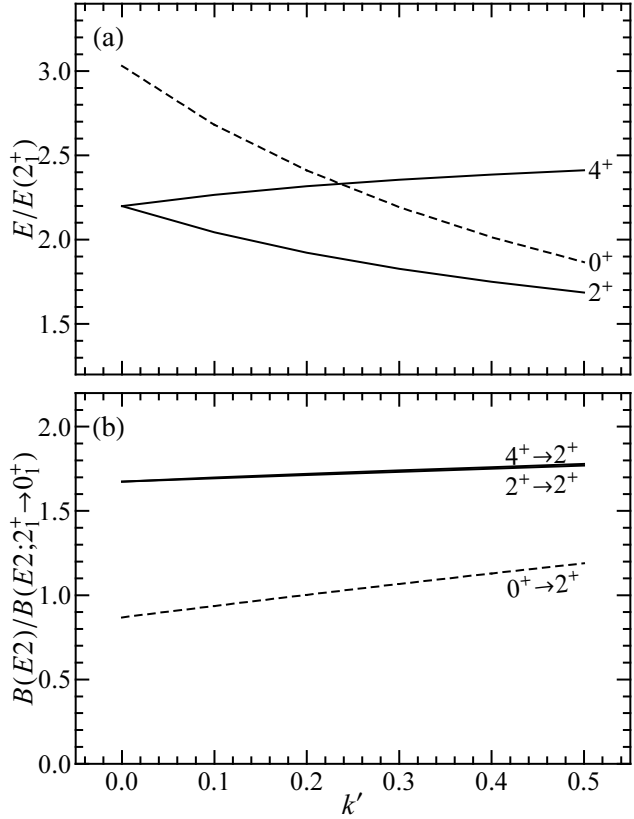}
\end{center}
%%\vspace{-12pt}
\caption
{Evolution of (a)~energies and (b)~$B(E2)$ strengths for the
$\grpe{5}$ model with $\beta$-dependent degeneracy breaking term
$k'\Lhat\cdot\Lhat/\beta^2$, for comparison with
Fig.~\ref{FigCoreEvolnkp}.  Observables involving only the $\xi\<=1$
family of levels [$E(4^+_1)$, $E(2^+_2)$,
$B(E2;4^+_1\rightarrow2^+_1)$, and $B(E2;2^+_2\rightarrow2^+_1)$] are
indicated by solid curves, while those involving the $\xi\<=2$ family
head [$E(0^+_{\xi=2})$ and $B(E2;0^+_{\xi=2}\rightarrow2^+_1)$] are
indicated by dashed curves.  Energies are normalized to $E(2^+_1)$ and
$B(E2)$ strengths to $B(E2;2^+_1\rightarrow0^+_1)$.
\label{FigCoreEvolnkpDep}
}
\end{figure}
% ---

Much more general interactions can be introduced while still retaining
an exactly solvable radial problem, and these are valuable to consider
since they provide flexibility in the treatment of physical
interactions.  Any term proportional to to $1/\beta^2$ can be absorbed
into the separaton constant, and $\beta$ dependences of the
form~(\ref{eqndavidson}) or~(\ref{eqnkratzer}) can also be
accomodated, so there is considerable freedom to include
terms
$\Lambdahat\circ\Lambdahat/\beta^2$, $\Lhat\cdot\Lhat/\beta^2$,
$(\Lambdahat\circ\Lambdahat)\beta^2$, $(\Lhat\cdot\Lhat)\beta^2$,
\textit{etc.}  For illustration, we consider 
\begin{equation}
\label{eqnHe5splitdep}
H = \tilde\pi \cdot \tilde\pi +V(\beta) + k'
\frac{\Lhat\cdot\Lhat}{\beta^2}.
\end{equation}
In this case, $C_2[\grpe{5}]\<=\tilde\pi \cdot \tilde\pi$ no longer
commutes with $H$, since $1/\beta^2$ is not translationally invariant.
Therefore, the eigenstates of $H$ are not eigenstates of
$C_2[\grpe{5}]$, and the label $\varepsilon$ of the
chain~(\ref{eqnchaine5}) no longer applies.  Proper labeling requires
the introduction of the $\grpso{2,1}$ algebra for the radial
problem~\cite{wybourne1974:groups}.  In the separation of variables,
the contribution of the additional angular momentum dependent term is
absorbed into the separation constant $\Lambda$ in~(\ref{eqnradial}),
so
\begin{equation}
\Lambda=\tau(\tau+3)+k'L(L+1).
\end{equation}
The eigenvalues and eigenfunctions are again given by
Eqs.~(\ref{eqneigen}) and~(\ref{eqnf}), but the order of the Bessel
functions is now determined by the separation constant $\Lambda$ as
given above, so $\nu(\tau,L)\<=[\tau(\tau+3)+k'L(L+1)+9/4]^{1/2}$.
The evolution of observables with respect to the degeneracy breaking
parameter $k'$ is shown in Fig.~\ref{FigCoreEvolnkpDep}.  The
evolution of the energy splittings [Fig.~\ref{FigCoreEvolnkpDep}(a)]
differs little in its qualitative features from that obtained for the
simpler degeneracy breaking term $k'\Lhat\cdot\Lhat$
(Fig.~\ref{FigCoreEvolnkpDep}), although a different numerical value
of the parameter $k'$ is needed to achieve the same magnitude effect
in the two cases (note the different horizontal scales of
Figs.~\ref{FigCoreEvolnkp} and~\ref{FigCoreEvolnkpDep}).  The
electromagnetic observables are only weakly dependent upon the
splitting interaction strength [Fig.~\ref{FigCoreEvolnkp}(b)].

More fundamental limitations in the use of the $\grpe{5}$ Hamiltonian
for realistic phenomenological analysis arise from the extreme nature
of the infinite square well potential.  The infinite depth \textit{per
se} is not a significant concern.  Even for finite well depth, the tails
of the radial wave functions outside the well wall are generally so
suppressed that the actual depth does not
matter in detail~\cite{caprio2002:fwell}.  However, the infinite rigidity of the wall
(infinite slope) introduces directly observable artifacts into the
spectroscopic predictions.  Most notably, the much greater energy
spacing scale between levels in the $\xi\<=2$ family than in the
$\xi\<=1$ family (Fig.~\ref{FigCoreLevels}) arises from the infinite
slope of the wall and disappears for soft potentials.  The detailed
mechanism is discussed in
Refs.~\cite{caprio2004:swell,caprio2005:axialsep}.  Comparisons of the
$\grpe{5}$ results with those for potentials involving various
polynomial dependences upon $\beta$ are given in
Refs.~\cite{zhang2001:gcm-random,caprio2003:diss,arias2003:quartic-e5,bonatsos2004:e5-monomial}.

\section{Transitional Bose-Fermi symmetry}
\label{sece54}

\subsection{Hamiltonian and solution}
\label{sece54soln}

In the previous section, we have presented an analytic solution for
the Bohr Hamiltonian which can be used in the description of even-even
nuclei near the critical point of the spherical to $\gamma$-unstable
transition.  In this section, we construct an analytic solution for
the Bohr Hamiltonian with coupling to an additional fermion.  The
study of dynamical symmetries of the coupled odd-mass system is much
more involved than for the core alone~\cite{iachello1991:ibfm}.  The
dynamical symmetries, called Bose-Fermi symmetries, can be constructed
only in special circumstances.  In the geometric formulation
considered here, the fermion must be confined to one or more
degenerate orbitals with suitable angular momenta.  Namely, these
orbitals must transform as a spinor representation of $\grpspin{5}$
(\textit{e.g.}, $j\<=3/2$~\cite{balantekin1981:ibfm-u64-os-pt}) or as
a tensor representation coupled to a spin in a pseudospin scheme
(\textit{e.g.}, $j\<=3/2$ and $5/2$, obtained as $2\otimes1/2$).
Furthermore, the core-fermion interaction must be of a special form,
such that it can be written in terms of Casimir operators of the
combined algebra of collective and single particle degrees of freedom.
Analytic solutions can then be constructed using a method similar to
that of the algebraic IBFM~\cite{iachello1991:ibfm}, but translated into
a differential framework.

To treat the coupled system, we start from the generic Hamiltonian
\begin{equation}
\label{eqnhbf}
H=H_{B}+H_{F}+V_{BF}.
\end{equation}%
The term $H_{B}$ is the Bohr Hamiltonian of Sec.~\ref{sece5},
describing the core in isolation.  We again restrict our attention to
the case of a $\gamma$-independent potential, and to the square well
in particular for transitional nuclei.  The Hamiltonian $H_{F}$ is
that of the single particle in a mean field, with eigenstates $\lvert
n\ell jm_{j}\rangle$ and eigenvalues which are simply the single
particle energies $\varepsilon _{n\ell jm_{j}}$.  For a set of
degenerate orbitals, this $H_{F}$ is simply a constant and can
henceforth be omitted.  The interaction $V_{BF}$ introduces couplings
between the collective coordinates and the single particle degrees of
freedom.

Here we consider the case of a single particle with $j\<=3/2$,
coupled to the nuclear quadrupole collective motion (rotations and vibrations
of a liquid drop).  The essential property which makes a solution
possible is that a particle with $j\<=3/2$ and components
$m_j\<=-3/2,\ldots,+3/2$ transforms as the representation $[1,0]$ of
$\grpsp{4}$~\cite{flowers1952:jj-coupling-part1,bayman1960:j32-sp4}.
However, $\grpsp{4}$ is isomorphic to $\grpso{5}$, providing a
relationship between the transformation properties of the fermion and
of the $\gamma$-soft core.

Specifically, a particle with $j\<=3/2$ is described by the subalgebra
chain
\begin{equation}
\label{eqnchainu4su}
\left \lvert
\begin{array}{ccccccccc}
\grpu{4}&\supset&\grpsu{4}&\supset&\grpsp{4}&\supset&\grpsu{2}&\supset&\grpspin{2}\\
{}[1,0,0,0]     &&      [1,0,0]    &&  [1,0] && [3] && m_j
\end{array}
\right\rangle,
\end{equation}
with the quantum numbers indicated.  Making use of the isomorphisms
$\grpsu{4}\<\approx\grpso{6}$, $\grpsp{4}\<\approx\grpso{5}$, and
$\grpsu{2}\<\approx\grpso{3}$, the particle can instead be
characterized by its transformation properties under the angular
momentum algebras $\grpso{n}$ ($n\<=6$, $5$, $3$, and $2$),
\begin{equation}
\label{eqnchainu4spin}
\left\lvert
\begin{array}{ccccccccc}
\grpu{4}&\supset&\grpspin{6}&\supset&\grpspin{5}&\supset&\grpspin{3}&\supset&\grpspin{2}\\
{}[1,0,0,0]     &&      (\tfrac12,\tfrac12,\tfrac12)    &&  (\tfrac12,\tfrac12) && \tfrac32 && m_j
\end{array}
\right\rangle,
\end{equation}
where the notation $\grpspin{n}$ is used for $\grpso{n}$ when spinor
representations are involved~\cite{iachello2006:liealg}.  To clarify
the common five-dimensional angular momentum structure for the core
and fermion, we explicitly note the $\grpspin{5}$ generators.  The
five-dimensional angular momentum operators for the core
[$\grpspinb{5}\<\equiv\grpso{5}$] are, from Sec.~\ref{sece5class},
$\Lambdahat^{(1)}_\mu \<= i\sqrt{2}\Galphapi{1}_\mu$ and
$\Lambdahat^{(3)}_\mu \<= i\sqrt{2}\Galphapi{3}_\mu$.  The
five-dimensional angular momentum operators for the fermion
[$\grpspinf{5}\<\equiv\grpsp{4}$] are, from Appendix~\ref{appqq},
$\Sigmahat^{(1)}_\mu \<= -\Gaa{1}_\mu$ and $\Sigmahat^{(3)}_\mu \<=
+\Gaa{3}_\mu$, where $a^\dagger$ and $\tilde{a}$ are the fermion
creation and annihilation operators.  A total five-dimensional angular
momentum algebra $\grpspinbf{5}$ is constructed from the sum
generators $\Lambdahatbf\<=\Lambdahat+\Sigmahat$.  The operators
$\Lambdahat$, $\Sigmahat$, and $\Lambdahatbf$ all transform as $(1,1)$
tensors under $\grpspinbf{5}$.

The $\grpe{5|4}$ description, proposed in
Ref.~\cite{iachello2005:geomsuper}, is obtained by taking the
core-fermion interaction to be of a special form, the five-dimensional
analogue of a spin-orbit interaction.  The interaction is given by the
$\grpspin{5}$ scalar product of the five-dimensional angular momenta,
\begin{equation}
\label{eqnLambdaSigma}
\Lambdahat\circ\Sigmahat = \Lambdahat^{(1)}\cdot\Sigmahat^{(1)} + \Lambdahat^{(3)}\cdot\Sigmahat^{(3)}.
\end{equation}
The Hamiltonian for the
coupled system is then
\begin{multline}
\label{eqnHcoupled}
H=- \frac{\hbar^2}{2B} \Biggl[
\frac{1}{\beta^4}
\frac{\partial}{\partial \beta}
\beta^4  \frac{\partial}{\partial \beta}
+
\frac{1}{\beta^2}
\Biggl(
\frac{1}{\sin 3\gamma} 
\frac{\partial}{\partial \gamma} \sin 3\gamma \frac{\partial}{\partial \gamma}
\\
 - \frac{1}{4}
\sum_\kappa \frac{\hat{L}_\kappa^{\prime2}}{\sin^2(\gamma -
\frac{2}{3} \pi \kappa
)}
\Biggr)
\Biggr]
%%\\
+V(\beta) + 2 k g(\beta) \Lambdahat\circ\Sigmahat,
\end{multline}
where we shall again set $\hbar^2/(2B)\<=1$.  The physical
interpretation of the five-dimensional spin-orbit interaction is
discussed further in Appendix~\ref{appqq}.

The eigenproblem for the Hamiltonian~(\ref{eqnHcoupled}) is separable
into a radial and an angular part, as in Eqs.~(\ref{eqnangular})
and~(\ref{eqnradial}), provided $g(\beta)\<=1$ or $1/\beta^2$.  The
interaction may be evaluated in terms of the Casimir
operators~(\ref{eqncasimirdefn}) of the total, core, and fermion
five-dimensional angular momentum algebras as
\begin{equation}
\label{eqnspinorbitcasimir}
\begin{aligned}
2\Lambdahat\circ\Sigmahat&=\Lambdahatbf\circ\Lambdahatbf-\Lambdahat\circ\Lambdahat-\Sigmahat\circ\Sigmahat\\
&=\tfrac{1}{2}\bigl[ C_2[\grpspinbf{5}]-C_2[\grpspinb{5}]-C_2[\grpspinf{5}]\bigr].
\end{aligned}
\end{equation}  The
choice $g(\beta)\<=1$ yields a Bose-Fermi dynamical symmetry in the
usual sense, since 
\begin{equation}
H=C_2[\grpe{5}] + \frac{k}{2}\bigl[ C_2[\grpspinbf{5}]-C_2[\grpspinb{5}]-C_2[\grpspinf{5}]\bigr].
\end{equation}
The interaction $\Lambdahat\circ\Sigmahat$
enforces a five-dimensional angular momentum coupling scheme in which
the states have both good $\grpso{5}$ angular momentum and good total
$\grpspinbf{5}$ angular momentum. 
The coupling of the core
representation $(\tau,0)$ and the fermion representation
$(\frac12,\frac12)$ gives two representations $(\tau_1,\frac12)$ with
$\tau_1\<=\tau\pm1/2$ for the total system, since\begin{equation}
\textstyle
(\tau,0) \otimes (\frac12,\frac12) = (\tau+\frac12,\frac12) \oplus
(\tau-\frac12,\frac12).
\end{equation}
The dynamical symmetry is characterized by
the subalgebra chain
%% \begin{equation}
%% \label{eqnchaine54}
%% \setlength{\arraycolsep}{0.5pt}
%% \left.
%% \begin{array}{ccccc}
%% &&\grpe{5}&\supset&\grpspinb{5}\\
%% &&\varepsilon&&(\tau,0)\\
%% \\
%% \grpu{4}&\supset&\grpspinf{6}&\supset&\grpspinf{5}\\
%% {}[1,0,0,0]     &&      (\tfrac12,\tfrac12,\tfrac12)    &&  (\tfrac12,\tfrac12)
%% \end{array}
%% \right\rbrace
%% \begin{array}{cccccc}
%% \\ % dummy for vert centering
%% \supset&\grpspinbf{5}&\supset&\grpspinbf{3}&\supset&\grpspinbf{2} ,\\
%% &(\tau_1,\tfrac12)&\nutt'&J&&M_J
%% \end{array}
%% \end{equation}
\begin{equation}
\label{eqnchaine54}
\left \lvert
\setlength{\arraycolsep}{0.5pt}
\begin{array}{ccccccccccccc}
\grpe{5}&\otimes&\grpu{4}&\supset&\grpspinb{5}&\otimes&\grpspinf{5}&
\supset&\grpspinbf{5}&\supset&\grpspinbf{3}&\supset&\grpspinbf{2} \\
\varepsilon && {}[1] &&(\tau,0)   && (\tfrac12,\tfrac12)&&(\tau_1,\tfrac12)&\nutt'&J&&M_J
\end{array}
\right\rangle,
\end{equation}
and the eigenstates are members of representations with the quantum
numbers indicated.  Alternatively, the choice $g(\beta)\<=1/\beta^2$
was considered in Ref.~\cite{iachello2005:geomsuper}.  In this case,
an analytic solution is still possible, much as for the generalized
degeneracy breaking interactions considered in Sec.~\ref{sece5gen}.
However, the states are not eigenstates of $C_2[\grpe{5}]$ and
therefore are no longer labeled by the quantum number $\varepsilon$ of
$\grpe{5}\otimes\grpu{4}$ in~(\ref{eqnchaine54}), much as discussed in
Sec.~\ref{sece5gen} for the $\grpe{5}$ case.

The angular states are obtained by the $\grpspin{5}$ tensor
coupling of core states $\lvert \tau \nutt LM_L\rangle\<\equiv\lvert
\Psi_{\tau\nutt{}LM_L}(\gamma,\vartheta)\rangle$ with single particle
states $\lvert \frac32 m_j\rangle$, according to the $\grpspin{5}\<\supset\grpspin{3}$ isoscalar
factors~\cite{kuyucak1982:diss,iachello1981:ibfm-spin6,vanisacker1987:spin6-spin5-isf,iachello1991:ibfm}.
The coupled state is
\begin{equation}
\label{eqnspin5state}
\lvert \tau \tau_1 J M_J \rangle
=
\sum_L \sum_{M_L,m_j} 
\ISFBF{\tau}{L}{\tau_1}{J}\ISF{L}{\frac32}{J}{M_L}{m_j}{M_J}
\lvert \tau LM_L\rangle\lvert {\textstyle \frac32} m_j\rangle,
\end{equation}
where multiplicity indices have been suppressed for simplicity.  The
calculation of the necessary isoscalar factors is discussed further in
Appendix~\ref{appisf}.

If the choice $g(\beta)\<=1$ is made for the core-fermion coupling
in~(\ref{eqnHcoupled}), the radial equation and its solutions are exactly
as in Sec.~\ref{sece5soln}.  The energies, however, include an
additional contribution $2k\langle\Lambdahat\circ\Sigmahat\rangle$
from the five dimensional spin orbit interaction, where
$\langle\Lambdahat\circ\Sigmahat\rangle$ is the eigenvalue of
$\Lambdahat\circ\Sigmahat$ acting on the angular
state~(\ref{eqnspin5state}).  The $\grpspin{5}$ eigenvalue
formula~(\ref{eqncasimireigen}) gives
\begin{equation}
\label{eqnspinorbiteigen}
\begin{aligned}
2\langle\Lambdahat\circ\Sigmahat\rangle&= 
\bigl[\tau_1(\tau_1+3)+\tfrac34\bigr]-\bigl[\tau(\tau+3)\bigr] -\bigl[\tfrac{10}4\bigr]\\
&=\begin{cases}
\tau & \tau_1=\tau+1/2\\
-(\tau+3) & \tau_1=\tau-1/2.
\end{cases}
\end{aligned}
\end{equation}
If instead the choice
$g(\beta)\<=1/\beta^2$ is made, the contribution of the core-fermion
coupling term is absorbed into the separation constant $\Lambda$ in
the radial equation~(\ref{eqnradial}), as
\begin{equation}
\begin{aligned}
\Lambda&=\tau(\tau+3)+2k\langle\Lambdahat\circ\Sigmahat\rangle\\
&=\tau(\tau+3)+k\bigl[\tau_1(\tau_1+3)-\tau(\tau+3)-\tfrac74\bigr].
\end{aligned}
\end{equation}
The eigenvalues and eigenfunctions are given, as in the even-even
case, by Eqs.~(\ref{eqneigen}) and~(\ref{eqnf}), but now the order
$\nu(\tau_1,\tau)\<=(\Lambda+9/4)^{1/2}$ of the Bessel functions is
determined by the separation constant as given in~(\ref{eqnsepe54}).
It therefore depends upon the strength $k$ of the core-fermion
coupling.  

In practice, introduction of the $1/\beta^2$ dependence
does not produce major qualitative changes in the predictions for
observables (see Sec.~\ref{sece5gen}).  Therefore, for the remainder
of this article, we shall restrict our attention to the simpler,
$\beta$-independent core-fermion coupling
$2k\Lambdahat\circ\Sigmahat$.  This choice allows the algebraic
classification scheme~(\ref{eqnchaine54}) to be retained.

The eigenstates of the $\grpe{5|4}$ Hamiltonian are specified by
the angular quantum numbers ($\tau_1$, $\tau$, $L$, and
$M_L$) together with the radial quantum number $\xi$
($\xi\<=1,2,\ldots$).  Each value of $\tau_1$ yields a
multiplet of degenerate states of various angular momenta, according
to the $\grpspin{5}\supset\grpspin{3}$ branching
rules~\cite{iachello1981:ibfm-spin6}.  The angular momentum contents
of the lowest $\grpspin{5}$ representations $(\tau_1,1/2)$ are
summarized for convenience in Table~\ref{TabSpin5Branching}.  For
$\tau_1\<\geq7/2$ the same angular momentum can occur more than once
within a $\grpspin{5}$ representation, and so a multiplicity index
$\nutt'$ is required.  The eigenstates may thus be denoted $\lvert \xi
\tau_1 \tau \nutt' L M_L \rangle$.  

\begin{figure}
\begin{center}
\includegraphics[width=0.9\hsize]{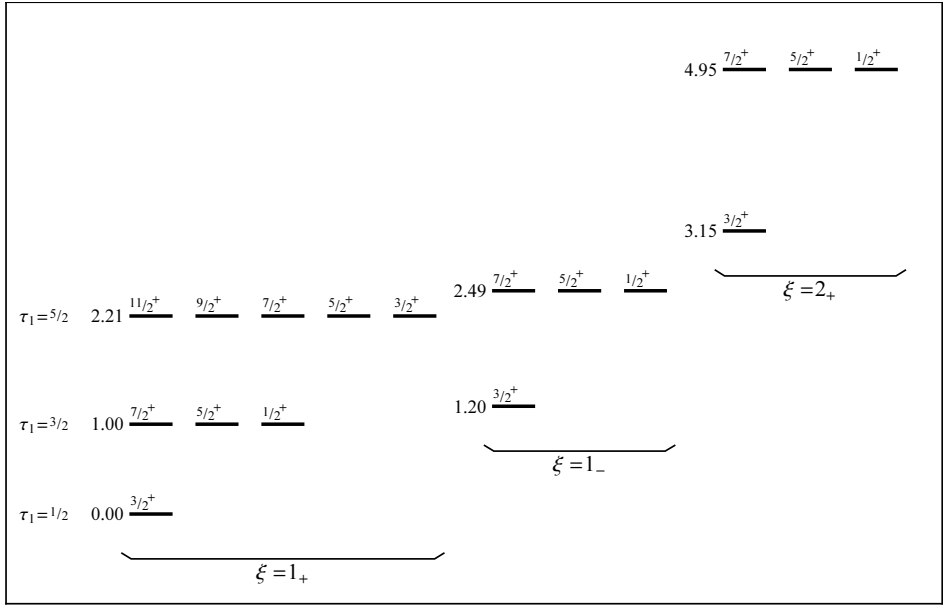}
\end{center}
%%\vspace{-12pt}
\caption
{Level scheme for the $\grpe{5|4}$ model for $k\<=-1/2$, showing the
excitation energies of the lowest $\tau_1$ multiplets of the ground
state, spin-flip excited ($\xi\<=1_-$), and first radially excited
($\xi\<=2_+$) families of levels.  All energies are normalized to
$E(7/2^+_1)$.  The $\xi\<=1_-$ family of levels was not included in
the corresponding figure of Ref.~\cite{iachello2005:geomsuper}.
\label{FigCoupledLevels}
}
\end{figure}

The addition of the core-fermion coupling degree of freedom implies
that there are \textit{two} fundamental types of excitation in the
$\grpe{5|4}$ system: \textit{radial} excitations, involving changes in
the radial quantum number $\xi$, and five-dimensional
\textit{spin-flip} excitations, involving the transition between
$\tau_1\<=\tau+1/2$ states and $\tau_1\<=\tau-1/2$ states.  (Note that
the spin-flip excitations were not addressed in
Ref.~\cite{iachello2005:geomsuper}.)  The excitation spectrum for the
$\grpe{5|4}$ model (with $k\<=-1/2$) is shown in
Fig.~\ref{FigCoupledLevels}.  A concise composite notation for the two
types of excitations is obtained by attaching a $\pm$ sign to the
$\xi$ label to indicate $\tau_1\<=\tau\pm1/2$, as suggested in
Ref.~\cite{fetea2006:135ba-beta-e54}.  That is, the families of levels
designated $\xi\<=1_+$, $2_+$, $\ldots$ have aligned $\grpspin{5}$
coupling ($\tau_1\<=\tau+1/2$), while those designated $\xi\<=1_-$,
$2_-$, $\ldots$ have anti-aligned $\grpspin{5}$ coupling
($\tau_1\<=\tau-1/2$).
% ---
% manually generated from TabSO5Branching as template
\begin{table}
\caption{Angular momentum contents of the representations $(\tau_1,\tfrac12)$ of $\grpspin{5}$, for $\tau_1\<\leq7/2$.}
\label{TabSpin5Branching}
\vspace{1ex}
\begin{tabular}{rl}
\hline
$\tau_1$&$J$\\
\hline
$1/2$&$3/2$\\
$3/2$&$7/2~~5/2~~1/2$\\
$5/2$&$11/2~~9/2~~7/2~~5/2~~3/2$\\
$7/2$&$15/2~~13/2~~11/2~~9/2~~9/2~~7/2~~5/2~~3/2$\\
\hline
\end{tabular}
\end{table}

% ---
% ---
% ---

\begin{figure}
\begin{center}
\includegraphics[width=0.6\hsize]{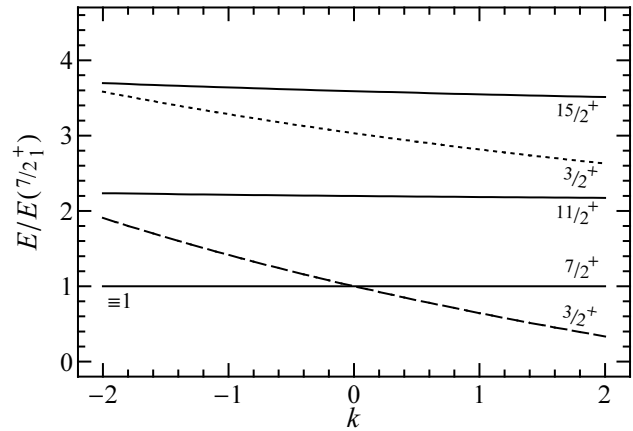}
\end{center}
%%\vspace{-12pt}
\caption
{Evolution of energies as functions of
the coupling parameter $k$, for the $\grpe{5|4}$ model.  Energies are
shown for the yrast $7/2^+$, $11/2^+$, and $15/2^+$ levels (solid curves),
the $\xi\<=1_-$ family head
$3/2^+_{\xi=1_-}$ (dashed curve), and the $\xi\<=2_+$ family
head $3/2^+_{\xi=2_+}$ (dotted
curve).  Energies are normalized to $E(7/2^+_1)$.
\label{FigCoupledEvolnk}
}
\end{figure}

The evolution of level energies which occurs with changing
core-fermion interaction strength $k$ is summarized in
Fig.~\ref{FigCoupledEvolnk}.  The coupling strength strongly affects
the energies of the excited ($\xi\<=1_-$ and $\xi\<=2_+$) families, as
discussed further in Sec.~\ref{sece54expt}.  The value $k\<=-1/2$ will
be used as a representative value for the remainder of this article,
since it yields reasonable excitation energies for use in
phenomenological interpretation, \textit{e.g.}, of $\mathrm{Ba}$
nuclei.  Electromagnetic transition strengths are independent of the
coupling strength.
% ---
% ---

\subsection{Electromagnetic transition strengths}
\label{sece54trans}

\subsubsection{General properties}

For the coupled odd-mass system, both collective (core) and fermionic
contributions are present in the electromagnetic transition strengths.
The transition operators of Sec.~\ref{sece5trans} must be extended to
contain fermionic terms, so
\begin{equation}
T=T_B+T_F,
\end{equation}
where $T_B$ acts only on the core and $T_F$ acts only on the 
fermion.  The $\grpspin{5}$ and, of course, $\grpspin{3}$ tensor
properties of the operators play an important role in defining
selection rules.  Selection rules on the quantum numers of both the
core ($\tau$ and $L$) and the coupled system ($\tau_1$ and $J$)
determine the overall pattern of allowed transitions. The tensor
properties of the relevant operators are summarized in
Table~\ref{TabTensors}.
% ---
% manually generated

\begin{table}
\caption{The $\grpspin{5}\<\supset\grpspin{3}$ tensor transformation properties of core and
fermionic operators.  A compound superscript notation such as
$\Galphaalpha{2,4}$ indicates that the $\grpspin{3}$ tensor products with both
angular momenta transform together as a single tensor under
$\grpspin{5}$.}
\label{TabTensors}
\vspace{1ex}
\renewcommand{\arraystretch}{1.3}
\begin{tabular}{llll}
\hline
%% \multicolumn{2}{c}{Operators}\\
%% \cline{1-2}
Core & Fermionic&$\grpspin{5}$&$\grpspin{3}$\\
\hline
$\Galphaalpha{0}$ & $\Gaa{0}$ & $(0,0)$ & $0$ \\
$\alpha$, $\tilde\pi$ & $\Gaa{2}$ & $(1,0)$ & $2$\\
$\Galphaalpha{2,4}$ &  & $(2,0)$ & $2,4$ \\
$\Lambdahat\propto\Galphapi{1,3}\,^a$& $\Sigmahat\propto\Gaa{1,3}\,^b$   & $(1,1)$ & $1,3$ \\
~ & $a^\dagger$, $\tilde{a}$ & $(\frac12,\frac12)$ & $\frac32$
\vspace{2pt}
\\
\hline
\end{tabular}
\begin{tablenotes}
\footnotesize
$^a$The operator $\Lhat$ is proportional to the component $\Lambdahat^{(1)}$ of this tensor.\\
$^b$The operator $\jhat$ is proportional to the component $\Sigmahat^{(1)}$ of this tensor.
\end{tablenotes}
\end{table}

% ---

The problem of computing transition strengths for the coupled
$\grpe{5|4}$ system can be simplified to that of calculating core
matrix elements and fermionic matrix elements separately.  First, the
$\grpspinbf{5}$-coupled states defined in~(\ref{eqnspin5state}) must
be decomposed in terms of states coupled at the $\grpspinbf{3}$ level,
using the $\grpspin{5}\<\supset\grpspin{3}$ isoscalar factors.  Then
the usual Racah reduction formulas~\cite{edmonds1960:am} for
angular-momentum coupled systems apply.  For a core operator
$T_B^{(\lambda)}$ of multipolarity $\lambda$, this yields
\begin{multline}
\langle \xi' \tau' \tau_1' J' \Vert
T_B^{(\lambda)}
\Vert \xi \tau \tau_1 J \rangle
=
\sum_{L',L}
\ISFBF{\tau'}{L'}{\tau_1'}{J'}
\ISFBF{\tau}{L}{\tau_1}{J}
\\\times
(-)^{L'+J+\lambda+3/2}
(2J'+1)^{1/2}(2J+1)^{1/2}
\sixj{L'}{J'}{\frac32}{J}{L}{\lambda}
\langle \xi' \tau' L' \Vert T_B^{(\lambda)} \Vert \xi \tau L \rangle.
\end{multline}
Here the core matrix element $\langle \xi' \tau' L' \Vert
T_B^{(\lambda)} \Vert \xi \tau \rangle$ is calculated as in
Sec.~\ref{sece5trans}, though now the Bessel function orders and
eigenvalues appearing in the radial integral~(\ref{eqnradialme}) are
the $\grpe{5|4}$ values from~(\ref{eqnsepe54}).   
Similarly, for a single particle operator $T_F^{(\lambda)}$ of multipolarity $\lambda$, 
\begin{multline}
\langle \xi' \tau' \tau_1' J' \Vert
T_F^{(\lambda)}
\Vert \xi \tau \tau_1 J \rangle
=
\sum_{L}
\ISFBF{\tau}{L}{\tau_1'}{J'}
\ISFBF{\tau}{L}{\tau_1}{J}
\\\times
(-)^{L+J'+\lambda+3/2}
(2J'+1)^{1/2}(2J+1)^{1/2}
\sixj{\frac32}{J'}{L}{J}{\frac32}{\lambda}\delta_{\xi'\xi}\delta_{\tau'\tau}
\langle {\textstyle\frac32} \Vert T_F^{(\lambda)} \Vert {\textstyle\frac32}  \rangle.
\end{multline}
The fermionic matrix element is given by
\(
\langle \frac32 \Vert \Gaa{\lambda} \Vert \frac32  \rangle
\<=
-(2\lambda+1)^{1/2}
\).

The angular matrix elements for the various transition operators are
common to all problems involving the coupling of a $\gamma$-soft core
to a $j\<=3/2$ fermion in the $\grpspin{5}$ coupling scheme.  These
generic angular matrix elements are tabulated for reference in
Appendix~\ref{appcoupledrme}.  Only the radial matrix elements depend
upon the specific choice of core potential (the square well in this
case) or possible $\beta$ dependence of the core-fermion coupling.  

\subsubsection{$E2$ transitions}

\begin{figure}
\begin{center}
\includegraphics[width=0.9\hsize]{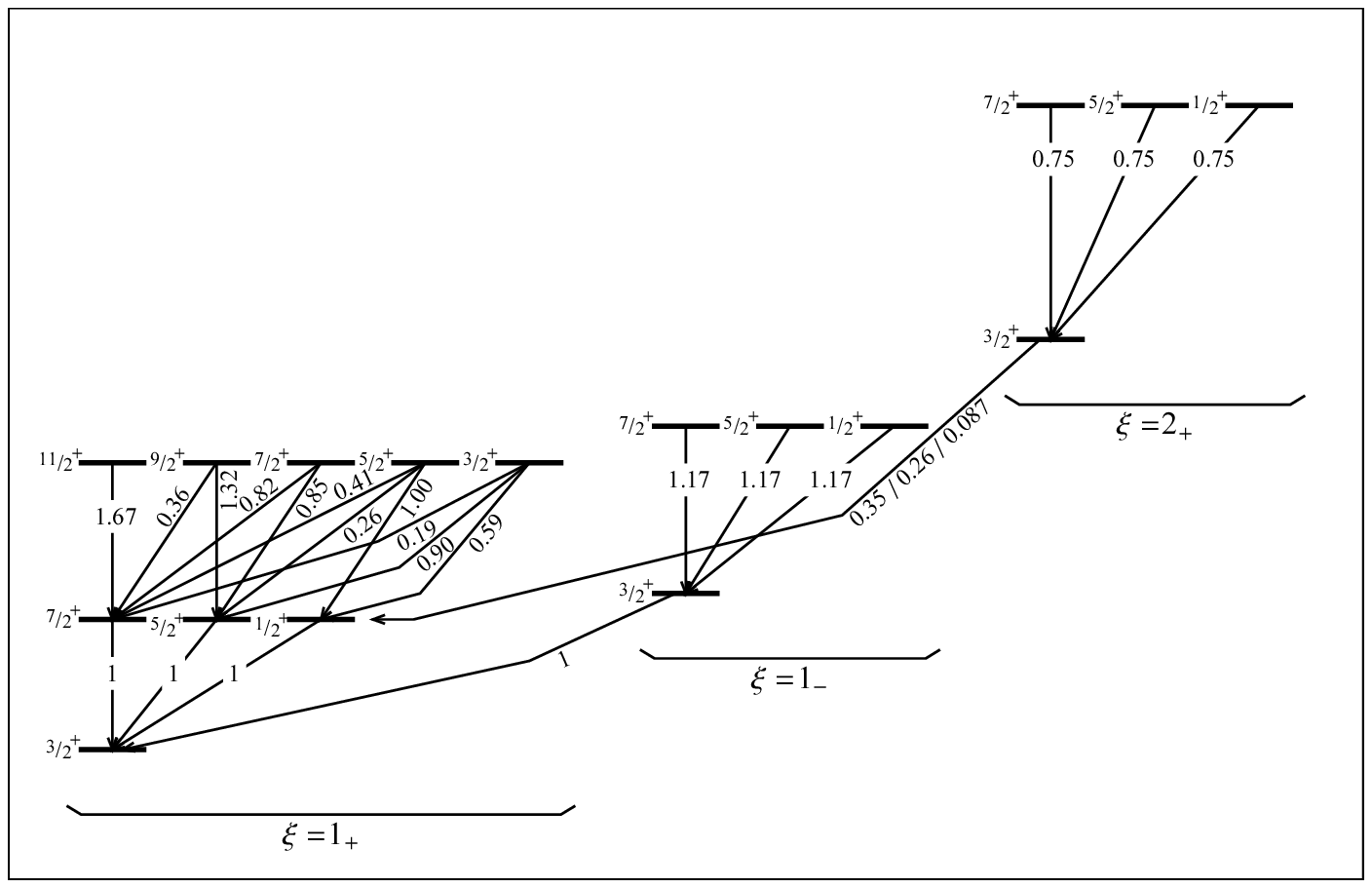}
\end{center}
%%\vspace{-12pt}
\caption
{$B(E2)$ strengths induced by the operator $t_{21} \alpha$, for the
$\grpe{5|4}$ model.  Transitions obey the selection
rules $\Delta
\tau\<=\pm1$ and $\Delta \tau_1\<=0$ or $\pm1$.  All transition strengths are normalized
relative to $B(E2;7/2^+_1\rightarrow3/2^+_1)$.  For absolute values,
strengths should be multiplied by $(0.07453) t_{21}^2 \beta_W^2$.
\label{FigCoupledAlpha}
}
\end{figure}

Incorporating the leading fermionic contribution into the $E2$
operator~(\ref{eqncorete2}) gives
\begin{equation}
\label{eqncoupledte2}
T^{(E2)}= t_{21} \alpha+ t_{22} \Galphaalpha{2} + t_{22}' \Gaa{2}.
\end{equation}
The leading order collective term $\alpha$ is a $(1,0)$ tensor
operator with respect both to $\grpso{5}$, as already noted, and to
$\grpspinbf{5}$.  The former property leads to the selection rule
$\Delta\tau\<=\pm1$, the latter to the selection rule
$\Delta\tau_1\<=0$ or
$\pm1$~\cite{kuyucak1982:diss,iachello1981:ibfm-spin6}.  The
$\Delta\tau\<=1$, $\Delta\tau_1\<=\pm1$ transitions connect states in
successive $\tau_1$ multiplets of the \textit{same} family of states
(\textit{e.g.}, producing transitions within the $\xi\<=1_-$ or
$\xi\<=1_+$ families).  The $\Delta\tau\<=1$,
$\Delta\tau_1\<=0$ transitions instead connect states of a given
$\tau_1$ multiplet in one family with those in the corresponding
$\tau_1$ multiplet of the spin-flip excited family (\textit{e.g.},
connecting the $\xi\<=1_-$ and $\xi\<=1_+$ families).  Intra-multiplet
transitions and quadrupole moments are forbidden for the leading-order
collective transition operator, as in the
even-even case.  The transition strengths resulting from the leading
order operator are shown in Fig.~\ref{FigCoupledAlpha}.
% ---
% ---
% ---
\begin{figure}
\begin{center}
\includegraphics[width=0.9\hsize]{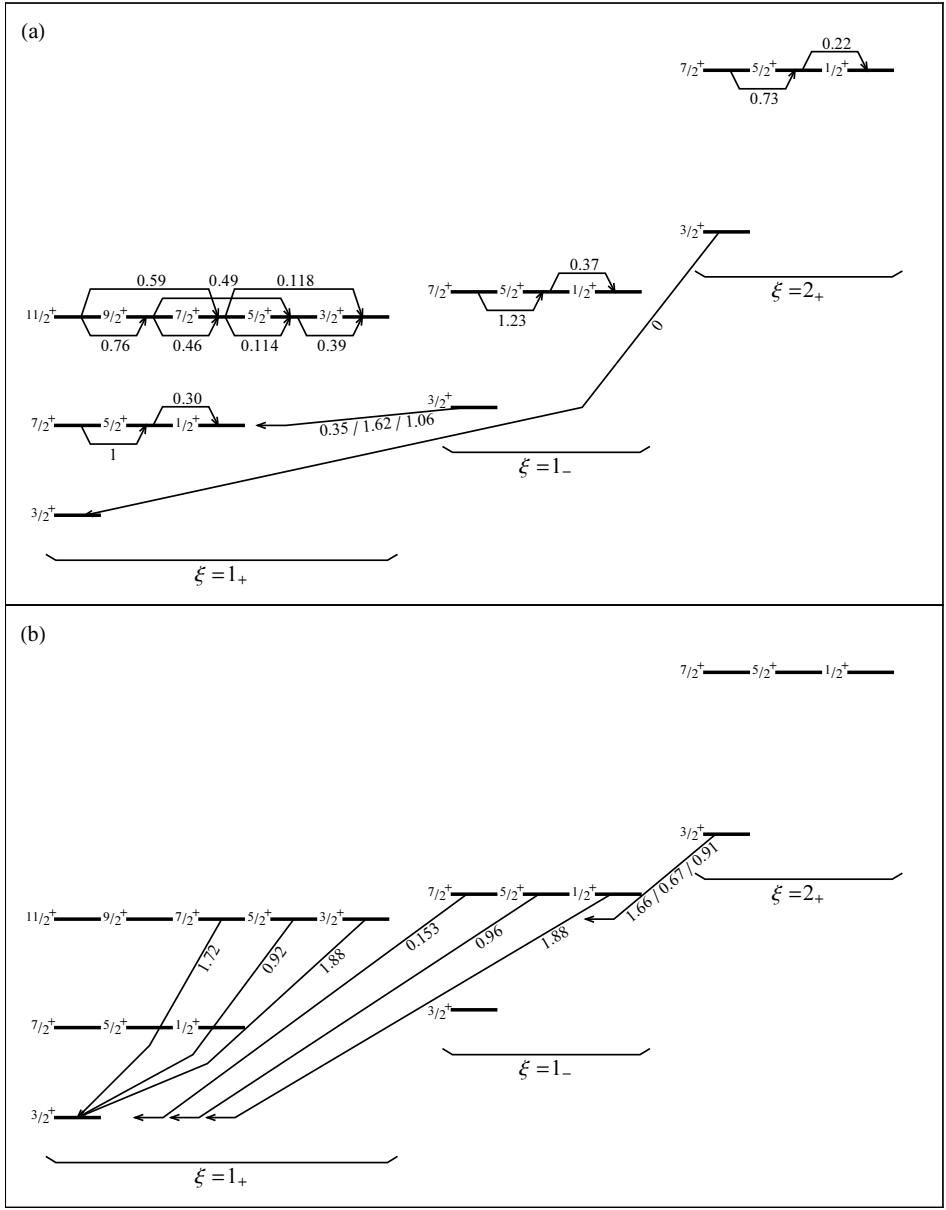}
\end{center}
%%\vspace{-12pt}
\caption
{$B(E2)$ strengths induced by the operator $t_{22}
(\alpha\times\alpha)^{(2)}$, for the $\grpe{5|4}$ model:
(a)~transitions with $\Delta
\tau\<=0$ (and $\Delta \tau_1\<=0$ or $\pm1$) and (b)~transitions with $\Delta
\tau\<=\pm2$ (and $\Delta \tau_1\<=\pm1$ or $\pm2$).  All transition strengths are normalized 
relative to $B(E2;7/2^+_1\rightarrow5/2^+_1)$.  For absolute values,
transition strengths should be multiplied by $(0.005194) t_{22}^2
\beta_W^4$.
\label{FigCoupledAlphaAlpha2Trans}
}
\end{figure}
% ---
% ---
\begin{figure}
\begin{center}
\includegraphics[width=0.9\hsize]{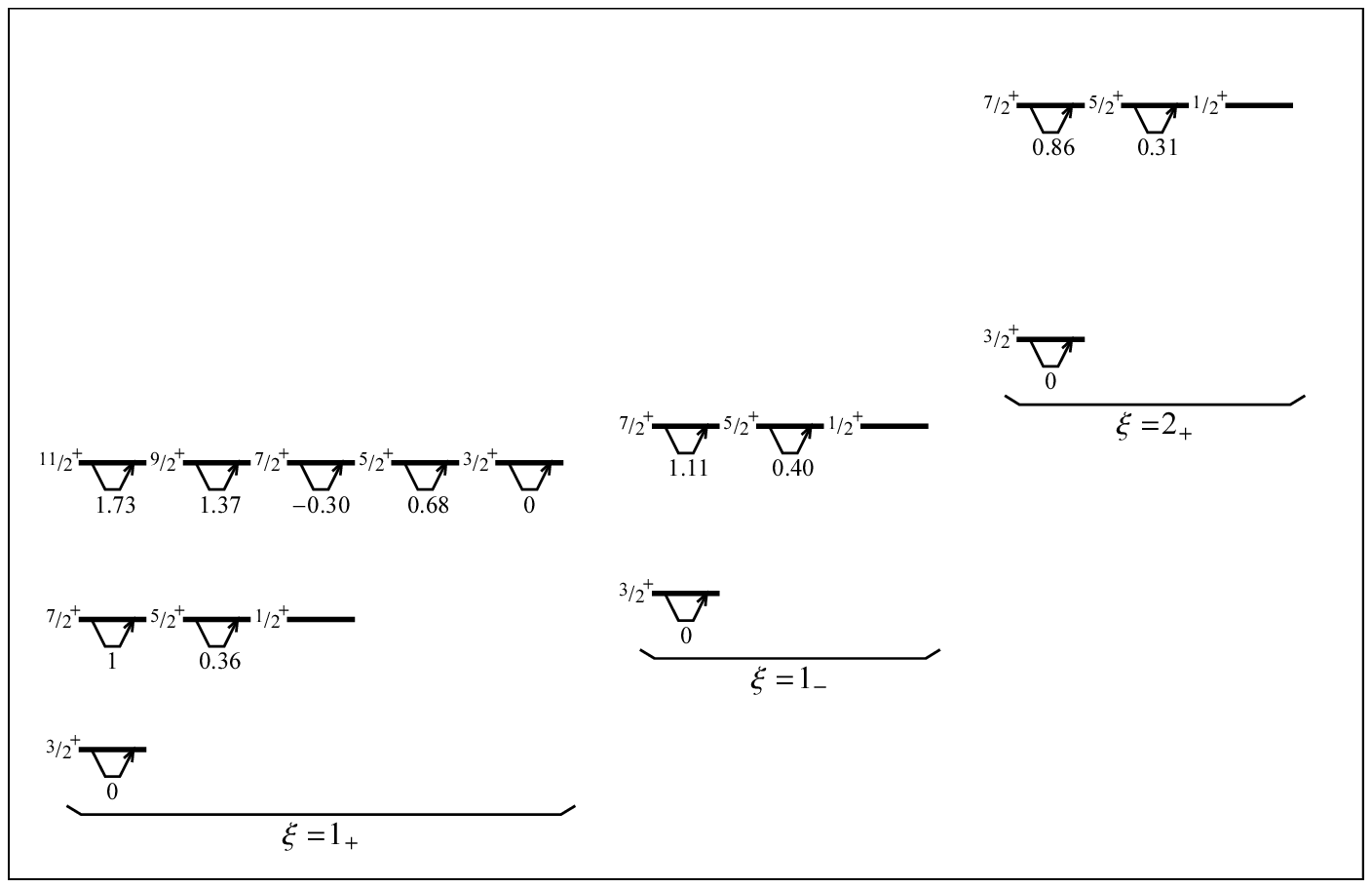}
\end{center}
%%\vspace{-12pt}
\caption
{Quadrupole moments induced by the operator $t_{22}
(\alpha\times\alpha)^{(2)}$, for the $\grpe{5|4}$ model.  All moments are normalized relative to $Q(7/2^+_1)$.  For
absolute values, moments should be multiplied by $(0.2124) t_{22} \beta_W^2$.
\label{FigCoupledAlphaAlpha2Moment}
}
\end{figure}
% ---
% ---
% may need to fix up in file to placement [t] to avoid doubling up
% with figure on a float page
% LaTeXTable style ElsartTiny
\begin{table}
\caption{Selected $B(E2)$ and $B(M1)$ values for transitions involving interference between core and fermionic contributions, for the $\grpe{5|4}$ model.}
\label{TabCoupledTrans}
\vspace{1ex}
\tiny
\begin{tabular}{rrrrrrrr}
\hline
$\xi$&$\tau_1$&$J$&$\xi'$&$\tau_1'$&$J'$&\multicolumn{1}{c}{$B(E2)$}&\multicolumn{1}{c}{$B(M1)$}\\
\hline
$1_+$&$3/2$&$7/2$&$1_+$&$3/2$&$5/2$&$[(-0.072)t_{22}\beta_W^2+(-0.717)t_{22}']^2$&$[(-0.293)t_{10}+(-0.414)t_{12}']^2$\\
$1_+$&$3/2$&$5/2$&$1_+$&$3/2$&$1/2$&$[(+0.040)t_{22}\beta_W^2+(-0.592)t_{22}']^2$&\\
$1_-$&$1/2$&$3/2$&$1_+$&$3/2$&$7/2$&$[(+0.042)t_{22}\beta_W^2+(-0.632)t_{22}']^2$&\\
$1_-$&$1/2$&$3/2$&$1_+$&$3/2$&$5/2$&$[(+0.092)t_{22}\beta_W^2+(+0.548)t_{22}']^2$&$[(+0.458)t_{10}+(+0.648)t_{12}']^2$\\
$1_-$&$1/2$&$3/2$&$1_+$&$3/2$&$1/2$&$[(-0.074)t_{22}\beta_W^2+(+0.316)t_{22}']^2$&$[(-0.387)t_{10}+(-0.548)t_{12}']^2$\\
\hline
\end{tabular}
\end{table}

% ---
% ---
% LaTeXTable style Elsart
\begin{table}
\caption{Selected $Q$ and $\mu$ values, including both core and fermionic contributions, for the $\grpe{5|4}$ model.}
\label{TabCoupledMoment}
\vspace{1ex}
\begin{tabular}{rrrr@{}lr@{}l}
\hline
$\xi$&$\tau_1$&$J$&\multicolumn{2}{c}{$Q(J)$}&\multicolumn{2}{c}{$\mu(J)$}\\
\hline
$1_+$&$1/2$&$3/2$&$$&$(-1.585)t_{22}'$&$$&$(-1.373)t_{12}'$\\
$1_+$&$3/2$&$7/2$&$(+0.212)t_{22}\beta_W^2+$&$(-1.585)t_{22}'$&$(+1.294)t_{10}+$&$(-1.373)t_{12}'$\\
$1_+$&$3/2$&$5/2$&$(+0.076)t_{22}\beta_W^2+$&$(+0.226)t_{22}'$&$(+1.017)t_{10}+$&$(-0.850)t_{12}'$\\
$1_+$&$3/2$&$1/2$&&&$(+0.647)t_{10}+$&$(+0.458)t_{12}'$\\
$1_-$&$1/2$&$3/2$&$$&$(+0.951)t_{22}'$&$(+0.777)t_{10}+$&$(-0.275)t_{12}'$\\
$2_+$&$1/2$&$3/2$&$$&$(-1.585)t_{22}'$&$$&$(-1.373)t_{12}'$\\
\hline
\end{tabular}
\end{table}

% ---

The second-order collective quadrupole operator $\Galphaalpha{2}$
induces $\Delta\tau\<=0$ ($\Delta\tau_1\<=0$ or $\pm1$) transitions,
as shown in Fig.~\ref{FigCoupledAlphaAlpha2Trans}(a), and yields
nonvanishing quadrupole moments, shown in
Fig.~\ref{FigCoupledAlphaAlpha2Moment}.  The operator also induces
$\Delta\tau\<=\pm2$ ($\Delta\tau_1\<=\pm1$ or $\pm2$) transitions,
shown in Fig.~\ref{FigCoupledAlphaAlpha2Trans}(b).  The fermionic
quadrupole operator $\Gaa{2}$ likewise contributes to $\Delta\tau\<=0$
($\Delta\tau_1\<=0$ or $\pm1$) transitions and to the quadrupole
moments.  Any analysis of transition strengths must therefore address
the interference between these terms.  The combined strengths for the
most experimentally relevant transitions are given in
Table~\ref{TabCoupledTrans}, and the combined quadrupole moments are
given in Table~\ref{TabCoupledMoment}.

\subsubsection{$M1$ transitions}

The magnetic dipole operator~(\ref{eqncoretm1}) with fermionic
contribution is
\begin{equation}
\label{eqncoupledtm1}
\begin{aligned}
T^{(M1)}&= t_{10} [i\Galphapi{1}] + t_{11}
[\alpha\times i\Galphapi{1}]^{(1)}+ t_{12}' \Gaa{1}\\
&=t_{10} \tfrac1{\sqrt{10}}  \Lhat + t_{11}\tfrac1{\sqrt{10}}
(\alpha\times\Lhat)^{(1)} - t_{12}'\tfrac1{\sqrt5}\jhat.
\end{aligned}
\end{equation}
The leading collective term is proportional to the collective angular
momentum operator $\Lhat$, and, similarly, the fermionic term is
proportional to the single particle angular momentum operator $\jhat$.
% ---
\begin{figure}
\begin{center}
\includegraphics[width=0.9\hsize]{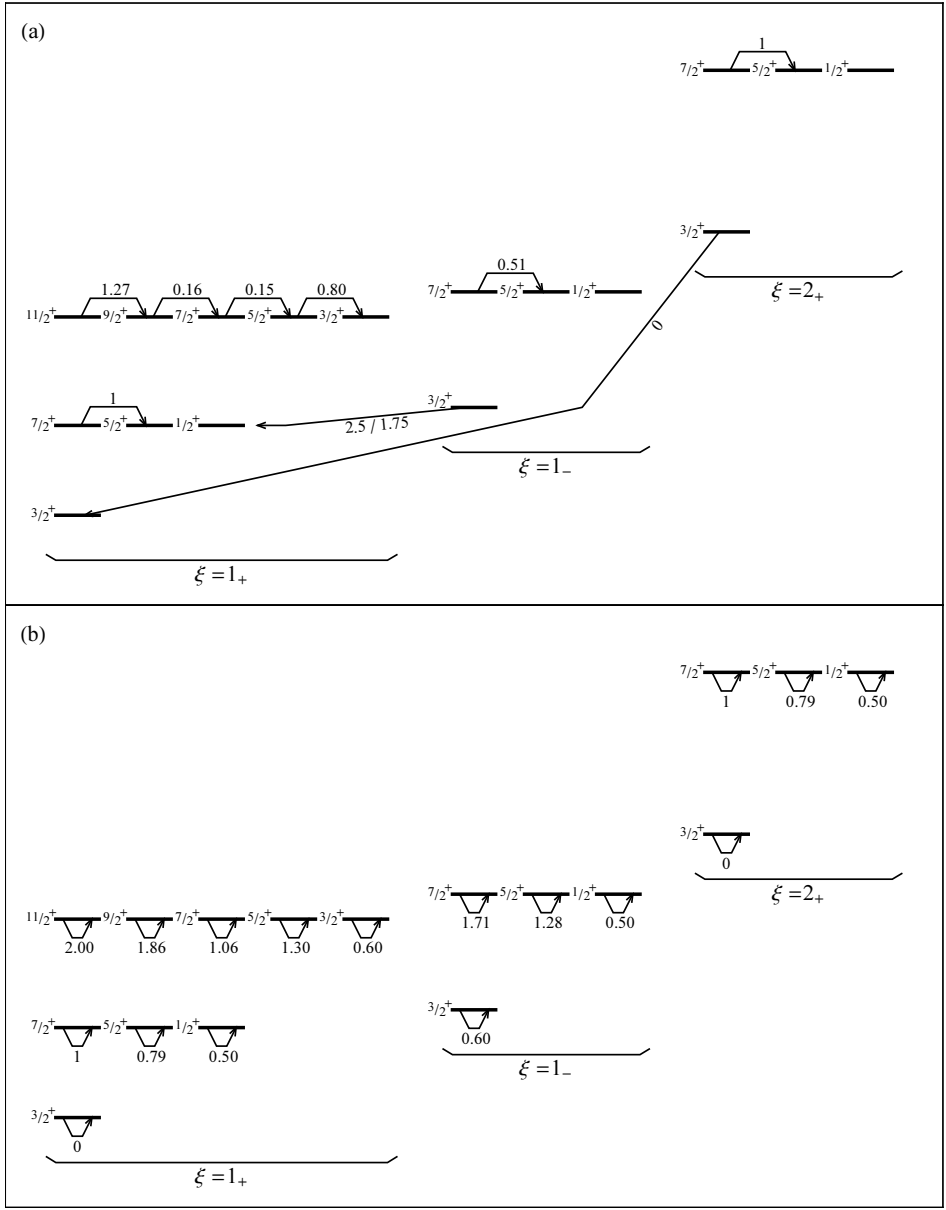}
\end{center}
%%\vspace{-12pt}
\caption
{(a)~$B(M1)$ strengths and (b)~dipole moments induced by the operator
$t_{10} i (\alpha\times\pi)^{(1)}$, for the $\grpe{5|4}$ model.
Transitions obey the selection rule $\Delta \tau\<=0$ and $\Delta
\tau_1\<=0$ or $\pm1$.  All transition strengths are normalized
relative to $B(M1;7/2^+_1\rightarrow5/2^+_1)$ and moments relative to
$\mu(7/2^+_1)$.  For absolute values, transition strengths should be
multiplied by $(0.08571) t_{10}^2$ and moments by $(1.294) t_{10}$.
\label{FigCoupledAlphaPi1}
}
\end{figure}
% ---

Though $\Lhat$ is a diagonal operator for the even-even system in
isolation (Sec~\ref{sece5trans}), it is no longer diagonal for the
coupled odd-mass system.  It can connect any two states which share
amplitudes for the same core wave function $\lvert
\xi \tau L\rangle$.  Transitions induced by $\Lhat$ obey the
selection rules 
$\Delta\tau\<=0$ and $\Delta\tau_1\<=0$ or $\pm1$.  The
transition strengths and dipole moments arising from this leading
collective term are shown in Fig.~\ref{FigCoupledAlphaPi1}.

The fermionic term $\jhat$ induces transitions and dipole moments
obeying the same selection rules as $\Lhat$, so, in general,
interference of the two operators is present.  The combination $\Jhat
= \Lhat + \jhat$ is the total angular momentum operator of the coupled
system, which is diagonal and has reduced matrix element
$[J(J+1)(2J+1)]^{1/2}$.  The matrix elements of the two operators
$\Lhat$ and $\jhat$ are therefore closely related.  In particular, for
all off-diagonal matrix elements (dipole transitions),
\(
\langle \xi'\tau'\tau_1'J' \Vert \Lhat \Vert \xi\tau\tau_1 J \rangle
\<= 
- \langle \xi'\tau'\tau_1'J' \Vert \jhat \Vert \xi\tau\tau_1 J \rangle
\).
Consequently, although interference of the core and fermionic
contributions to $T^{(M1)}$ affects the overall normalization of the
$\Delta\tau\<=0$ dipole transition strengths, which are proportional
to $(t_{10}+\sqrt{2}t_{12}')^2$, it does not affect the relative
strengths of different transitions (Table~\ref{TabCoupledTrans}).  For
diagonal matrix elements (dipole moments), interference affects the
values in a less trivial fashion (Table~\ref{TabCoupledMoment}).

The higher-order collective term $(\alpha\times\Lhat)^{(1)}$ in the
dipole operator yields transitions obeying the selection rule
$\Delta\tau\<=\pm1$.  The most experimentally accessible
$\Delta\tau\<=\pm1$ transitions are those between the various
low-lying $\tau_1\<=3/2$ and $\tau_1\<=1/2$ states.  However, the
matrix elements of $(\alpha\times\Lhat)^{(1)}$ between these low-lying
states all vanish, as a result of the vanishing
$2^+\<\leftrightarrow0^+$ core transitions seen in
Fig.~\ref{FigCoreAlphaAlphaPi11}.  This limits the phenomenological
relevance of the higher-order transition operator.  Transition
strengths can be calculated if needed from the angular matrix elements
in Table~\ref{TabCoupledRMEAlphaAlphaPi11}.

\subsection{Comparison with experimental data}
\label{sece54expt}
% ---
\begin{figure}
\begin{center}
\includegraphics[width=0.6\hsize]{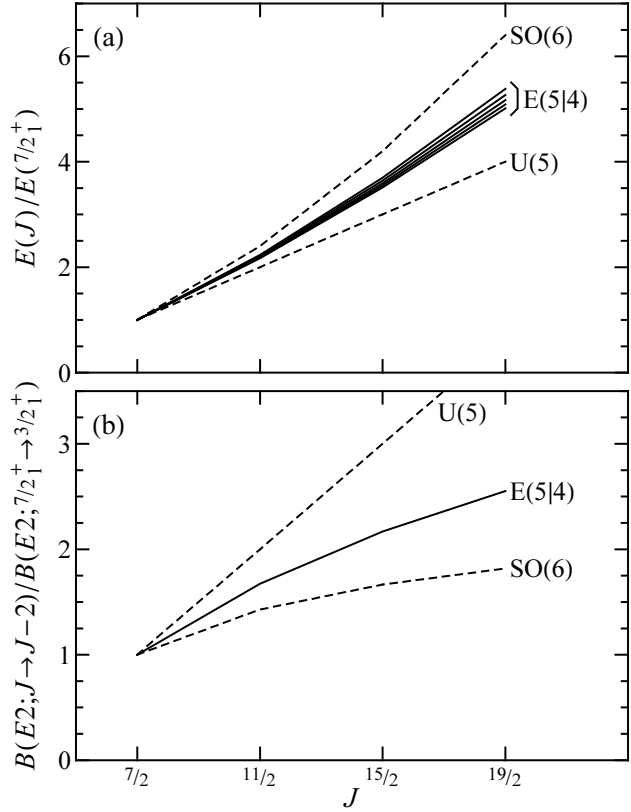}
\end{center}
%%\vspace{-12pt}
\caption
{(a)~Energies and (b)~$B(E2)$ strengths as functions of angular
momentum, for the yrast states ($\xi\<=1$, $\tau_1\<=\tau+1/2$,
$J\<=2\tau_1+1/2$) in the $\grpe{5|4}$ model.  Energies are normalized
to $E(3/2^+_1)$ and $B(E2)$ strengths to
$B(E2;7/2^+_1\rightarrow3/2^+_1)$.  The energies depend upon the
coupling constant $k$ and are shown here for $k\<=-2$, $-1$, $0$,
$+1$, and $+2$ (solid curves, uppermost to lowermost).  Values for the
$\grpu{5}$ and $\grpso{6}$ dynamical symmetries in the large boson
number limit are shown for comparison (dashed curves).
\label{FigCoupledYrast}
}
\end{figure}
% ---

A comparison of the $\grpe{5|4}$ description with experimental data
for odd-mass nuclei is more difficult than the comparison of the
$\grpe{5}$ description with even-even nuclei, because the occurence of
other single particle orbitals near the $j\<=3/2$ orbital can
considerably perturb the spectrum and electromagnetic transition
strengths.  Much as $\grpso{5}$ symmetry dictates the multiplet structure
and branching patterns for the $\grpe{5}$ predictions
(Sec.~\ref{sece5expt}), the basic form of the $\grpe{5|4}$ predictions
is a consequence of the $\grpspin{5}$ coupling scheme, rather than of
the $\grpe{5|4}$ square well potential \textit{per se}.  Therefore, we
concentrate here upon those distinguishing observables which vary
along the $\grpu{5}$--$\grpso{6}$ transition.

However, the $\grpspin{5}$ coupling scheme is itself an important
aspect of the model to test.  In the IBFM, the $\grpu{5}$ symmetry
limit is treated in the $\grpspin{5}$ coupling scheme, but the
$\grpso{6}$ limit is instead treated in a $\grpspin{6}$ coupling
scheme~\cite{iachello1981:ibfm-spin6}.  While the $\grpspin{5}$
coupled states considered in the present work are $\tau$ eigenstates,
the $\grpspin{6}$ coupled states are near-equal $\tau\<=\tau_1\pm1/2$
admixtures.  The leading order $E2$ operator ($\Delta\tau\<=\pm1$)
therefore yields nonvanishing matrix elements between members of a
$\tau_1$ multiplet (see, \textit{e.g.}, Fig.~5 of
Ref.~\cite{iachello1981:ibfm-spin6}), by connecting the
$\tau\<=\tau_1\pm1/2$ components of the different states.

The observables most sensitive to the $\grpu{5}$--$\grpso{6}$
structural transition are the yrast properties and the energies and
decay properties of the spin-flip ($\xi\<=1_-$) and radial
($\xi\<=2_+$) excitations.  The yrast energies and $B(E2)$ strengths
are shown in Fig.~\ref{FigCoupledYrast}.  Note that there is a
moderate dependence of the yrast energies
[Fig.~\ref{FigCoupledYrast}(a)] upon the core-fermion coupling
strength $k$.  Since $2\langle\Lambdahat\circ\Sigmahat\rangle\<=\tau$
for the ground state family, a positive contribution from this term
($k\<>0$) produces a more linear [$\grpu{5}$-like] yrast energy curve.
The spin-flip excitation energy is, naturally, strongly dependent upon
the strength of the five-dimensional spin-orbit interaction
(Fig.~\ref{FigCoupledEvolnk}).  The spin-flip excitation varies from
near degeneracy with the ground state to near degeneracy with the
$\tau_1\<=5/2$ multiplet over the range of $k$ values considered
($-2\<\leq k \<\leq +2$).  It therefore serves as the main basis for
determining the coupling strength $k$ within the model.  For $k\<=0$,
the weak-coupling limit is recovered, and the $3/2^+$ state
($\tau_1\<=1/2$) of the $\xi\<=1_-$ excitation forms a degenerate
multiplet with the $7/2^+$, $5/2^+$, and $1/2^+$ states
($\tau_1\<=3/2$) of the $\xi\<=1_+$ ground state family.  The energy
of the radial excitation is somewhat less sensitive to $k$, varying
from $\rtrim\sim2.6$ to $3.6$ times the ground state $\tau_1\<=3/2$
multiplet energy over the same range of $k$ values
(Fig.~\ref{FigCoupledEvolnk}).

Only limited data are presently available for comparison with the
$\grpe{5|4}$ predictions.  The nucleus $^{135}\mathrm{Ba}$ has been
considered as a candidate for description by the $\grpe{5|4}$ model in
Ref.~\cite{fetea2006:135ba-beta-e54}.  Another possible candidate is
the nucleus $^{63}\mathrm{Cu}$, previously considered as a good
example of $\grpspin{5}$ symmetry for a $j\<=3/2$ particle coupled to
a $\grpu{5}$ boson core~\cite{bijker1984:ibfm-part2-u5}.  The
available energies and transition strengths are consistent with either a
$\grpu{5}$ interpretation or an $\grpe{5|4}$ interpretation, since,
unfortunately, they test only the common $\grpspin{5}$ symmetry.  In
order to assess whether or not $\grpe{5|4}$ might be a better
description, one must experimentally determine the properties of the
states in the $\xi\<=2_+$ family, in particular the energy of the
$J\<=3/2$ ($\tau_1\<=1/2$) state and its $E2$ transition strengths to
the members of the $\tau_1\<=3/2$ multiplet of the ground state family.

\subsection{Generalizations of the $\grpe{5|4}$ description}
\label{sece54gen}

\begin{figure}
\begin{center}
\includegraphics[width=0.9\hsize]{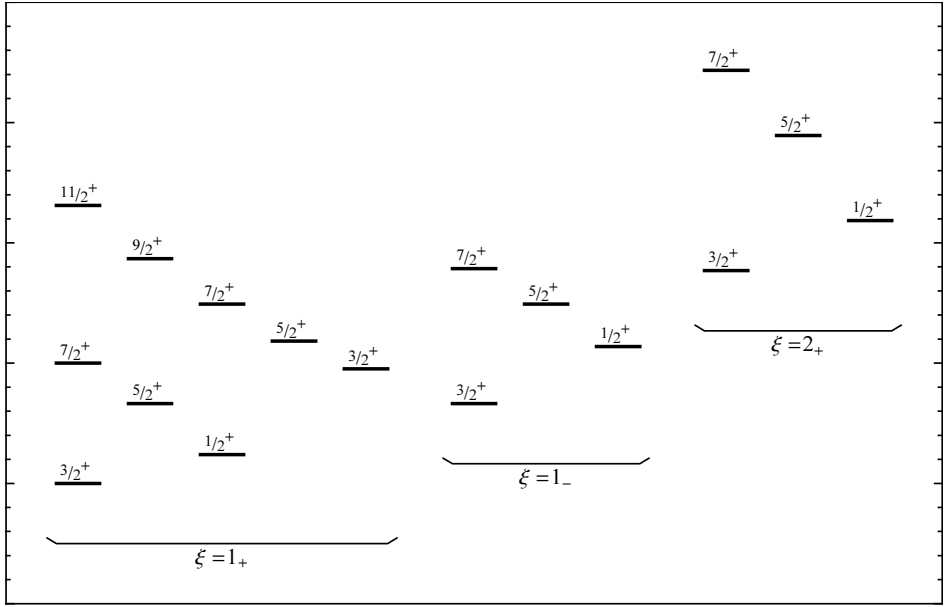}
\end{center}
%%\vspace{-12pt}
\caption
{Level scheme for the $\grpe{5|4}$ model, with core-fermion coupling
parameter value $k\<=-1/2$ and angular momentum
degeneracy breaking parameter value $k'\<=1/2$.  All
energies are normalized to $E(7/2^+_1)$.
\label{FigCoupledLevelsSplit}
}
\end{figure}

For odd-mass nuclei, the energy splitting of the $\grpspin{5}$
multiplets is typically much more dramatic than the splitting of
$\grpso{5}$ multiplets discussed in Sec.~\ref{sece5gen}.  It must
therefore be considered carefully in the analysis, and it cannot be
separated from the determination of $k$, since any significant
spin-dependent perturbation to the energies also substantially
modifies the yrast and excited family energy ratios considered above.
The generalized dynamical symmetry Hamiltonian for the subalgebra
chain~(\ref{eqnchaine54}) is of the form
\begin{multline}
\label{eqnHe54splitcasimir}
H = C_2[\grpe{5}]
+ c'' C_2[\grpspinb{5}] + c''' C_2[\grpspinf{5}] 
\\
+ c C_2[\grpspinbf{5}] + c' C_2[\grpspinbf{3}],
\end{multline}
or, equivalently, since $\Lambdahat\circ\Sigmahat$ is just a linear combination
of $C_2[\grpspinbf{5}]$, $C_2[\grpspinb{5}]$, and
$C_2[\grpspinf{5}]$,
\begin{equation}
\label{eqnHe54split}
H = \tilde\pi \cdot \tilde\pi+V(\beta)
+ 2 k {\Lambdahat\circ\Sigmahat}
+ k''{\Lambdahat\circ\Lambdahat}
+ k'''{\Sigmahat\circ\Sigmahat}
+ k' {\Jhat\cdot\Jhat}.
\end{equation}
These coupling terms are all diagonal in the states described
by~(\ref{eqnchaine54}), and they affect only the energies, not eigenstates.
A level scheme for representative values of $k$ and $k'$ is shown in
Fig.~\ref{FigCoupledLevelsSplit}.
% ---
% ---

Alternatively, as in Sec.~\ref{sece5gen}, a limited variety of more
general $\beta$ dependences can be included in the degeneracy breaking
terms while still preserving the analytically solvable radial
equation,
\textit{e.g.},
\begin{equation}
\label{eqnHe54splitdep}
H = \tilde\pi \cdot \tilde\pi+V(\beta)
+ 2 k \frac{\Lambdahat\circ\Sigmahat}{\beta^2} 
+ k''\frac{\Lambdahat\circ\Lambdahat}{\beta^2}
+ k'''\frac{\Sigmahat\circ\Sigmahat}{\beta^2}
+ k' \frac{\Jhat\cdot\Jhat}{\beta^2}.
\end{equation}
For angular momentum degeneracy breaking, we need only consider $k'$
(and $k$) nonzero, giving
\begin{equation}
\label{eqnsepe54}
\Lambda=\tau(\tau+3)+k\bigl[\tau_1(\tau_1+3)-\tau(\tau+3)-\tfrac74\bigr]
+k' J(J+1).
\end{equation}

\section{Conclusion}
\label{secconcl}

In this article, we have considered solutions to the Bohr equation
with a square well potential [$\grpe{5}$] and for the equation obtained
by coupling this to an additional fermion by a five-dimensional
spin-orbit interaction [$\grpe{5|4}$].  These are examples of new
classes of dynamical symmetry and Bose-Fermi dynamical symmetry.
These symmetries are useful descriptions of even-even and odd-mass
nuclei near the critical point of the spherical [$\grpu{5}$] to
$\gamma$-unstable [$\grpso{6}$] phase transition in nuclei.

The results described here provide benchmarks for nuclei near the
critical point of the $\grpu{5}$--$\grpso{6}$ phase transition and can
be used as a basis for comparison with experiment.  However, since the
Hamiltonian exhibits $\grpso{5}$ symmetry (for even-even nuclei) or
$\grpspinbf{5}$ symmetry (for odd-mass nuclei) throughout this
transition, in either the algebraic or geometric formulations of the
problem, many observables are relatively insensitive to the
transition.  This makes comparison with experiment challenging.
Nonetheless, nuclei with properties close to those of the $\grpe{5}$
predictions have been
identified~\cite{casten2000:134ba-e5,arias2001:134ba-e5,frank2001:104ru-e5,zamfir2002:102pd-beta,zhang2002:108pd-e5},
and preliminary comparisons have been carried out for the $\grpe{5|4}$
description~\cite{fetea2006:135ba-beta-e54}.

The approach discussed here may be extended in several directions.
The situation considered in Sec.~\ref{sece54}, of coupling to a
$j\<=3/2$ particle, is relevant only to a restricted set of nuclei.
Beyond this, the more general possible combinations of angular momenta
for the fermion orbitals must also be considered.  The case of
coupling to orbitals with $j\<=1/2$, $3/2$, and $5/2$, termed
$\grpe{5|12}$, has been treated by Alonso, Arias, and
Vitturi~\cite{alonsoETAL-INPREP}.  Another important possibility to
consider is the embedding of the present Bose-Fermi dynamical symmetry
[$\grpe{5}\otimes\grpu{4}$] within a supersymmetry.  Within the
differential formulation, the related problem of a harmonic oscillator
with spin-orbit interaction has already been explored (in three
dimensions) by Balantekin~\cite{balantekin1992:osp12-shell}.  In this
case, an $\grposp{1|2}$ supersymmetry is present.  A problem of great
phenomenological relevance is the extension of the present approach to
the transition between spherical [$\grpu{5}$] and axially-symmetric
deformed [$\grpsu{3}$] structure.  While the problem can be solved by
numerical diagonalization of the Bohr
Hamiltonian~\cite{caprio2005:axialsep}, any analytic solution appears
to require substantial approximations~\cite{iachello2001:x5}.
Finally, the general concept of a dynamical symmetry based on the
square well potential, for application near the critical point of a
phase transition, may be applied to a broad variety of systems other
than quadrupole deformed nuclei.  The pairing phase
transition~\cite{clark2006:e2-pairing} and the linear-bent phase
transition in
molecules~\cite{iachello2003:floppy-molecules,perezbernal-INPREP}
are two examples.

\begin{ack}
This work was supported in part by the US DOE under grant
DE-FG02-91ER-40608.  We wish to thank R.~F.~Casten and N.~V.~Zamfir
for providing the first evidence for $\grpe{5}$ structure and
M.~S.~Fetea and collaborators for performing the first comparison of
$\grpe{5|4}$ with experimental data.  Discussions with A.~Vitturi,
J.~M.~Arias, and R.~Bijker are
gratefully acknowledged.
\end{ack}

\appendix

\section{Algebraic properties and the 
quadrupole-quadrupole core-fermion interaction}
\label{appqq}

In this appendix, we establish the relationship between the
quadrupole-quadrupole core-fermion interaction used in interacting
boson fermion model (IBFM) studies~\cite{iachello1991:ibfm} and the
five-dimensional spin-orbit interaction $\Lambdahat \circ \Sigmahat$
used in the present $\grpe{5|4}$ analysis.  This is especially of
interest to provide grounds for comparison with the recent study of
the $\grpu{5}$--$\grpso{6}$ transition in the IBFM by Alonso
\textit{et al.}~\cite{alonso2005:ibfm-gsoft-phase}.  In the process,
the basic definitions and relations are established for several
algebraic quantities used in the main text.

First, let us review the algebraic properties of the $\grpspin{6}$
algebra and its subalgebras in Racah
form~\cite{kuyucak1982:diss,iachello1981:ibfm-spin6,iachello1987:ibm,iachello1991:ibfm}.
The bosonic $\grpso{6}$ algebra and fermionic $\grpsu{4}$ algebra are
isomorphic, so we adopt the more uniform notation
$\grpspinb{6}\<\equiv\grpso{6}$ and $\grpspinf{6}\<\equiv\grpsu{4}$.
The commutation relations of these algebras are brought into identical
form if the generators are normalized as
\begin{equation}
\begin{aligned}
B^{(1)}&=\Gdd{1}& A^{(1)}&=-\frac{1}{\sqrt2}\Gaa{1}\\
B^{(2)}&=\Gsdds{2}&  A^{(2)}&=\Gaa{2}\\
B^{(3)}&=\Gdd{3}& A^{(3)}&=\frac{1}{\sqrt2}\Gaa{3}.
\end{aligned}
\end{equation}
The IBFM quadrupole operators are the quadrupole generators,
$\Qhat_B\equiv B^{(2)}$ and $\Qhat_F\equiv A^{(2)}$.  The algebra
$\grpspinbf{6}$ has generators
\begin{equation}
\label{eqndefng}
G^{(\lambda)}=B^{(\lambda)}+A^{(\lambda)}.
\end{equation}
The subalgebras $\grpspinb{5}\<\equiv\grpso{5}$,
$\grpspinf{5}\<\equiv\grpsp{4}$, and
$\grpspinbf{5}$ are obtained by
omitting the quadrupole generators $B^{(2)}$, $A^{(2)}$, and
$G^{(2)}$.  The subalgebras $\grpspinb{3}\<\equiv\grpso{3}$,
$\grpspinf{3}\<\equiv\grpsu{2}$, and
$\grpspinbf{3}$ are obtained by
further omitting the octupole generators $B^{(3)}$, $A^{(3)}$, and
$G^{(3)}$.  The Casimir operators for the algebras are
\begin{equation}
\label{eqncasimirdefn}
\begin{aligned}
C_2[\grpspin{6}]&=2G^{(2)}\cdot G^{(2)} +4 [G^{(1)}\cdot
G^{(1)}+G^{(3)}\cdot G^{(3)}]
\\&=2G^{(2)}\cdot G^{(2)} +4 G\circ G\\
C_2[\grpspin{5}]&=4 [G^{(1)}\cdot
G^{(1)}+G^{(3)}\cdot G^{(3)}]
\\&=4 G\circ G\\
C_2[\grpspin{3}]&=20 G^{(1)}\cdot
G^{(1)},
\end{aligned}
\end{equation}
where $G$ here generically represents the generator $B$, $A$, or $G$
as appropriate for $\grpspinb{n}$, $\grpspinf{n}$, or $\grpspinbf{n}$.
The eigenvalues of these Casimir operators are
\begin{equation}
\label{eqncasimireigen}
\begin{aligned}
\langle C_2[\grpspin{6}] \rangle_{(\sigma_1,\sigma_2,\sigma_3)}
&=2[\sigma_1(\sigma_1+4)+\sigma_2(\sigma_2+2)+\sigma_3^2]\\
\langle C_2[\grpspin{5}] \rangle_{(\tau_1,\tau_2)}
&=2[\tau_1(\tau_1+3)+\tau_2(\tau_2+1)]\\
\langle C_2[\grpspin{3}] \rangle_{(J)}&=2[J(J+1)].
\end{aligned}
\end{equation}
Comparing~(\ref{eqncasimirdefn}) and~(\ref{eqncasimireigen}), we see
that the conventional normalizations of the three-dimensional angular
momentum operators are obtained by setting
$\Lhat\<\equiv\sqrt{10}B^{(1)}$ and $\jhat\<\equiv\sqrt{10}A^{(1)}$,
so $\langle\Lhat\cdot\Lhat\rangle\<=L(L+1)$ and
$\langle\jhat\cdot\jhat\rangle\<=j(j+1)$.  Similarly, chosing the
normalizations of the five-dimensional angular momentum operators as
$\Lambdahat^{(\lambda)}\<\equiv\sqrt2B^{(\lambda)}$ and
$\Sigmahat^{(\lambda)}\<\equiv\sqrt2A^{(\lambda)}$ yields
$\langle\Lambdahat\circ\Lambdahat\rangle\<=\tau(\tau+3)$ and
$\langle\Sigmahat\circ\Sigmahat\rangle\<=\tau_1(\tau_1+3)+\tau_2(\tau_2+1)$.

Let us now return to the core-fermion interaction.  The spin-orbit interaction $\Lambdahat\circ\Sigmahat$ only involves $\grpspin{5}$
Casimir operators, as shown in~(\ref{eqnspinorbitcasimir}).  The
quadrupole-quadrupole interaction of the IBFM is, by the above definitions,
\begin{equation}
\Qhat_B\cdot \Qhat_F = \frac12[G^{(2)}\cdot G^{(2)} -
B^{(2)}\cdot B^{(2)} - A^{(2)}\cdot A^{(2)}],
\end{equation}
Adding to this the five-dimensional spin-orbit interaction 
\begin{equation}
\Lambdahat\circ\Sigmahat  = G\circ G -
B\circ B - A\circ A,
\end{equation}
and recognizing the $\grpspin{6}$ Casimir
operators~(\ref{eqncasimirdefn}), we have
\begin{equation}
\label{eqnqqlscasimir}
\Qhat_B\cdot \Qhat_F + \Lambdahat\circ\Sigmahat = \frac14 \bigl[
C_2[\grpspinbf{6}] - C_2[\grpspinb{6}] - C_2[\grpspinf{6}]
\bigr].
\end{equation}
Both $\Qhat_B \cdot \Qhat_F$ and $\Lambdahat\circ\Sigmahat$ are
$\grpspinbf{5}$ scalars, so both yield degenerate $\grpspin{5}$
multiplets of levels (with good $\tau_1$) when used as the
core-fermion interaction in the Hamiltonian.  However, they differ in
physical content in an important fashion. The interaction
$\Lambdahat\circ\Sigmahat$ induces a pure $\grpspin{5}$ coupling
scheme (where the levels have good $\tau$ as well as $\tau_1$).  The interaction
$\Qhat_B \cdot
\Qhat_F$ additionally contains $\grpspin{6}$ Casimir
operators, as seen from~(\ref{eqnqqlscasimir}), so it tends to induce
a hybrid of the $\grpspin{5}$ and $\grpspin{6}$ coupling schemes
(admixed $\tau\<=\tau_1\pm1/2$).  [This may loosely be interpreted as
mixing the spin-aligned and spin-flip states.  However, such a simple
two-state mixing description strictly holds only in the $\grpso{6}$
limit.]  Consequently, IBFM predictions obtained with the $\Qhat_B
\cdot \Qhat_F$ interaction
do not exhibit the $\Delta\tau$ selection rules encountered in the
$\grpe{5|4}$ predictions.  This leads to significant qualitative
differences between the results of
Ref.~\cite{alonso2005:ibfm-gsoft-phase} and the present $\grpe{5|4}$
predictions.

\section{\boldmath$\grpspin{5}\<\supset\grpspin{3}$ isoscalar factors}
\label{appisf}

The calculation of electromagnetic transition observables for the
$\grpe{5|4}$ model (Sec.~\ref{sece54}) is heavily dependent upon the
calculation of the $\grpspin{5}\<\supset\grpspin{3}$ isoscalar factors
involved in the construction of the coupled state core-fermion
state~(\ref{eqnspin5state}).  While the relevant isoscalar factors
have been encountered before on several
occasions~\cite{kuyucak1982:diss,iachello1981:ibfm-spin6,vanisacker1987:spin6-spin5-isf,iachello1991:ibfm}
in the context of the IBFM, considerable phase differences
are present among the reported values.  In this appendix, we briefly
clarify the phase degrees of freedom present in the determination of
the $\grpspin{5}\<\supset\grpspin{3}$ isoscalar factors and establish
a consistent set applicable to the geometric framework for the
$\grpe{5|4}$ calculations.

The symmetric representations $(\tau,0)$ of
$\grpspinb{5}$ and the spinor fundamental representation
$(\frac12,\frac12)$ of $\grpspinf{5}$ are coupled to yield
representations $(\tau\pm\frac12,\frac12)$ of $\grpspinbf{5}$ according to the
$\grpspin{5}\<\supset\grpspin{3}$ isoscalar factors
\begin{equation}
\label{eqnisf}
\ISFBF{\tau}{\nutt\,L}{\tau_1}{\nuttp\,J},
\end{equation}
where $\tau_1\<=\tau\pm1/2$.  The multiplicity indices $\nutt$ and
$\nuttp$ are only necessary for $\tau\<\geq6$ and $\tau_1\<\geq7/2$,
respectively, and so they are suppressed in~(\ref{eqnspin5state}) and
in the following discussion.

The isoscalar factors are usually
obtained~\cite{kuyucak1982:diss,iachello1981:ibfm-spin6,vanisacker1987:spin6-spin5-isf,iachello1991:ibfm}
by embedding the $\grpspin{5}$ algebras within $\grpspin{6}$ algebras,
as outlined in Appendix~\ref{appqq}, and applying Racah's method.  The
matrix element of $C_2[\grpspinbf{6}]=2G^{(2)}\cdot G^{(2)} +
C_2[\grpspin{5}]$ between a coupled state and an uncoupled state can
be evaluated in two different ways: directly by using the eigenvalue
formula~(\ref{eqncasimireigen}) for $C_2[\grpspin{6}]$ or indirectly
by evaluating the matrix element of $G^{(2)}\cdot G^{(2)}$ and using
the eigenvalue formula for $C_2[\grpspin{5}]$.  Equating the results
obtained by these two methods yields a system of equations involving
the isoscalar factors, the core matrix elements $\langle\tau L \Vert
B^{(2)} \Vert \tau' L' \rangle$, and the trivial fermionic matrix
element $\langle\tfrac32 \Vert A^{(2)} \Vert \tfrac32
\rangle\<=-\sqrt{5}$ (see
Refs.~\cite{kuyucak1982:diss,iachello1981:ibfm-spin6,vanisacker1987:spin6-spin5-isf,iachello1991:ibfm}
for details).  
Namely, isoscalar factors involving the same values of $\tau_1$ and
$J$ but different values of $\tau$ and $L$ are related by
\begin{multline}
\label{eqnisfsystem}
0=\ISFBF{\tau}{L}{\tau_1}{J}
-2\sqrt5\sum_{L'}(-)^{3/2+L'+J}\sixj{L}{\frac32}{J}{\frac32}{L'}{2}
\langle\tau L \Vert B^{(2)} \Vert \tau' L' \rangle
\\\times
\ISFBF{\tau'}{L'}{\tau_1}{J},
\end{multline}
where $\tau'\<\equiv\tau_1\mp1/2$ when $\tau\<=\tau_1\pm1/2$.  The full system is obtained by allowing $\tau$ to range over the
values $\tau_1\pm1/2$ and $L$ to range over the angular momentum
values contained in the $\grpso{5}$ representation $(\tau,0)$.  

All isoscalar factors of the same $\tau_1$ and $J$ are found
simultanously, to within an overall normalization factor and sign (or
phase), by solution of the linear homogeneous system of equations
given by~(\ref{eqnisfsystem}).  This is equivalent to solution for the
null vector of the coefficient matrix of the system, which is easily
carried out by Gaussian elimination.  The normalization of
the family of isoscalar factors sharing the same $\tau_1$ and $J$
values follows from the condition
\begin{equation}
\label{eqnisfnorm}
\sum_L \ISFBF{\tau'}{L}{\tau_1'}{J'} \ISFBF{\tau}{L}{\tau_1}{J} =
\delta_{\tau'\tau} \delta_{\tau_1'\tau_1}\delta_{J'J},
\end{equation}
but the overall sign 
is arbitrary, corresponding to the freedom in defining the sign of the
coupled state $\lvert \tau \tau_1 J \rangle$
in~(\ref{eqnspin5state}).  

The \textit{relative} signs of isoscalar factors sharing the same
$\tau_1$ and $J$ values are in principle fixed by the
system~(\ref{eqnisfsystem}).  However, comparison of the isoscalar
factors computed in
Refs.~\cite{kuyucak1982:diss,iachello1981:ibfm-spin6,vanisacker1987:spin6-spin5-isf,iachello1991:ibfm}
shows apparent sign discrepancies even among sets of isoscalar factors
sharing the same $\tau_1$ and $J$.  The differences arise from
differing signs for the matrix elements of $B^{(2)}\<=\Gsdds{2}$
appearing in~(\ref{eqnisfsystem}).  These in turn follow from
the phase freedom in defining the
$\grpspinb{5}$ basis states $\lvert \tau
L\rangle$~\cite{arima1979:ibm-o6} and, ultimately, from the arbitrary
phases of the $\grpu{5}$ basis states used in defining the matrix
elements of $d^\dagger$,
\textit{i.e.},
$a_{L'L}(n_d)$~\cite{arima1976:ibm-u5,arima1979:ibm-o6}.  
Physical
quantities, such as transition matrix elements, for the coupled system
are calculated
correctly \textit{only} if the same phase convention for the $\grpspinb{5}$ basis
states is used consistently both in the construction of the
isoscalar factors and in the calculation of physical quantities.
Consequently, the isoscalar factors for use in the present work cannot
be taken verbatim from any of the
Refs.~\cite{kuyucak1982:diss,iachello1981:ibfm-spin6,vanisacker1987:spin6-spin5-isf,iachello1991:ibfm}
but rather must be recalculated with phases consistent with the
present core $\grpspinb{5}$ basis functions, the $\Psi_{\tau\nutt{}LM_L}(\gamma,\vartheta)$.

The form of~(\ref{eqnisfsystem}) appropriate to the present geometric
framework for the $\grpe{5|4}$ problem is obtained by replacing the
bosonic operator $B^{(2)}$ with its classical limit $\alpha$.  Then
the matrix elements $\langle\tau L
\Vert \alpha \Vert \tau' L' \rangle$ are evaluated using the explicit
expressions for the wave functions
$\Psi_{\tau\nutt{}LM_L}(\gamma,\vartheta)$ as discussed in
Sec.~\ref{sece5trans}.  The resulting isoscalar factors are given in
Table~\ref{TabISF}.  These isoscalar factors have phases consistent
with the core reduced matrix elements of Appendix~\ref{appcorerme}.
They are used in the calculation of the coupled system reduced
matrix elements of Appendix~\ref{appcoupledrme}.
% ---
% LaTeXTable style Elsart
\begin{table}
\caption{The $\grpspin{5}\supset\grpspin{3}$ isoscalar factors~(ISFs), tabulated for $\tau_1\leq7/2$ and omitting degenerate $J$ values (see text). A square root sign is to be understood over the magnitude of each isoscalar factor.}
\label{TabISF}
\vspace{1ex}
\begin{tabular}{rrrrr@{\qquad\qquad}rrrrr}
\hline
$\tau_1$&$J$&$\tau$&$L$&ISF&$\tau_1$&$J$&$\tau$&$L$&ISF\\
\hline
$1/2$&$3/2$&$0$&$0$&$1$&$5/2$&$11/2$&$3$&$4$&$-8/99$\\
$1/2$&$3/2$&$1$&$2$&$1$&$5/2$&$11/2$&$3$&$6$&$91/99$\\
$3/2$&$1/2$&$1$&$2$&$1$&$7/2$&$3/2$&$3$&$0$&$2/3$\\
$3/2$&$5/2$&$1$&$2$&$1$&$7/2$&$3/2$&$3$&$3$&$1/3$\\
$3/2$&$7/2$&$1$&$2$&$1$&$7/2$&$5/2$&$3$&$3$&$11/18$\\
$3/2$&$1/2$&$2$&$2$&$-1$&$7/2$&$5/2$&$3$&$4$&$7/18$\\
$3/2$&$5/2$&$2$&$2$&$25/49$&$7/2$&$7/2$&$3$&$3$&$11/18$\\
$3/2$&$5/2$&$2$&$4$&$24/49$&$7/2$&$7/2$&$3$&$4$&$-7/18$\\
$3/2$&$7/2$&$2$&$2$&$-4/49$&$7/2$&$11/2$&$3$&$4$&$91/99$\\
$3/2$&$7/2$&$2$&$4$&$45/49$&$7/2$&$11/2$&$3$&$6$&$8/99$\\
$5/2$&$3/2$&$2$&$2$&$1$&$7/2$&$13/2$&$3$&$6$&$1$\\
$5/2$&$5/2$&$2$&$2$&$24/49$&$7/2$&$15/2$&$3$&$6$&$1$\\
$5/2$&$5/2$&$2$&$4$&$-25/49$&$7/2$&$3/2$&$4$&$2$&$1$\\
$5/2$&$7/2$&$2$&$2$&$45/49$&$7/2$&$5/2$&$4$&$2$&$12/77$\\
$5/2$&$7/2$&$2$&$4$&$4/49$&$7/2$&$5/2$&$4$&$4$&$-65/77$\\
$5/2$&$9/2$&$2$&$4$&$1$&$7/2$&$7/2$&$4$&$2$&$-45/154$\\
$5/2$&$11/2$&$2$&$4$&$1$&$7/2$&$7/2$&$4$&$4$&$-52/385$\\
$5/2$&$3/2$&$3$&$0$&$1/3$&$7/2$&$7/2$&$4$&$5$&$63/110$\\
$5/2$&$3/2$&$3$&$3$&$-2/3$&$7/2$&$11/2$&$4$&$4$&$-56/605$\\
$5/2$&$5/2$&$3$&$3$&$-7/18$&$7/2$&$11/2$&$4$&$5$&$9/55$\\
$5/2$&$5/2$&$3$&$4$&$11/18$&$7/2$&$11/2$&$4$&$6$&$90/121$\\
$5/2$&$7/2$&$3$&$3$&$7/18$&$7/2$&$13/2$&$4$&$5$&$-10/77$\\
$5/2$&$7/2$&$3$&$4$&$11/18$&$7/2$&$13/2$&$4$&$6$&$9/55$\\
$5/2$&$9/2$&$3$&$3$&$-11/90$&$7/2$&$13/2$&$4$&$8$&$272/385$\\
$5/2$&$9/2$&$3$&$4$&$49/198$&$7/2$&$15/2$&$4$&$6$&$-4/55$\\
$5/2$&$9/2$&$3$&$6$&$104/165$&$7/2$&$15/2$&$4$&$8$&$51/55$\\
\hline
\end{tabular}
\end{table}

% ---

Note that the procedure for calculation of isoscalar factors described
above is easily generalized to the case where degenerate $J$ values
arise in the $\grpspin{5}$ representation $(\tau_1,\frac12)$.  The
multiplicity of $J$ is equal to the dimension of the null space of the
coefficient matrix of~(\ref{eqnisfsystem}).  The null space is found by
Gaussian elimination, as usual, and the isoscalar factors are obtained
as the entries of an orthonormal set of basis vectors for the null
space.  The (arbitrary) choice of orthonormal basis determines the
meaning of the multiplicity index $\nuttp$ for the coupled state
$\lvert \tau \tau_1 \nutt{} J
\rangle$.

\clearpage

\section{Core angular reduced matrix elements}
\label{appcorerme}

Angular reduced matrix elements (RMEs) for the even-even core, as described
in Sec.~\ref{sece5trans}, are tabulated in
Tables~\ref{TabCoreRMEAlphatau1}--\ref{TabCoreRMEAlphaAlphaPi11tau1}.
The matrix elements are valid for the angular solutions of
$\gamma$-soft problems in general, not just for the $\grpe{5}$ case
considered in this article.  They also provide exact expressions for
the matrix elements of transition operators in the large-$N$ limit of
the IBM $\grpso{6}$ dynamical symmetry.  All reduced matrix elements
are normalized according to the Wigner-Eckart theorem convention of
Racah~\cite{edmonds1960:am}.

\clearpage

% LaTeXTable style Elsart
\begin{table}
\caption{Angular reduced matrix elements $\langle L'\lVert \alpha\rVert L\rangle$, between basis states related by $\Delta\<\tau=1$, tabulated for $\tau\<\leq5$.  A square root sign is to be understood over the magnitude of each matrix element.}
\label{TabCoreRMEAlphatau1}
\vspace{1ex}
\begin{tabular}{rrrrr@{\qquad}rrrrr}
\hline
$\tau$&$L$&$\tau'$&$L'$&RME&$\tau$&$L$&$\tau'$&$L'$&RME\\
\hline
$1$&$2$&$0$&$0$&$1$&$4$&$8$&$3$&$6$&$68/11$\\
$2$&$2$&$1$&$2$&$-10/7$&$5$&$2$&$4$&$2$&$-10/7$\\
$2$&$4$&$1$&$2$&$18/7$&$5$&$2$&$4$&$4$&$45/91$\\
$3$&$0$&$2$&$2$&$1/3$&$5$&$4$&$4$&$2$&$135/77$\\
$3$&$3$&$2$&$2$&$5/3$&$5$&$4$&$4$&$4$&$15552/55055$\\
$3$&$3$&$2$&$4$&$2/3$&$5$&$4$&$4$&$5$&$63/55$\\
$3$&$4$&$2$&$2$&$11/7$&$5$&$4$&$4$&$6$&$441/1573$\\
$3$&$4$&$2$&$4$&$-10/7$&$5$&$5$&$4$&$4$&$189/65$\\
$3$&$6$&$2$&$4$&$13/3$&$5$&$5$&$4$&$5$&$-66/65$\\
$4$&$2$&$3$&$0$&$2/3$&$5$&$5$&$4$&$6$&$4/13$\\
$4$&$2$&$3$&$3$&$5/6$&$5$&$6$&$4$&$4$&$3213/1573$\\
$4$&$2$&$3$&$4$&$7/22$&$5$&$6$&$4$&$5$&$68/77$\\
$4$&$4$&$3$&$3$&$26/15$&$5$&$6$&$4$&$6$&$-1224/605$\\
$4$&$4$&$3$&$4$&$-182/121$&$5$&$6$&$4$&$8$&$-256/5005$\\
$4$&$4$&$3$&$6$&$-64/1815$&$5$&$7$&$4$&$5$&$360/91$\\
$4$&$5$&$3$&$3$&$21/10$&$5$&$7$&$4$&$6$&$9/13$\\
$4$&$5$&$3$&$4$&$21/22$&$5$&$7$&$4$&$8$&$102/91$\\
$4$&$5$&$3$&$6$&$52/55$&$5$&$8$&$4$&$6$&$323/65$\\
$4$&$6$&$3$&$4$&$390/121$&$5$&$8$&$4$&$8$&$-102/65$\\
$4$&$6$&$3$&$6$&$-182/121$&$5$&$10$&$4$&$8$&$105/13$\\
\hline
\end{tabular}
\end{table}

% LaTeXTable style Elsart
\begin{table}
\caption{Angular reduced matrix elements $\langle L'\lVert (\alpha\times\alpha)^{(2)}\rVert L\rangle$, between basis states related by $\Delta\<\tau=0$, tabulated for $\tau\<\leq5$.  A square root sign is to be understood over the magnitude of each matrix element.}
\label{TabCoreRMEAlphaAlpha2tau0}
\vspace{1ex}
\begin{tabular}{rrrrr@{\qquad}rrrrr}
\hline
$\tau$&$L$&$\tau'$&$L'$&RME&$\tau$&$L$&$\tau'$&$L'$&RME\\
\hline
$1$&$2$&$1$&$2$&$20/49$&$4$&$8$&$4$&$8$&$15504/5915$\\
$2$&$2$&$2$&$2$&$-20/441$&$5$&$2$&$5$&$2$&$-4/245$\\
$2$&$2$&$2$&$4$&$-64/441$&$5$&$2$&$5$&$4$&$-96/2695$\\
$2$&$4$&$2$&$2$&$-64/441$&$5$&$4$&$5$&$2$&$-96/2695$\\
$2$&$4$&$2$&$4$&$440/441$&$5$&$4$&$5$&$4$&$1527752/8152375$\\
$3$&$3$&$3$&$3$&$0$&$5$&$4$&$5$&$5$&$32/1375$\\
$3$&$3$&$3$&$4$&$24/77$&$5$&$4$&$5$&$6$&$2176/33275$\\
$3$&$4$&$3$&$3$&$-24/77$&$5$&$5$&$5$&$4$&$-32/1375$\\
$3$&$4$&$3$&$4$&$5760/65219$&$5$&$5$&$5$&$5$&$-1188/11375$\\
$3$&$4$&$3$&$6$&$-2496/9317$&$5$&$5$&$5$&$6$&$4896/13475$\\
$3$&$6$&$3$&$4$&$-2496/9317$&$5$&$5$&$5$&$7$&$-256/3185$\\
$3$&$6$&$3$&$6$&$2340/1331$&$5$&$6$&$5$&$4$&$2176/33275$\\
$4$&$2$&$4$&$2$&$20/441$&$5$&$6$&$5$&$5$&$-4896/13475$\\
$4$&$2$&$4$&$4$&$160/5733$&$5$&$6$&$5$&$6$&$247572/8152375$\\
$4$&$4$&$4$&$2$&$160/5733$&$5$&$6$&$5$&$7$&$2176/13475$\\
$4$&$4$&$4$&$4$&$-356168/4099095$&$5$&$6$&$5$&$8$&$-21888/67375$\\
$4$&$4$&$4$&$5$&$32/195$&$5$&$7$&$5$&$5$&$-256/3185$\\
$4$&$4$&$4$&$6$&$-896/5577$&$5$&$7$&$5$&$6$&$-2176/13475$\\
$4$&$5$&$4$&$4$&$-32/195$&$5$&$7$&$5$&$7$&$8228/9555$\\
$4$&$5$&$4$&$5$&$132/455$&$5$&$7$&$5$&$8$&$1292/3675$\\
$4$&$5$&$4$&$6$&$32/91$&$5$&$8$&$5$&$6$&$-21888/67375$\\
$4$&$6$&$4$&$4$&$-896/5577$&$5$&$8$&$5$&$7$&$-1292/3675$\\
$4$&$6$&$4$&$5$&$-32/91$&$5$&$8$&$5$&$8$&$16524/16625$\\
$4$&$6$&$4$&$6$&$324/715$&$5$&$8$&$5$&$10$&$-128/285$\\
$4$&$6$&$4$&$8$&$-2176/5915$&$5$&$10$&$5$&$8$&$-128/285$\\
$4$&$8$&$4$&$6$&$-2176/5915$&$5$&$10$&$5$&$10$&$1012/285$\\
\hline
\end{tabular}
\end{table}

% LaTeXTable style Elsart
\begin{table}
\caption{Angular reduced matrix elements $\langle L'\lVert (\alpha\times\alpha)^{(2)}\rVert L\rangle$, between basis states related by $\Delta\<\tau=2$, tabulated for $\tau\<\leq5$.  A square root sign is to be understood over the magnitude of each matrix element.}
\label{TabCoreRMEAlphaAlpha2tau2}
\vspace{1ex}
\begin{tabular}{rrrrr@{\qquad}rrrrr}
\hline
$\tau$&$L$&$\tau'$&$L'$&RME&$\tau$&$L$&$\tau'$&$L'$&RME\\
\hline
$2$&$2$&$0$&$0$&$-2/7$&$5$&$2$&$3$&$4$&$-1/11$\\
$3$&$0$&$1$&$2$&$-2/21$&$5$&$4$&$3$&$3$&$144/385$\\
$3$&$3$&$1$&$2$&$10/21$&$5$&$4$&$3$&$4$&$432/1331$\\
$3$&$4$&$1$&$2$&$-22/49$&$5$&$4$&$3$&$6$&$-66978/605605$\\
$4$&$2$&$2$&$2$&$160/441$&$5$&$5$&$3$&$3$&$-3/5$\\
$4$&$2$&$2$&$4$&$-50/441$&$5$&$5$&$3$&$4$&$3/11$\\
$4$&$4$&$2$&$2$&$-260/441$&$5$&$5$&$3$&$6$&$32/5005$\\
$4$&$4$&$2$&$4$&$-416/4851$&$5$&$6$&$3$&$4$&$-1020/1331$\\
$4$&$5$&$2$&$4$&$2/3$&$5$&$6$&$3$&$6$&$-1088/9317$\\
$4$&$6$&$2$&$4$&$-130/231$&$5$&$7$&$3$&$6$&$810/1001$\\
$5$&$2$&$3$&$0$&$-4/21$&$5$&$8$&$3$&$6$&$-646/1001$\\
$5$&$2$&$3$&$3$&$5/21$&&&&&\\
\hline
\end{tabular}
\end{table}

% LaTeXTable style Elsart
\begin{table}
\caption{Angular reduced matrix elements $\langle L'\lVert [\alpha\times i(\alpha\times\tilde\pi)^{(1)}]^{(1)}\rVert L\rangle$, between basis states related by $\Delta\<\tau=1$, tabulated for $\tau\<\leq5$.  A square root sign is to be understood over the magnitude of each matrix element.}
\label{TabCoreRMEAlphaAlphaPi11tau1}
\vspace{1ex}
\begin{tabular}{rrrrr@{\qquad}rrrrr}
\hline
$\tau$&$L$&$\tau'$&$L'$&RME&$\tau$&$L$&$\tau'$&$L'$&RME\\
\hline
$2$&$2$&$1$&$2$&$3/10$&$5$&$4$&$4$&$4$&$-3888/17875$\\
$3$&$3$&$2$&$2$&$-2/5$&$5$&$4$&$4$&$5$&$-1134/1375$\\
$3$&$3$&$2$&$4$&$-3/10$&$5$&$5$&$4$&$4$&$-3402/1625$\\
$3$&$4$&$2$&$4$&$11/10$&$5$&$5$&$4$&$5$&$297/250$\\
$4$&$2$&$3$&$3$&$-1/5$&$5$&$5$&$4$&$6$&$-21/65$\\
$4$&$4$&$3$&$3$&$-39/50$&$5$&$6$&$4$&$5$&$-51/55$\\
$4$&$4$&$3$&$4$&$637/550$&$5$&$6$&$4$&$6$&$918/275$\\
$4$&$5$&$3$&$4$&$-189/275$&$5$&$7$&$4$&$6$&$-324/325$\\
$4$&$5$&$3$&$6$&$-273/275$&$5$&$7$&$4$&$8$&$-1377/650$\\
$4$&$6$&$3$&$6$&$273/110$&$5$&$8$&$4$&$8$&$2907/650$\\
$5$&$2$&$4$&$2$&$3/10$&&&&&\\
\hline
\end{tabular}
\end{table}

\clearpage

\section{Coupled system reduced matrix elements}
\label{appcoupledrme}

Angular reduced matrix elements (RMEs) for the coupled odd-mass system, as
described in Sec.~\ref{sece54trans}, are taulated in
Tables~\ref{TabCoupledRMEAlpha}--\ref{TabCoupledRMEParticle1}.
The matrix elements are applicable to problems involving a
$\gamma$-soft core coupled to a $j\<=3/2$ particle in general, not
just to the $\grpe{5|4}$ case considered in this article.  All reduced matrix elements
are normalized according to the Wigner-Eckart theorem convention of
Racah~\cite{edmonds1960:am}.

In Tables~\ref{TabCoupledRMEAlpha} and~\ref{TabCoupledRMEAlphaAlpha2},
certain redundant matrix element values have been suppressed for
conciseness.  In Table~\ref{TabCoupledRMEAlpha}, only the
$\Delta\tau_1\<=1$ matrix elements within the $\tau_1\<=\tau+1/2$
family of states are tabulated, since those within the
$\tau_1\<=\tau-1/2$ family are identical.  In
Table~\ref{TabCoupledRMEAlphaAlpha2}, only the $\Delta\tau_1\<=0$ or
$2$ matrix elements within the $\tau_1\<=\tau+1/2$ family of states
are tabulated, since those within the $\tau_1\<=\tau-1/2$ family are
identical.  Also, only the $\Delta\tau_1\<=1$ matrix elements with
$\Delta\tau\<=0$ are tabulated, since those with $\Delta\tau\<=2$ are
identical.

\clearpage
% LaTeXTable style ElsartTiny
\begin{table}
\caption{Angular reduced matrix elements $\langle J'\lVert \alpha\rVert J\rangle$ between $\grpspin{5}$ coupled states, for $\tau_1\<\leq5/2$.   A square root sign is to be understood over the magnitude of each matrix element.}
\label{TabCoupledRMEAlpha}
\vspace{1ex}
\tiny
\begin{tabular}{rrrrrrr@{\qquad}rrrrrrr}
\hline
$\tau$&$\tau_1$&$J$&$\tau'$&$\tau_1'$&$J'$&RME&$\tau$&$\tau_1$&$J$&$\tau'$&$\tau_1'$&$J'$&RME\\
\hline
$1$&$1/2$&$3/2$&$0$&$1/2$&$3/2$&$4/5$&$3$&$5/2$&$9/2$&$2$&$5/2$&$7/2$&$32/11907$\\
$2$&$3/2$&$1/2$&$1$&$3/2$&$5/2$&$6/35$&$3$&$5/2$&$9/2$&$2$&$5/2$&$9/2$&$-125/2673$\\
$2$&$3/2$&$5/2$&$1$&$3/2$&$1/2$&$6/35$&$3$&$5/2$&$9/2$&$2$&$5/2$&$11/2$&$-832/2673$\\
$2$&$3/2$&$5/2$&$1$&$3/2$&$5/2$&$-12/1715$&$3$&$5/2$&$11/2$&$2$&$5/2$&$7/2$&$128/8505$\\
$2$&$3/2$&$5/2$&$1$&$3/2$&$7/2$&$-576/1715$&$3$&$5/2$&$11/2$&$2$&$5/2$&$9/2$&$832/2673$\\
$2$&$3/2$&$7/2$&$1$&$3/2$&$5/2$&$576/1715$&$3$&$5/2$&$11/2$&$2$&$5/2$&$11/2$&$364/1485$\\
$2$&$3/2$&$7/2$&$1$&$3/2$&$7/2$&$120/343$&$2$&$3/2$&$1/2$&$1$&$1/2$&$3/2$&$2/5$\\
$3$&$5/2$&$3/2$&$2$&$5/2$&$3/2$&$-4/45$&$2$&$3/2$&$5/2$&$1$&$1/2$&$3/2$&$6/5$\\
$3$&$5/2$&$3/2$&$2$&$5/2$&$5/2$&$-64/2205$&$2$&$3/2$&$7/2$&$1$&$1/2$&$3/2$&$8/5$\\
$3$&$5/2$&$3/2$&$2$&$5/2$&$7/2$&$32/441$&$3$&$5/2$&$3/2$&$2$&$3/2$&$1/2$&$2/5$\\
$3$&$5/2$&$5/2$&$2$&$5/2$&$3/2$&$64/2205$&$3$&$5/2$&$3/2$&$2$&$3/2$&$5/2$&$-30/49$\\
$3$&$5/2$&$5/2$&$2$&$5/2$&$5/2$&$121/46305$&$3$&$5/2$&$3/2$&$2$&$3/2$&$7/2$&$-32/245$\\
$3$&$5/2$&$5/2$&$2$&$5/2$&$7/2$&$-11552/83349$&$3$&$5/2$&$5/2$&$2$&$3/2$&$1/2$&$-36/35$\\
$3$&$5/2$&$5/2$&$2$&$5/2$&$9/2$&$-1375/11907$&$3$&$5/2$&$5/2$&$2$&$3/2$&$5/2$&$-90/343$\\
$3$&$5/2$&$7/2$&$2$&$5/2$&$3/2$&$32/441$&$3$&$5/2$&$5/2$&$2$&$3/2$&$7/2$&$-726/1715$\\
$3$&$5/2$&$7/2$&$2$&$5/2$&$5/2$&$11552/83349$&$3$&$5/2$&$7/2$&$2$&$3/2$&$5/2$&$400/343$\\
$3$&$5/2$&$7/2$&$2$&$5/2$&$7/2$&$63368/416745$&$3$&$5/2$&$7/2$&$2$&$3/2$&$7/2$&$-384/343$\\
$3$&$5/2$&$7/2$&$2$&$5/2$&$9/2$&$-32/11907$&$3$&$5/2$&$9/2$&$2$&$3/2$&$5/2$&$110/49$\\
$3$&$5/2$&$7/2$&$2$&$5/2$&$11/2$&$128/8505$&$3$&$5/2$&$9/2$&$2$&$3/2$&$7/2$&$30/49$\\
$3$&$5/2$&$9/2$&$2$&$5/2$&$5/2$&$-1375/11907$&$3$&$5/2$&$11/2$&$2$&$3/2$&$7/2$&$24/7$\\
\hline
\end{tabular}
\end{table}

% LaTeXTable style ElsartTiny
\begin{table}
\caption{Angular reduced matrix elements $\langle J'\lVert (\alpha\times\alpha)^{(2)}\rVert J\rangle$ between $\grpspin{5}$ coupled states, for $\tau_1\<\leq5/2$.   A square root sign is to be understood over the magnitude of each matrix element.}
\label{TabCoupledRMEAlphaAlpha2}
\vspace{1ex}
\tiny
\begin{tabular}{rrrrrrr@{\qquad}rrrrrrr}
\hline
$\tau$&$\tau_1$&$J$&$\tau'$&$\tau_1'$&$J'$&RME&$\tau$&$\tau_1$&$J$&$\tau'$&$\tau_1'$&$J'$&RME\\
\hline
$1$&$1/2$&$3/2$&$1$&$1/2$&$3/2$&$0$&$3$&$5/2$&$9/2$&$3$&$5/2$&$11/2$&$416/2079$\\
$2$&$3/2$&$1/2$&$2$&$3/2$&$5/2$&$12/245$&$3$&$5/2$&$11/2$&$3$&$5/2$&$7/2$&$-1024/6615$\\
$2$&$3/2$&$5/2$&$2$&$3/2$&$1/2$&$12/245$&$3$&$5/2$&$11/2$&$3$&$5/2$&$9/2$&$-416/2079$\\
$2$&$3/2$&$5/2$&$2$&$3/2$&$5/2$&$120/2401$&$3$&$5/2$&$11/2$&$3$&$5/2$&$11/2$&$1664/1485$\\
$2$&$3/2$&$5/2$&$2$&$3/2$&$7/2$&$2592/12005$&$1$&$3/2$&$1/2$&$1$&$1/2$&$3/2$&$4/35$\\
$2$&$3/2$&$7/2$&$2$&$3/2$&$5/2$&$-2592/12005$&$1$&$3/2$&$5/2$&$1$&$1/2$&$3/2$&$-60/343$\\
$2$&$3/2$&$7/2$&$2$&$3/2$&$7/2$&$960/2401$&$1$&$3/2$&$7/2$&$1$&$1/2$&$3/2$&$64/1715$\\
$3$&$5/2$&$3/2$&$3$&$5/2$&$3/2$&$0$&$2$&$5/2$&$3/2$&$2$&$3/2$&$1/2$&$-4/315$\\
$3$&$5/2$&$3/2$&$3$&$5/2$&$5/2$&$-160/3087$&$2$&$5/2$&$3/2$&$2$&$3/2$&$5/2$&$4/5145$\\
$3$&$5/2$&$3/2$&$3$&$5/2$&$7/2$&$-64/3087$&$2$&$5/2$&$3/2$&$2$&$3/2$&$7/2$&$-1024/15435$\\
$3$&$5/2$&$5/2$&$3$&$5/2$&$3/2$&$160/3087$&$2$&$5/2$&$5/2$&$2$&$3/2$&$1/2$&$-32/2205$\\
$3$&$5/2$&$5/2$&$3$&$5/2$&$5/2$&$9610/64827$&$2$&$5/2$&$5/2$&$2$&$3/2$&$5/2$&$-15376/108045$\\
$3$&$5/2$&$5/2$&$3$&$5/2$&$7/2$&$-48/2401$&$2$&$5/2$&$5/2$&$2$&$3/2$&$7/2$&$-5476/324135$\\
$3$&$5/2$&$5/2$&$3$&$5/2$&$9/2$&$-110/1029$&$2$&$5/2$&$7/2$&$2$&$3/2$&$5/2$&$-32/194481$\\
$3$&$5/2$&$7/2$&$3$&$5/2$&$3/2$&$-64/3087$&$2$&$5/2$&$7/2$&$2$&$3/2$&$7/2$&$256/7203$\\
$3$&$5/2$&$7/2$&$3$&$5/2$&$5/2$&$48/2401$&$2$&$5/2$&$9/2$&$2$&$3/2$&$5/2$&$-880/27783$\\
$3$&$5/2$&$7/2$&$3$&$5/2$&$7/2$&$-87616/2917215$&$2$&$5/2$&$9/2$&$2$&$3/2$&$7/2$&$-60/343$\\
$3$&$5/2$&$7/2$&$3$&$5/2$&$9/2$&$8464/83349$&$2$&$5/2$&$11/2$&$2$&$3/2$&$7/2$&$64/1323$\\
$3$&$5/2$&$7/2$&$3$&$5/2$&$11/2$&$-1024/6615$&$3$&$5/2$&$3/2$&$1$&$1/2$&$3/2$&$-8/35$\\
$3$&$5/2$&$9/2$&$3$&$5/2$&$5/2$&$-110/1029$&$3$&$5/2$&$5/2$&$1$&$1/2$&$3/2$&$-288/1715$\\
$3$&$5/2$&$9/2$&$3$&$5/2$&$7/2$&$-8464/83349$&$3$&$5/2$&$7/2$&$1$&$1/2$&$3/2$&$-144/343$\\
$3$&$5/2$&$9/2$&$3$&$5/2$&$9/2$&$1750/2673$&&&&&&&\\
\hline
\end{tabular}
\end{table}

% LaTeXTable style ElsartTiny
\begin{table}
\caption{Angular reduced matrix elements $\langle J'\lVert (a^\dag\times\tilde{a})^{(2)}\rVert J\rangle$ between $\grpspin{5}$ coupled states, for $\tau_1\<\leq5/2$.   A square root sign is to be understood over the magnitude of each matrix element.}
\label{TabCoupledRMEParticle2}
\vspace{1ex}
\tiny
\begin{tabular}{rrrrrrr@{\qquad}rrrrrrr}
\hline
$\tau$&$\tau_1$&$J$&$\tau'$&$\tau_1'$&$J'$&RME&$\tau$&$\tau_1$&$J$&$\tau'$&$\tau_1'$&$J'$&RME\\
\hline
$0$&$1/2$&$3/2$&$0$&$1/2$&$3/2$&$-5$&$3$&$5/2$&$3/2$&$3$&$5/2$&$3/2$&$-49/45$\\
$1$&$3/2$&$1/2$&$1$&$3/2$&$5/2$&$-21/10$&$3$&$5/2$&$3/2$&$3$&$5/2$&$5/2$&$-16/45$\\
$1$&$3/2$&$5/2$&$1$&$3/2$&$1/2$&$-21/10$&$3$&$5/2$&$3/2$&$3$&$5/2$&$7/2$&$8/9$\\
$1$&$3/2$&$5/2$&$1$&$3/2$&$5/2$&$3/35$&$3$&$5/2$&$5/2$&$3$&$5/2$&$3/2$&$16/45$\\
$1$&$3/2$&$5/2$&$1$&$3/2$&$7/2$&$144/35$&$3$&$5/2$&$5/2$&$3$&$5/2$&$5/2$&$121/3780$\\
$1$&$3/2$&$7/2$&$1$&$3/2$&$5/2$&$-144/35$&$3$&$5/2$&$5/2$&$3$&$5/2$&$7/2$&$-2888/1701$\\
$1$&$3/2$&$7/2$&$1$&$3/2$&$7/2$&$-30/7$&$3$&$5/2$&$5/2$&$3$&$5/2$&$9/2$&$-1375/972$\\
$2$&$5/2$&$3/2$&$2$&$5/2$&$3/2$&$9/5$&$3$&$5/2$&$7/2$&$3$&$5/2$&$3/2$&$8/9$\\
$2$&$5/2$&$3/2$&$2$&$5/2$&$5/2$&$144/245$&$3$&$5/2$&$7/2$&$3$&$5/2$&$5/2$&$2888/1701$\\
$2$&$5/2$&$3/2$&$2$&$5/2$&$7/2$&$-72/49$&$3$&$5/2$&$7/2$&$3$&$5/2$&$7/2$&$15842/8505$\\
$2$&$5/2$&$5/2$&$2$&$5/2$&$3/2$&$-144/245$&$3$&$5/2$&$7/2$&$3$&$5/2$&$9/2$&$-8/243$\\
$2$&$5/2$&$5/2$&$2$&$5/2$&$5/2$&$-363/6860$&$3$&$5/2$&$7/2$&$3$&$5/2$&$11/2$&$224/1215$\\
$2$&$5/2$&$5/2$&$2$&$5/2$&$7/2$&$2888/1029$&$3$&$5/2$&$9/2$&$3$&$5/2$&$5/2$&$-1375/972$\\
$2$&$5/2$&$5/2$&$2$&$5/2$&$9/2$&$1375/588$&$3$&$5/2$&$9/2$&$3$&$5/2$&$7/2$&$8/243$\\
$2$&$5/2$&$7/2$&$2$&$5/2$&$3/2$&$-72/49$&$3$&$5/2$&$9/2$&$3$&$5/2$&$9/2$&$-6125/10692$\\
$2$&$5/2$&$7/2$&$2$&$5/2$&$5/2$&$-2888/1029$&$3$&$5/2$&$9/2$&$3$&$5/2$&$11/2$&$-10192/2673$\\
$2$&$5/2$&$7/2$&$2$&$5/2$&$7/2$&$-15842/5145$&$3$&$5/2$&$11/2$&$3$&$5/2$&$7/2$&$224/1215$\\
$2$&$5/2$&$7/2$&$2$&$5/2$&$9/2$&$8/147$&$3$&$5/2$&$11/2$&$3$&$5/2$&$9/2$&$10192/2673$\\
$2$&$5/2$&$7/2$&$2$&$5/2$&$11/2$&$-32/105$&$3$&$5/2$&$11/2$&$3$&$5/2$&$11/2$&$4459/1485$\\
$2$&$5/2$&$9/2$&$2$&$5/2$&$5/2$&$1375/588$&$1$&$3/2$&$1/2$&$1$&$1/2$&$3/2$&$-2/5$\\
$2$&$5/2$&$9/2$&$2$&$5/2$&$7/2$&$-8/147$&$1$&$3/2$&$5/2$&$1$&$1/2$&$3/2$&$-6/5$\\
$2$&$5/2$&$9/2$&$2$&$5/2$&$9/2$&$125/132$&$1$&$3/2$&$7/2$&$1$&$1/2$&$3/2$&$-8/5$\\
$2$&$5/2$&$9/2$&$2$&$5/2$&$11/2$&$208/33$&$2$&$5/2$&$3/2$&$2$&$3/2$&$1/2$&$-2/5$\\
$2$&$5/2$&$11/2$&$2$&$5/2$&$7/2$&$-32/105$&$2$&$5/2$&$3/2$&$2$&$3/2$&$5/2$&$30/49$\\
$2$&$5/2$&$11/2$&$2$&$5/2$&$9/2$&$-208/33$&$2$&$5/2$&$3/2$&$2$&$3/2$&$7/2$&$32/245$\\
$2$&$5/2$&$11/2$&$2$&$5/2$&$11/2$&$-273/55$&$2$&$5/2$&$5/2$&$2$&$3/2$&$1/2$&$36/35$\\
$1$&$1/2$&$3/2$&$1$&$1/2$&$3/2$&$9/5$&$2$&$5/2$&$5/2$&$2$&$3/2$&$5/2$&$90/343$\\
$2$&$3/2$&$1/2$&$2$&$3/2$&$5/2$&$15/14$&$2$&$5/2$&$5/2$&$2$&$3/2$&$7/2$&$726/1715$\\
$2$&$3/2$&$5/2$&$2$&$3/2$&$1/2$&$15/14$&$2$&$5/2$&$7/2$&$2$&$3/2$&$5/2$&$-400/343$\\
$2$&$3/2$&$5/2$&$2$&$3/2$&$5/2$&$-15/343$&$2$&$5/2$&$7/2$&$2$&$3/2$&$7/2$&$384/343$\\
$2$&$3/2$&$5/2$&$2$&$3/2$&$7/2$&$-720/343$&$2$&$5/2$&$9/2$&$2$&$3/2$&$5/2$&$-110/49$\\
$2$&$3/2$&$7/2$&$2$&$3/2$&$5/2$&$720/343$&$2$&$5/2$&$9/2$&$2$&$3/2$&$7/2$&$-30/49$\\
$2$&$3/2$&$7/2$&$2$&$3/2$&$7/2$&$750/343$&$2$&$5/2$&$11/2$&$2$&$3/2$&$7/2$&$-24/7$\\
\hline
\end{tabular}
\end{table}

% LaTeXTable style ElsartTiny
\begin{table}
\caption{Angular reduced matrix elements $\langle J'\lVert i(\alpha\times\tilde\pi)^{(1)}\rVert J\rangle$ between $\grpspin{5}$ coupled states, for $\tau_1\<\leq5/2$. For reduced matrix elements of $\hat{L}$, multiply by $\sqrt{10}$.  A square root sign is to be understood over the magnitude of each matrix element.}
\label{TabCoupledRMEAlphaPi1}
\vspace{1ex}
\tiny
\begin{tabular}{rrrrrrr@{\qquad}rrrrrrr}
\hline
$\tau$&$\tau_1$&$J$&$\tau'$&$\tau_1'$&$J'$&RME&$\tau$&$\tau_1$&$J$&$\tau'$&$\tau_1'$&$J'$&RME\\
\hline
$0$&$1/2$&$3/2$&$0$&$1/2$&$3/2$&$0$&$2$&$3/2$&$7/2$&$2$&$3/2$&$7/2$&$20736/1715$\\
$1$&$3/2$&$1/2$&$1$&$3/2$&$1/2$&$3/5$&$3$&$5/2$&$3/2$&$3$&$5/2$&$3/2$&$128/75$\\
$1$&$3/2$&$5/2$&$1$&$3/2$&$5/2$&$363/175$&$3$&$5/2$&$3/2$&$3$&$5/2$&$5/2$&$56/225$\\
$1$&$3/2$&$5/2$&$1$&$3/2$&$7/2$&$24/35$&$3$&$5/2$&$5/2$&$3$&$5/2$&$3/2$&$-56/225$\\
$1$&$3/2$&$7/2$&$1$&$3/2$&$5/2$&$-24/35$&$3$&$5/2$&$5/2$&$3$&$5/2$&$5/2$&$38809/4725$\\
$1$&$3/2$&$7/2$&$1$&$3/2$&$7/2$&$144/35$&$3$&$5/2$&$5/2$&$3$&$5/2$&$7/2$&$4/63$\\
$2$&$5/2$&$3/2$&$2$&$5/2$&$3/2$&$24/25$&$3$&$5/2$&$7/2$&$3$&$5/2$&$5/2$&$-4/63$\\
$2$&$5/2$&$3/2$&$2$&$5/2$&$5/2$&$72/175$&$3$&$5/2$&$7/2$&$3$&$5/2$&$7/2$&$54080/5103$\\
$2$&$5/2$&$5/2$&$2$&$5/2$&$3/2$&$-72/175$&$3$&$5/2$&$7/2$&$3$&$5/2$&$9/2$&$308/3645$\\
$2$&$5/2$&$5/2$&$2$&$5/2$&$5/2$&$48387/8575$&$3$&$5/2$&$9/2$&$3$&$5/2$&$7/2$&$-308/3645$\\
$2$&$5/2$&$5/2$&$2$&$5/2$&$7/2$&$36/343$&$3$&$5/2$&$9/2$&$3$&$5/2$&$9/2$&$235225/8019$\\
$2$&$5/2$&$7/2$&$2$&$5/2$&$5/2$&$-36/343$&$3$&$5/2$&$9/2$&$3$&$5/2$&$11/2$&$392/495$\\
$2$&$5/2$&$7/2$&$2$&$5/2$&$7/2$&$71824/15435$&$3$&$5/2$&$11/2$&$3$&$5/2$&$9/2$&$-392/495$\\
$2$&$5/2$&$7/2$&$2$&$5/2$&$9/2$&$44/315$&$3$&$5/2$&$11/2$&$3$&$5/2$&$11/2$&$13000/297$\\
$2$&$5/2$&$9/2$&$2$&$5/2$&$7/2$&$-44/315$&$1$&$3/2$&$1/2$&$1$&$1/2$&$3/2$&$3/5$\\
$2$&$5/2$&$9/2$&$2$&$5/2$&$9/2$&$1681/99$&$1$&$3/2$&$5/2$&$1$&$1/2$&$3/2$&$-21/25$\\
$2$&$5/2$&$9/2$&$2$&$5/2$&$11/2$&$72/55$&$2$&$5/2$&$3/2$&$2$&$3/2$&$1/2$&$3/5$\\
$2$&$5/2$&$11/2$&$2$&$5/2$&$9/2$&$-72/55$&$2$&$5/2$&$3/2$&$2$&$3/2$&$5/2$&$3/7$\\
$2$&$5/2$&$11/2$&$2$&$5/2$&$11/2$&$1248/55$&$2$&$5/2$&$5/2$&$2$&$3/2$&$5/2$&$-288/343$\\
$1$&$1/2$&$3/2$&$1$&$1/2$&$3/2$&$24/25$&$2$&$5/2$&$5/2$&$2$&$3/2$&$7/2$&$-1521/1715$\\
$2$&$3/2$&$1/2$&$2$&$3/2$&$1/2$&$3/5$&$2$&$5/2$&$7/2$&$2$&$3/2$&$5/2$&$-216/343$\\
$2$&$3/2$&$5/2$&$2$&$3/2$&$5/2$&$1875/343$&$2$&$5/2$&$7/2$&$2$&$3/2$&$7/2$&$64/343$\\
$2$&$3/2$&$5/2$&$2$&$3/2$&$7/2$&$120/343$&$2$&$5/2$&$9/2$&$2$&$3/2$&$7/2$&$-11/7$\\
$2$&$3/2$&$7/2$&$2$&$3/2$&$5/2$&$-120/343$&&&&&&&\\
\hline
\end{tabular}
\end{table}

% LaTeXTable style ElsartTiny
\begin{table}
\caption{Angular reduced matrix elements $\langle J'\lVert [\alpha\times i(\alpha\times\tilde\pi)^{(1)}]^{(1)}\rVert J\rangle$ between $\grpspin{5}$ coupled states, for $\tau_1\<\leq5/2$.   A square root sign is to be understood over the magnitude of each matrix element.}
\label{TabCoupledRMEAlphaAlphaPi11}
\vspace{1ex}
\tiny
\begin{tabular}{rrrrrrr@{\qquad}rrrrrrr}
\hline
$\tau$&$\tau_1$&$J$&$\tau'$&$\tau_1'$&$J'$&RME&$\tau$&$\tau_1$&$J$&$\tau'$&$\tau_1'$&$J'$&RME\\
\hline
$1$&$1/2$&$3/2$&$0$&$1/2$&$3/2$&$0$&$1$&$3/2$&$1/2$&$0$&$1/2$&$3/2$&$0$\\
$2$&$3/2$&$1/2$&$1$&$3/2$&$1/2$&$-3/50$&$1$&$3/2$&$5/2$&$0$&$1/2$&$3/2$&$0$\\
$2$&$3/2$&$5/2$&$1$&$3/2$&$5/2$&$363/3430$&$2$&$5/2$&$3/2$&$1$&$3/2$&$1/2$&$-3/50$\\
$2$&$3/2$&$5/2$&$1$&$3/2$&$7/2$&$12/343$&$2$&$5/2$&$3/2$&$1$&$3/2$&$5/2$&$21/250$\\
$2$&$3/2$&$7/2$&$1$&$3/2$&$5/2$&$48/8575$&$2$&$5/2$&$5/2$&$1$&$3/2$&$5/2$&$4356/42875$\\
$2$&$3/2$&$7/2$&$1$&$3/2$&$7/2$&$-288/8575$&$2$&$5/2$&$5/2$&$1$&$3/2$&$7/2$&$288/8575$\\
$3$&$5/2$&$3/2$&$2$&$5/2$&$3/2$&$16/375$&$2$&$5/2$&$7/2$&$1$&$3/2$&$5/2$&$-108/1715$\\
$3$&$5/2$&$3/2$&$2$&$5/2$&$5/2$&$-36/875$&$2$&$5/2$&$7/2$&$1$&$3/2$&$7/2$&$648/1715$\\
$3$&$5/2$&$5/2$&$2$&$5/2$&$3/2$&$112/1125$&$2$&$5/2$&$9/2$&$1$&$3/2$&$7/2$&$0$\\
$3$&$5/2$&$5/2$&$2$&$5/2$&$5/2$&$-146689/2315250$&$2$&$3/2$&$1/2$&$1$&$1/2$&$3/2$&$-3/50$\\
$3$&$5/2$&$5/2$&$2$&$5/2$&$7/2$&$512/15435$&$2$&$3/2$&$5/2$&$1$&$1/2$&$3/2$&$-3/70$\\
$3$&$5/2$&$7/2$&$2$&$5/2$&$5/2$&$-18/1715$&$3$&$5/2$&$3/2$&$2$&$3/2$&$1/2$&$-8/75$\\
$3$&$5/2$&$7/2$&$2$&$5/2$&$7/2$&$40328/6251175$&$3$&$5/2$&$3/2$&$2$&$3/2$&$5/2$&$8/105$\\
$3$&$5/2$&$7/2$&$2$&$5/2$&$9/2$&$-154/18225$&$3$&$5/2$&$5/2$&$2$&$3/2$&$5/2$&$3844/15435$\\
$3$&$5/2$&$9/2$&$2$&$5/2$&$7/2$&$5632/127575$&$3$&$5/2$&$5/2$&$2$&$3/2$&$7/2$&$36481/154350$\\
$3$&$5/2$&$9/2$&$2$&$5/2$&$9/2$&$16129/80190$&$3$&$5/2$&$7/2$&$2$&$3/2$&$5/2$&$-1024/5145$\\
$3$&$5/2$&$9/2$&$2$&$5/2$&$11/2$&$36/275$&$3$&$5/2$&$7/2$&$2$&$3/2$&$7/2$&$95048/138915$\\
$3$&$5/2$&$11/2$&$2$&$5/2$&$9/2$&$16/2475$&$3$&$5/2$&$9/2$&$2$&$3/2$&$7/2$&$-55/1134$\\
$3$&$5/2$&$11/2$&$2$&$5/2$&$11/2$&$-832/7425$&&&&&&&\\
\hline
\end{tabular}
\end{table}

% LaTeXTable style ElsartTiny
\begin{table}
\caption{Angular reduced matrix elements $\langle J'\lVert (a^\dag\times\tilde{a})^{(1)}\rVert J\rangle$ between $\grpspin{5}$ coupled states, for $\tau_1\<\leq5/2$. For reduced matrix elements of $\hat{j}$, multiply by $-\sqrt{5}$.   A square root sign is to be understood over the magnitude of each matrix element.}
\label{TabCoupledRMEParticle1}
\vspace{1ex}
\tiny
\begin{tabular}{rrrrrrr@{\qquad}rrrrrrr}
\hline
$\tau$&$\tau_1$&$J$&$\tau'$&$\tau_1'$&$J'$&RME&$\tau$&$\tau_1$&$J$&$\tau'$&$\tau_1'$&$J'$&RME\\
\hline
$0$&$1/2$&$3/2$&$0$&$1/2$&$3/2$&$-3$&$2$&$3/2$&$7/2$&$2$&$3/2$&$7/2$&$-18/1715$\\
$1$&$3/2$&$1/2$&$1$&$3/2$&$1/2$&$3/10$&$3$&$5/2$&$3/2$&$3$&$5/2$&$3/2$&$1/75$\\
$1$&$3/2$&$5/2$&$1$&$3/2$&$5/2$&$-507/350$&$3$&$5/2$&$3/2$&$3$&$5/2$&$5/2$&$112/225$\\
$1$&$3/2$&$5/2$&$1$&$3/2$&$7/2$&$48/35$&$3$&$5/2$&$5/2$&$3$&$5/2$&$3/2$&$-112/225$\\
$1$&$3/2$&$7/2$&$1$&$3/2$&$5/2$&$-48/35$&$3$&$5/2$&$5/2$&$3$&$5/2$&$5/2$&$6241/9450$\\
$1$&$3/2$&$7/2$&$1$&$3/2$&$7/2$&$-162/35$&$3$&$5/2$&$5/2$&$3$&$5/2$&$7/2$&$8/63$\\
$2$&$5/2$&$3/2$&$2$&$5/2$&$3/2$&$-3/25$&$3$&$5/2$&$7/2$&$3$&$5/2$&$5/2$&$-8/63$\\
$2$&$5/2$&$3/2$&$2$&$5/2$&$5/2$&$144/175$&$3$&$5/2$&$7/2$&$3$&$5/2$&$7/2$&$-4418/25515$\\
$2$&$5/2$&$5/2$&$2$&$5/2$&$3/2$&$-144/175$&$3$&$5/2$&$7/2$&$3$&$5/2$&$9/2$&$616/3645$\\
$2$&$5/2$&$5/2$&$2$&$5/2$&$5/2$&$243/17150$&$3$&$5/2$&$9/2$&$3$&$5/2$&$7/2$&$-616/3645$\\
$2$&$5/2$&$5/2$&$2$&$5/2$&$7/2$&$72/343$&$3$&$5/2$&$9/2$&$3$&$5/2$&$9/2$&$6241/16038$\\
$2$&$5/2$&$7/2$&$2$&$5/2$&$5/2$&$-72/343$&$3$&$5/2$&$9/2$&$3$&$5/2$&$11/2$&$784/495$\\
$2$&$5/2$&$7/2$&$2$&$5/2$&$7/2$&$-59858/15435$&$3$&$5/2$&$11/2$&$3$&$5/2$&$9/2$&$-784/495$\\
$2$&$5/2$&$7/2$&$2$&$5/2$&$9/2$&$88/315$&$3$&$5/2$&$11/2$&$3$&$5/2$&$11/2$&$13/1485$\\
$2$&$5/2$&$9/2$&$2$&$5/2$&$7/2$&$-88/315$&$1$&$3/2$&$1/2$&$1$&$1/2$&$3/2$&$6/5$\\
$2$&$5/2$&$9/2$&$2$&$5/2$&$9/2$&$-289/198$&$1$&$3/2$&$5/2$&$1$&$1/2$&$3/2$&$-42/25$\\
$2$&$5/2$&$9/2$&$2$&$5/2$&$11/2$&$144/55$&$2$&$5/2$&$3/2$&$2$&$3/2$&$1/2$&$6/5$\\
$2$&$5/2$&$11/2$&$2$&$5/2$&$9/2$&$-144/55$&$2$&$5/2$&$3/2$&$2$&$3/2$&$5/2$&$6/7$\\
$2$&$5/2$&$11/2$&$2$&$5/2$&$11/2$&$-351/55$&$2$&$5/2$&$5/2$&$2$&$3/2$&$5/2$&$-576/343$\\
$1$&$1/2$&$3/2$&$1$&$1/2$&$3/2$&$-3/25$&$2$&$5/2$&$5/2$&$2$&$3/2$&$7/2$&$-3042/1715$\\
$2$&$3/2$&$1/2$&$2$&$3/2$&$1/2$&$3/10$&$2$&$5/2$&$7/2$&$2$&$3/2$&$5/2$&$-432/343$\\
$2$&$3/2$&$5/2$&$2$&$3/2$&$5/2$&$3/686$&$2$&$5/2$&$7/2$&$2$&$3/2$&$7/2$&$128/343$\\
$2$&$3/2$&$5/2$&$2$&$3/2$&$7/2$&$240/343$&$2$&$5/2$&$9/2$&$2$&$3/2$&$7/2$&$-22/7$\\
$2$&$3/2$&$7/2$&$2$&$3/2$&$5/2$&$-240/343$&&&&&&&\\
\hline
\end{tabular}
\end{table}

\clearpage

% Bibliography created with apsrevm.bst
\providecommand{\ELSEVIER}{}
\ELSEVIER\newcommand{\identity}[1]{{#1}}

%bibliography{apsrevm_elsevier,master,theory,expt,data,misc,mc,geomsuper2_special,books,proc}

\end{document}